**Please cite as\*:**
Dilley, L., Welna, W. & Foster, F. (2021/2022). QAnon Propaganda on Twitter as Information Warfare: Influencers, Networks, and Narratives. Accepted Sept. 16, 2021 at *Frontiers in Communication*, 6:707595, doi: 10.3389/fcomm.2021.707595 [Archived Oct. 23, 2021 at https://web.archive.org/web/20211023213819/https://www.frontiersin.org/articles/10.3389/fcomm.2021.707595/abstract ]. *ARXIV*. Retrieved on <date> from <URL>.

# QAnon Propaganda on Twitter as Information Warfare: Influencers, Networks, and Narratives

*Laura Dilley \*, William Welna and Faith Foster*

*Department of Communicative Sciences and Disorders, College of Communication Arts and Sciences, Michigan State University, East Lansing, MI, United States*

QAnon refers to a set of far-right, conspiratorial ideologies that have risen in popularity in the U.S. and globally since their initial promotion in 2017 on the 4 chan Internet message board. A central narrative element of QAnon has been that a powerful group of elite, liberal members of the Democratic Party engage in morally reprehensible practices, but that former U.S. President Donald J. Trump was prosecuting them. Five studies investigated the influence and network connectivity of accounts promoting QAnon propaganda on Twitter from August 2020 through January 2021. Selection of Twitter accounts emphasized online influencers and "persons-of-interest" known or suspected of participation in QAnon propaganda promotion activities. Evidence of large-scale coordination among accounts promoting QAnon propaganda was observed, demonstrating the first rigorous, quantitative evidence of "astroturfing" in QAnon propaganda promotion on Twitter, as opposed to strictly "grassroots" activities of citizens acting independently. Furthermore, evidence was obtained supporting that networks of extreme far-right adherents engaged in organized QAnon propaganda promotion, as revealed by network overlap among accounts promoting 1) far-right extremist (e.g., anti-Semitic) content and insurrectionist themes; 2) New Age, occult, and "esoteric" themes; and 3) Internet puzzle games like Cicada 3301 and other "alternative reality games." Based on well-grounded theories and empirical findings from the social sciences, it is argued that QAnon propaganda on Twitter in the months preceding the 2020 U.S. Presidential election likely reflected joint participation of multiple actors, including nation-states like Russia and corporate entities, in innovative misuse of social media toward undermining democratic processes by promoting "magical" thinking, ostracism of Democrats and liberals, and salience of white extinction narratives that are common among otherwise ideologically diverse groups on the extreme far-right.

# INTRODUCTION

Conspiracy theories have long played a role in U.S. politics (Hofstadter, 1964; Uscinski, 2017); however, such theories' influence on discourse in politics has become increasingly visible and divisive in recent years. A bizarre conspiracy that emerged during the 2020 U.S. presidential election cycle has become known as "QAnon," which refers to a set of conspiratorial ideologies especially popular on the far-right that was first promoted on the 4 chan message board starting in 2017 (Amarasingam &

\*See F. Grassi (2022) for more information.

Argentino, 2020; Cosentino, 2020; Aliapoulios et al., 2021). A core narrative of QAnon is that a powerful group of elite, Satan-worshipping Democrats and liberals engage in practices such as pedophilia and cannibalism, but that this "Deep State" cabal was being secretly prosecuted by Donald Trump (Friedberg and Donovan, 2020; Vrzal, 2020; Miller, 2021).

This paper presents five studies on QAnon propaganda's spread through the social media platform Twitter circa the 2020 U.S. Presidential election. These studies provide one of the first quantitative treatments of QAnon's spreading on social media, thereby complementing qualitative journalistic treatments available to date. Our results reveal the complexity of the media information and narrative environments in which this propaganda proliferated. Notably, we demonstrate clear, rigorous quantitative evidence that a sophisticated digital "astroturfing" campaign aimed at promoting QAnon propaganda and influencers on Twitter was well underway by 2020. We further identify multiple lines of evidence of Russian involvement QAnon propaganda promotion on Twitter.

## Historical Context and Spread of QAnon Conspiracy Narratives: Pizzagate and the Trump-Russia Scandal

QAnon's central narratives are rooted in anti-Semitic tropes (Greenspan, 2020; Vrzal, 2020), as well as the so-called "Pizzagate" conspiracy (Robb, 2017), which has been called "one of the most absurd and creative fictional political narratives circulating during the 2016 U.S. presidential election cycle" (Cosentino, 2020). Presaging themes of QAnon, the 2016 Pizzagate conspiracy alleged that a Washington, D.C.-area pizzeria was the operational base of a ring of pedophiles which included prominent Democratic Party members like Hillary Clinton and John Podesta, who supposedly also engaged in Satanic rites and cannibalism (Robb, 2017; Tuters et al., 2018; DiResta et al., 2019; Cosentino, 2020). QAnon entailed the added narrative element that Donald Trump was a hero who was secretly combating the "Deep State" by working to save abused children and to punish pedophile Democrats in a coming day of reckoning known as "the Storm".

Events around the 2016 election and 2020 attempted reelection of Donald Trump to the U.S. presidency are central to understanding narratives of QAnon and Pizzagate and how they spread. The actions of Russia and Russian operatives are especially relevant since sophisticated Russian influence campaigns were underway in both these elections (Mueller report; DiResta et al., 2019; Golovchenko et al., 2020; Lukito, 2020; Smith et al., 2021; Watts, 2018). These influence operations targeted multiple social media platforms, including Twitter, YouTube, and others (Golovchenko et al., 2020), as well as organizations such as the National Rifle Association (Fandos, 2017). Russian Twitter accounts began boosting QAnon in November 2017, less than a month after the first "Q drop"—a term referring to the anonymous 4 chan posts promoting QAnon narratives allegedly made by a government insider with a security clearance who became known as "Q." Russian Twitter accounts also interacted with and promoted YouTuber Tracy "Beanz"

Diaz, an early QAnon conspiracy publicist (Zadrozny & Collins, 2018), as early as April 2017 (Menn, 2020b). Further, a Russian spear-phishing attack was responsible for exfiltration of emails of Clinton campaign manager John Podesta, which provided raw narrative elements (e.g., discussions of pizza) for Pizzagate/QAnon; these were distributed by Wikileaks, a crime for which founder Julian Assange was later jailed. (Note that, a counternarrative about how Wikileaks obtained Democratic National Committee (DNC) emails was the "Seth Rich" conspiracy; this conspiracy alleged that Seth Rich, a DNC staffer who had been found murdered in an apparent robbery, had instead leaked the emails, as opposed to Russian hackers obtaining them. The Seth Rich conspiracy was promoted by prominent Republicans, including wealth manager Ed Butowsky, as well as Fox News, which later retracted the associated story.)

The present studies focused on influencers in spreading QAnon propaganda, a broad range of whom advanced QAnon's narrative reach from 2016 onward on Twitter and other platforms. One such influencer was Trump himself, who promoted the narrative of a "Deep State" and refused to denounce QAnon narratives (Dickson, 2020). Other Trump affiliates advancing QAnon included: Trump's former National Security Advisor Michael Flynn (Sollenberger, 2020); Roger Stone, a longtime Trump friend and former campaign manager who advanced the Pizzagate, Seth Rich, and QAnon conspiracy narratives; and Jason Sullivan, a Stone employee and creator of a Twitter botnet that boosted Republican propaganda (Porter, 2020). Multiple former U.S. government officials, some with ties to intelligence, also advanced Pizzagate and QAnon, including Bill Binney, Robert David Steele, and Tony Shaffer (Kamouni, 2017; Hananoki, 2020; HBO, 2021).

Multiple far-right media outlets also advanced QAnon and Trump, including the Epoch Times[1], One America News Network (Palmer, 2021), and InfoWars (Kamouni, 2017). Other neo-fascist and/or militia organizations promoting Trump and QAnon included the all-male organization Proud Boys and the Oath Keepers (Dougherty, 2018; Polantz, 2021). These narratives culminated in January 6, 2021, violent insurrection at the U.S. Capitol, where not only were the Proud Boys and Oath Keepers heavily represented but also many QAnon supporters were attested (Stahl, 2021; Tollefson, 2021).

## Disinformation, the Internet, and Digital Astroturfing

QAnon represents true disinformation, to the extent that it was spread knowingly by as-yet largely unknown actors within a larger campaign structure (Starbird, 2019). Terms like disinformation, propaganda, and information warfare all reflect the ability of the information realm to impose interpretations and projections of a physical realm and its supposed "realities" onto the cognitive realm of a target audience (Simons, 2020). The ability of QAnon propaganda to influence targets to adopt geopolitical stances preferred by disinformation actors makes it a sophisticated form of strategic communication that is especially potent when paired with the Internet (Maréchal, 2017; see also; Carr, 2016; Price, 2015). As such, online activities have become a new arena for political dominance and soft power (Simons, 2020). The single greatest development in information warfare, and conflict in general, is argued to be the growth of *cyber*, the so-called "Fifth Domain" of warfare, after land, sea, air, and space (Scott, 2020). New uses of social media for propaganda, deceptive persuasion, and information warfare are an active area of research (Bakir et al., 2019; Bakir, 2020; Scott, 2020).

---

[1]The Rachel Maddow Show, Aug. 20, 2019. https://www.msnbc.com/rachel-maddow/watch/pro-trump-conspiracy-theories-pay-off-for-anti-china-group-66712645854

The rise of the Internet has made more potent a form of disinformation known as political astroturfing, which is defined as a centrally coordinated disinformation campaign in which participants pretend to be ordinary citizens acting independently as part of "grassroots" campaigns. Digital astroturfing on the Internet is a newer phenomenon which is more effective in achieving campaign coordinators' goals and in spreading disinformation (Kovic et al., 2018; Zerback et al., 2021). While QAnon has been previously claimed by some journalists to be a fully grassroots movement, this study presents evidence that a significant level of QAnon activity represented a digital astroturfing effort involving central, topdown orchestration intended to give the appearance of broader support.

The above suggests what makes a campaign "disinformation" might pertain to the *false impression of independent popular support*, consistent with some accounts spreading disinformation being non-genuine actors (Paletz et al., 2019; Keller et al., 2020). Identifying astroturfing campaigns is challenging without ground truth information; Keller et al. (2020) provide evidence that coordinated group activity is a robust hallmark of astroturfing. They note that past research's preeminent focus on automated accounts misses its target because reports indicate that astroturfing campaigns are at least partially run by actual humans. Keller et al. (2020) argue that "similar behavior among a group of managed accounts is a stronger signal of a disinformation campaign than 'bot like' individual behavior" (p. 257). The methodology used in our studies prioritized discovery of digital astroturfing and presented an important complement to studies that involve larger-scale datasets but which typically do not consider nuances of social constructs like prestige. (See Jackson et al., 2021.; Smith, 2020 for some initial unpublished, non-peer-reviewed findings using such large-scale approaches.) Previewing our findings, our studies of QAnon Twitter accounts reveal evidence of substantial coordination across accounts, consistent with political astroturfing.

## Fascism and the Alt-Right in QAnon: Connections to Metaphysics/the Occult, "Magical Thinking", and Alternative Reality Games

QAnon and Pizzagate have been especially popular both among people affiliated with the American ethnonationalist far-right (the so-called "Alt-Right"; Cosentino, 2020), as well as among evangelical Christians (Argentino, 2020; General & Naik, 2021). Across the Alt-Right, the closely related "Groyper" movement, and the Euro-American far-right more broadly, overarching narratives of fear of white extinction and white replacement are common (Cosentino, 2020; Bhatt, 2021). Bhatt emphasizes that these narratives are associated with "metaphysical themes that are deployed in contemporary fascism [including] occultist ideas of nature and vitalism … and ideas about cosmic destiny" (p. 27).

The potential of metaphysical and occultist themes to signal contemporary fascist ideology provided motivation for examining the juxtaposition of such elements with QAnon propaganda. Consistent with Bhatt (2021) and Edel (2021a, b), we observed early in investigations of QAnon on Twitter an apparent juxtaposition of QAnon content with various metaphysical themes, New Age ideas, and occultism. Further, Scott (2020) proposed that the use of metaphysical themes and occult content on social media might reflect novel information warfare strategies, further motivating the study of such elements and their relation to QAnon on Twitter. In particular, Scott proposed that employing "magical" deception tactics involving metaphysical and occult content has occasionally been used historically in warfare tactics. As such, encouraging "magical" thinking and mindsets through magic-inspired deception tactics grounded in cognitive psychology is a new way for nation-states to exert tech-enabled soft power (cf. Hutchinson, 2006; Kuhn et al., 2008; Pailhès et al., 2020).

Finally, our studies were motivated by connections reported between QAnon and so-called "alternative reality games" (ARGs; Kinsella, 2011; Kristiansen, 2014; Andjelkovic, 2021) and "live action role play" (i.e., LARPs; Tuters, 2019). Individuals who promoted Internet puzzle and ARG Cicada 3301, which has been called the "Web's greatest mystery" (Kushner, 2015), were involved in early promotion of QAnon narratives; individuals implicated in this regard include Thomas Schoenberger and his associate Lisa Clapier (Bicks, 2021; Edel 2021a, b; Vice News, 2021). Schoenberger and Clapier promoted the puzzle, organized groups of puzzle solvers and content creators, and groomed and enlisted some participants to post as Q. Clapier was a QAnon influencer, promoting QAnon heavily from the Twitter account @SnowWhite7Iam and the website QSuperHeroes.com. @SnowWhite7IAm also promoted the QAnon-related hashtags #FollowTheWhiteRabbit and #FollowSnowWhite from early 2018. The third area for recruitment is alluded to by Clapier's @SnowWhite7IAM Twitter handle: the fascist "I Am" theosophy cult of Guy and Edna Ballard (Whitsel, 1989) with which both Schoenberger and Clapier are connected[2]. This cult is known to promote fascist ideology and promote brainwashing techniques. The connection to the "I Am" theosophy cult thus constitutes a further connection to far-right fascism relevant to understanding QAnon narratives and who spread them.

---

[2]The patron saint of the "I Am" cult is St. Germain; Thomas Schoenberger has repeatedly favored this as a self-appellation (Edel, 2021a; Edel, 2021b)

## The Present Studies

In these studies, we constructed rich Twitter datasets using specific account inclusion criteria, toward investigating the hypotheses outlined above and informing understanding of influential actors promoting QAnon propaganda and their networks. Study 1 was an initial investigation of QAnon influencer accounts that aimed to produce quantitative data on these accounts' influence and connectivity. Study 2 was a larger investigation of over 1,000 accounts that tested the hypothesis that QAnon themes were systematically juxtaposed with New Age and "esoteric" themes and tested for evidence of astroturfing manifesting as coordination among accounts (cf. Keller et al., 2020). Study 3 used the case of an anti-Semitic, extremist account to test the extent to which accounts promoting or interacting with this extremist also promoted QAnon content and/or were coordinated with one another. Finally, Studies 4 and 5 presented separate Twitter samples of networks around accounts associated with QAnon influencers discussed in the media, toward systematically investigating influences of the far-right in QAnon Twitter networks. These studies further tested the extent to which QAnon-related Twitter accounts were networked with accounts espousing interest in alternative reality games, such as Internet puzzle Cicada 3301 (Andjelkovic, 2021).

## STUDY 1

Study 1 was an initial study of QAnon-related Twitter propaganda accounts, focusing on "influencers," broadly defined, and their networks (Mattan et al., 2017; Sohn and Choi, 2019). We sought to obtain a baseline characterization of dimensions of influence of selected "influencer" accounts for further study, as well as provide initial data about these select online QAnon actors, toward hypothesizing means and motives for promoting QAnon. We predicted that many pro-QAnon accounts would show Twitter behaviors inconsistent with grassroots activities but consistent with astroturfing (Keller et al., 2020). Further, we tested whether QAnon-related propaganda activity on Twitter was consistent with boosting the political standing of Donald Trump and his party and/or denigrate rivals.

## MATERIALS AND METHODS

### Selection of Twitter Accounts

Study 1 involved selecting 128 Twitter accounts according to specific *inclusion criteria* that were developed based on observations of QAnon-related Twitter activities from May through August 2020, supplemented by information about QAnon from journalistic sources. Throughout the study, additional details such as technical network data, pilot study results, and/or additional selection criteria can be found in the Supplemental Material document. **Supplementary Table S1** lists accounts which were foci of selection criteria across studies.

Study 1 inclusion criteria prioritized identification and discovery of account and network characteristics of select candidate "QAnon Twitter influencers" for initial quantitative study toward informing later study analyses and procedures. In this and other studies, we were interested especially in accounts for which convergent evidence from media sources suggested a tie to real-life QAnon promotion and/or means of conducting or furthering astroturfing campaigns, such as connections to wealthy or powerful individuals, corporations (especially media companies), and/or governments, with the goal of refining historical understanding of the QAnon campaign. Our selection approach brought to bear an informed social science perspective to develop criteria for studying QAnon account influence and prestige which considered the multi-faceted, dynamic nature of these constructs (Marin and Wellman, 2011; Mattan et al., 2017). The *inclusion criteria* for Study 1 were as follows: 1) The account routinely promoted QAnon hashtags (e.g., #QAnon, #WWG1WGA, #SaveTheChildren, and/or cryptocurrency and finance-related tags often observed within QAnon-related accounts like #XRP, #NESARA, #GESARA) and/or QAnon themes (e.g., depictions of Democrats as Satan-worshippers) in its bio or tweets while also showing evidence of online influence via one or more metrics (e.g., high follower counts or frequently being retweeted, or evidence of self-promotion, such as linking a YouTube channel or TikTok videos featuring the user); 2) The account displayed some QAnon involvement via its own content or via retweeting, or being retweeted by, QAnon-themed accounts while also depicting eye-catching visual content, especially magical, esoteric, or New Age themes, that seemed designed to catch attention and was hypothesized to have used professional graphics preparation; 3) The account was associated with a person discussed in mainstream and/or social media for being known or suspected to have furthered or promoted QAnon (e.g., "primary accounts"; see below); and/or 4) The account was mentioned, retweeted by, and/or followed by an account matching 1), 2), and/or 3) while also itself meeting at least one of criteria (1)–(3).

Finally, a subset of 15 selected accounts was designated post hoc to be *primary accounts* of particular interest for judged potential to shed light on QAnon propaganda means, methods, and motives (described in **Supplementary Tables S2 and S6**) for additional exploratory analyses. The primary accounts were as follows: @JasonSullivan_, @RQueenInc, @CodeMonkeyZ, @GrrrGraphics, @OSSRobertSteele, @ETheFriend, @OzRevealed, @3Days3Nights, @TrustThePlan_, @SnowWhite7IAm, @VeteransAlways_, @AmborellaWWG1, @MrDeeds1111, and @JudithRose91 (see **Supplementary Tables S2, S6**). Several accounts were associated with individuals interviewed for QAnon documentaries, while the others displayed continuous QAnon and/or QAnon-centric narrative promotion activities, while also being heavily followed and/or retweeted and/or engaged in recruitment (e.g., "follow trains"), viral video promotion, and/or organizing of other users for QAnon. All primary accounts except @OSSRobertSteele had over 5k followers at the time.

### Data Acquisition and Analysis

The Twitter Application Programming Interface (API) was accessed using NodeXL Pro v. 1.0.1.439 (Smith et al., 2010) on

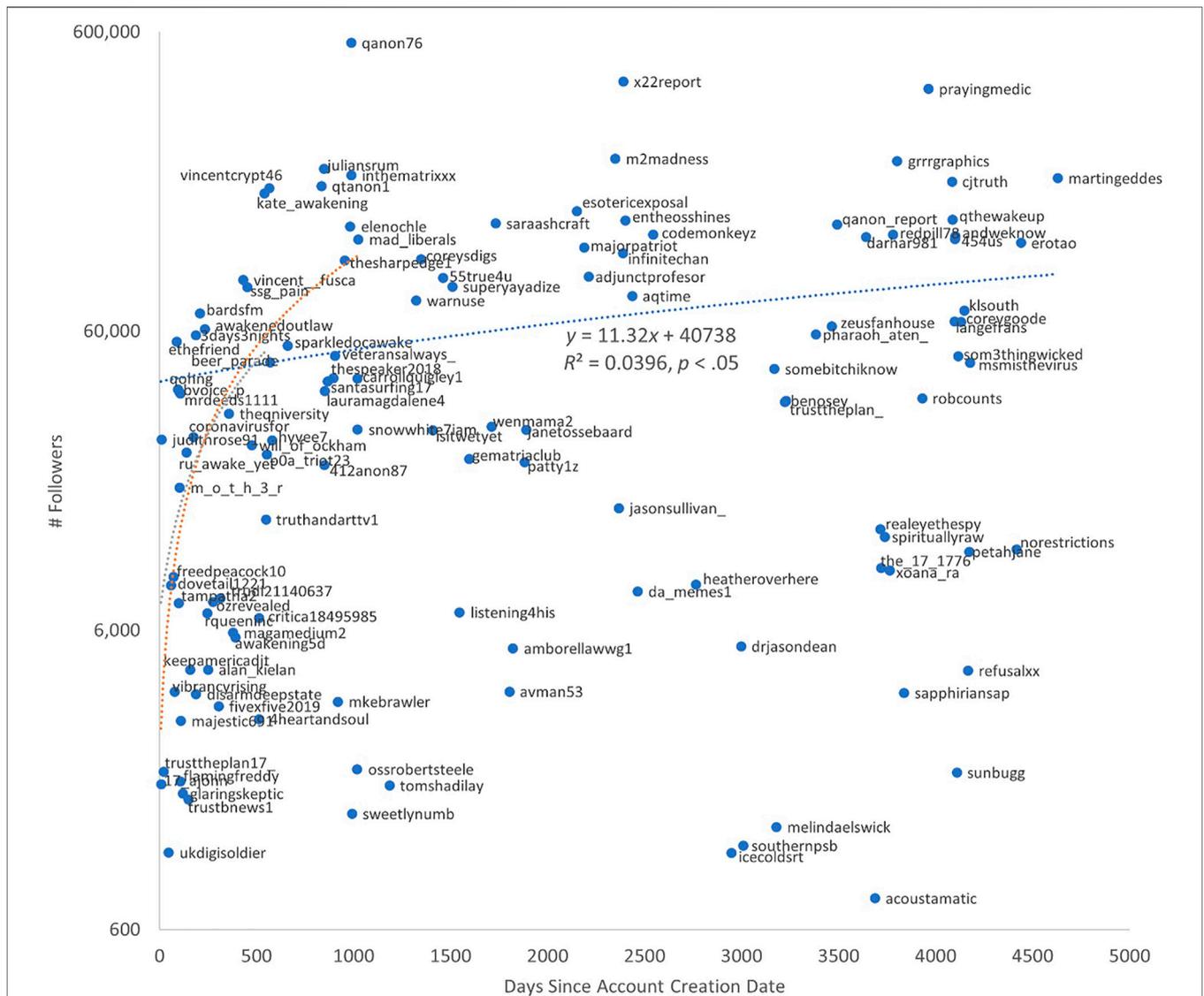

**FIGURE 1** | Number of followers vs account age (in days) for Study 1 based on $N = 128$ accounts. Accounts with fewer than 600 followers are not shown ($N = 5$). Number of followers is shown on a logarithmic scale. Linear best-fit lines show as curved for the log-linear scale. The blue dotted line shows the best-fitting equation based on all 128 accounts. The orange line plots $y = 103.1x + 2,108$, which is the best-fitting line for accounts created within the prior 3 years, $r(61) = 0.43$, $p < 0.001$. The grey line plots the equation $y = 81.9x + 6,683$, which is the best-fitting line for accounts created within the prior 18 months, $r(37) = 0.37$, $p < 0.05$.

August 25, 2020, to download and analyze public Twitter data from the 128 selected accounts. Account property documentation was supplemented with screenshots and web archiving software (Hunchly v. 2.2.2 and Internet archive sites).

Given that there is no single means of assessing Twitter account influence that applies in all situations (Sutton et al., 2015; Ben Gibson et al., 2020), we aimed to provide quantitative data on multiple influence metrics for these QAnon-related accounts. We focused on identifying anomalies relative to typical real-world influence patterns. In particular, we calculated *account statistics* (i.e., distributions of number of followers, follower acquisition rate, follower:followed ratio, and tweet activity/rate), *eigenvector centrality* (a measure of an account's influence on its neighbors within a single network), *betweenness centrality* (a measure of the extent to which an account serves as a "bridge" between putative subnetworks through its inclusion in shortest paths to other nodes in network samples), and *network top items* (e.g., top 10 most frequently mentioned accounts and hashtags).

## Results

### Account Influence Metrics

We first considered results for multiple metrics of account influence (see also **Supplementary Table S4**)[3].

---

[3]Twitter data from all five studies may be found at https://osf.io/rztes.

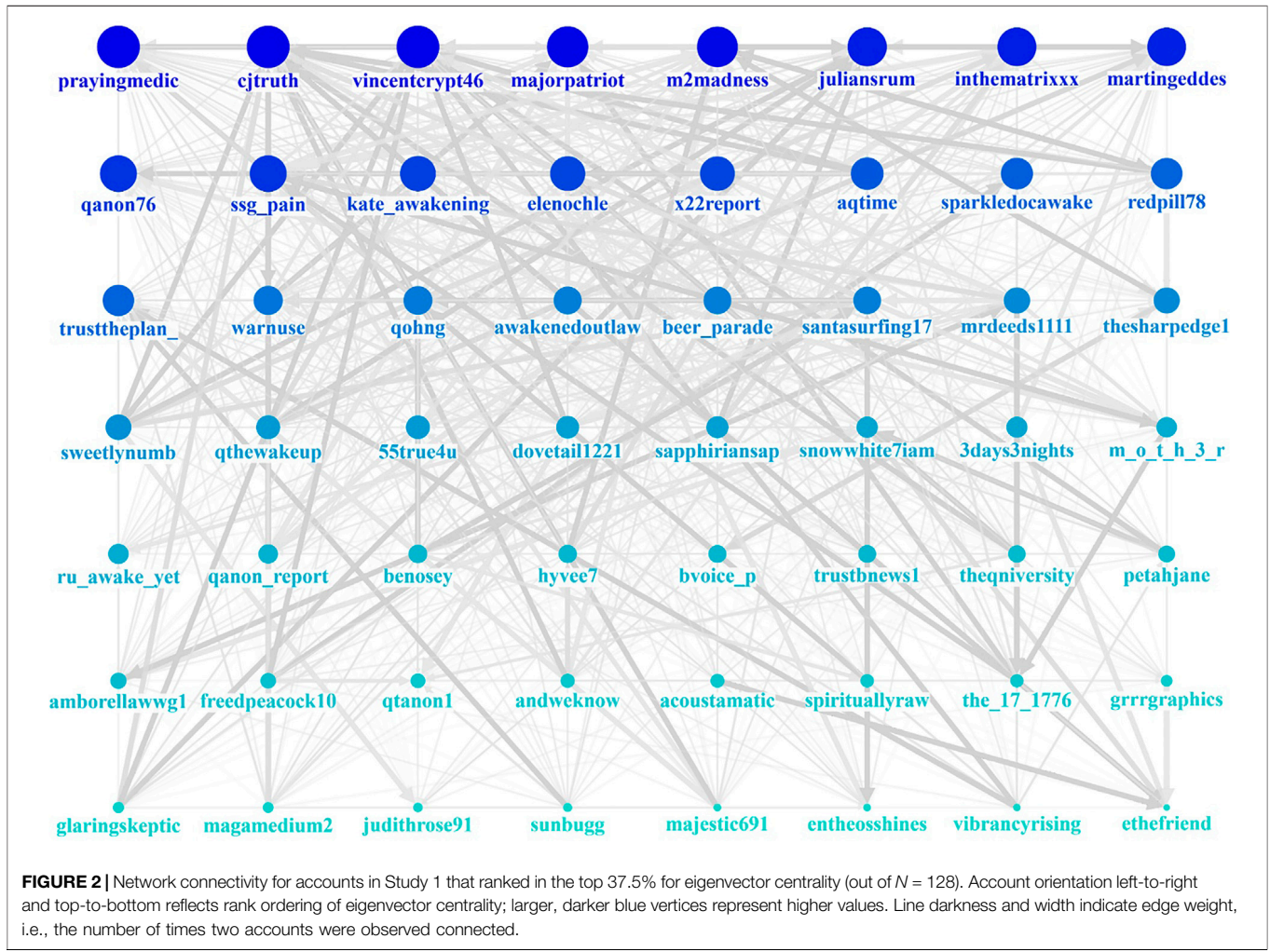

**FIGURE 2** | Network connectivity for accounts in Study 1 that ranked in the top 37.5% for eigenvector centrality (out of N = 128). Account orientation left-to-right and top-to-bottom reflects rank ordering of eigenvector centrality; larger, darker blue vertices represent higher values. Line darkness and width indicate edge weight, i.e., the number of times two accounts were observed connected.

1) Number of followers. Accounts had a mean of M = 60,486 followers (range: 110–552,787; median = 28,197). Accounts in the top 10 had over 150,000 followers. One account, @QAnon76, had over 500,000 followers.
2) Follower acquisition rate. Follower acquisition rate was calculated from the number of followers and the account age (i.e., number of followers/days since account creation). Accounts had been active for an average of 1,745 days (range: 7–4,628; median = 1,148). Accounts acquired a mean of 92 followers/day (range: 0.1–2,521; median = 28). Among accounts acquiring followers the fastest were several of "primary" interest, including @JudithRose91, who gained 2,521 followers/day; @ETheFriend, who gained 637 followers/day; @MrDeeds1111, who gained 371 followers/day; and @3Days3Nights, who gained 313 followers/day. **Figure 1** shows that, overall, numbers of followers were significantly correlated with account age, $r$ (126) = 0.20, $p$ = 0.024. This relationship was stronger for accounts less than 3 years old, $r$ (61) = 0.43, $p < 0.001$. Notably, for accounts older than 3 years, there was a non-significant relationship between account age and number of followers, $r$ (63) = 0.10, $p$ = 0.47.

3) Follower:followed ratio. The mean follower:followed ratio was 329:1 (range = 0.2:1–17,842:1; median 5.5:1). @X22Report showed the highest ratio at 17,842:1. The next highest ratios were for @InfiniteChan and @IsItWetYet, accounts connected to the 8 chan and 8kun message boards where the elusive "Q" posted messages, as well as to Jim and Ron Watkins—the father/son owner/operator team for those Internet boards. Primary account @ETheFriend was next, followed by @TrustThePlan17_—an account following just one other account, @TrustThePlan_, which was one of 15 primary accounts; @TrustThePlan17_, the apparent backup account for @TrustThePlan_, had gained 2,024 followers in just 20 days since account creation (about 100 followers/day). See **Supplementary Table S4**.
4) Tweet activity. Accounts had tweeted M = 50,113 times (range: 45–616,358; median = 14,368). Accounts in the top 10 had tweeted over 100,000 times.
5) Tweet rate. Accounts tweeted M = 53 tweets/day (range: 0.2–578; median = 14). The top 10 accounts tweeted over 170 times per day. @Tampatha2 was most active, at nearly 600 tweets/day.

6) Eigenvector centrality. **Figure 2** shows accounts top-ranked for eigenvector centrality and their network connections. The top 10 of these were as follows: @PrayingMedic, @CJTruth, @VincentCrypt46, @MajorPatriot, @M2Madness, @JuliansRum, @InTheMatrixxx, @MartinGeddes, @QAnon76, and @SSG_Pain. Accounts were densely interconnected.

We calculated bivariate correlations between pairs of metrics across eigenvector centrality, followers/day, follower:followed ratio, total tweets, and tweets/day. Eigenvector centrality was significantly correlated with number of followers ($r = 0.52$, $p < 0.001$) and followers/day ($r = 0.20$, $p < 0.05$). Also, follower: followed ratio was significantly correlated with number of followers ($r = 0.37$, $p < 0.001$), along with tweets/day vs total tweets ($r = 0.46$, $p < 0.001$). All remaining pairs were non-significant ($p$'s $> 0.05$).

### Network Top Items

QAnon was central to this data, as revealed by hashtags like #QAnon, #SaveTheChildren, and #WWG1WGA ("where we go one, we go all"). Another hashtag referencing QAnon was #Gematria, which refers to numerology. #Unrig refers to a philosophy which QAnon proponent @OSSRobertSteele (Robert David Steele) has advanced for election reform. The hashtag #PushingUpDaisies was used prolifically by a single account, @Tampatha2, in a series of tweets naming a politician or celebrity (e.g., "Barack Hussein Obama", John Podesta), followed by the phrase "Arrested and Executed #PushingUpDaisies." Top mentioned and replied-to accounts also indicated prominence of QAnon, as evidenced by recognizable QAnon influencers being in top-ranked spots—@SSG_Pain, @VincentCrypt46, @JuliansRum, @ETheFriend, and @Elenochle, where the latter two of these were primary accounts; all of these were influential based on eigenvector analyses reported in 2.2.1.

Network top items further confirmed that a focus of QAnon-related Twitter propaganda was to boost Donald Trump and his party and/or denigrate rivals. @realDonaldTrump was the top replied-to and top mentioned account. Other accounts among the top 10 were Trump-affiliated—including @POTUS, @GenFlynn, and @DonaldJTrumpJr—along with Republican party activist @realJamesWoods and Trump opponent @JoeBiden. Further consistent with the hypothesis, top hashtags included #MAGA, #Trump 2020, and #ObamaGate. Finally, several right-leaning media outlets were among the most influential domains, including disinformation outlets like The Gateway Pundit and Breitbart (Bovet & Makse, 2019).

### Network Influence and Account Features of "Primary" Accounts

Considering now features of the 15 primary accounts, as predicted, we found that many of these accounts were highly influential within the entire network. Using the metric of eigenvector centrality, the most influential primary accounts were all in the upper quartile: @Elenochle (91st percentile), @TrustThePlan_ (87th percentile), @MrDeeds1111 (83rd percentile), and @SnowWhite7IAm (77th percentile). At the other end of the spectrum were @JasonSullivan_ (30th percentile), @OzRevealed (13th percentile), and @OSSRobertSteele (sixth percentile).

We recalculated network statistics specifically for the 15 primary accounts (see Figure S1, Supplementary Materials). Illustrating the distinctive pattern of network influence among accounts in this smaller "insider" network, eigenvector centrality rank for the 128-account analysis did not predict that for the 15-account primary network analysis, $r(13) = 0.21$, $p > 0.05$. Within the network of 15 primary accounts, @ETheFriend ranked as the most influential, followed by @JasonSullivan_, @CodeMonkeyZ, @OzRevealed, @RQueenInc, and @GrrrGraphics.

## DISCUSSION

This sample of 128 QAnon-related accounts revealed multiple remarkable features. A substantial proportion of accounts—23%—had over 100,000 followers, a fact more noteworthy due to the typically young age of accounts. For accounts younger than 3 years, there was a strong correlation between account age and numbers of followers. However, for accounts older than 3 years, account age did not significantly predict the number of followers. This provides some initial evidence of astroturfing within the QAnon propaganda campaign.

Further consistent with astroturfing, many accounts gained followers at astonishing rates. For instance, @JudithRose91 had been in existence for 10 days and in that time had gained ~2,500 followers/day. This account had been selected as an account of primary interest in part for posting "slick," edited viral TikTok videos of herself and for retweeting influential QAnon accounts like @TrustThePlan_ and @ETheFriend. Across our sample, numbers of followers typically dwarfed numbers of followed accounts, with the mean ratio (329:1) being higher than the still-remarkable median (5.5:1). Even the lowest quartile ranked for follower:followed ratio revealed 50% more followers for every account followed, i.e., 1.5:1. @X22Report had the highest ratio, with 17,842 followers for every followed account.

We selected 15 accounts as of "primary" interest as part of conducting an exploratory, post-hoc analysis to clarify metrics of connectivity and influence for just those 15 accounts, toward shedding further light on actors and methods in promotion of QAnon propaganda. Of these 15, @Elenochle was most influential relative to our full network sample, based on eigenvector centrality.

This account further published a "patriot" newsletter and assisted organizing videos submitted during the #TakeTheOath campaign, which former NSA Advisor Michael Flynn participated in (Sollenberger, 2020). @TrustThePlan_ was next highest in influence across the network sample. This account had a notably high follower:followed ratio of 9,471:1[5]. Two accounts displaying "otherworldly" themes and eye-catching visuals, @MrDeeds1111 and @SnowWhite7IAm, were next most influential.

Finally, a network analysis consisting of only 15 primary accounts showed @ETheFriend—a Twitter account that has been discussed in mainstream media—had top-ranked influence; @ETheFriend was less than 90 days old but had acquired 637 followers/day with a follower:followed ratio of 2,518:1. Next most influential was @JasonSullivan_, a Roger Stone associate promoting himself as "The Wizard of Twitter" who operated a network of bots promoting Republican propaganda through Spring of 2020. Ranked next for influence were as follows: @CodeMonkeyZ, account for Ron Watkins, who ran the 8kun message board; @OzRevealed, which was in frequent two-way Twitter conversations with @JasonSullivan_; @RQueenInc, account for 8 chan owner Jim Watkins; and @GrrrGraphics, account for Ben Garrison, a pro-Trump political cartoonist frequently retweeted by @JasonSullivan_. Although former CIA official Robert David Steele has been the subject of media attention for his role in promoting QAnon propaganda, @OSSRobertSteele showed relatively limited influence in this Twitter data.

# STUDY 2

Study 2 aimed to characterize the Twitter behaviors of a larger number and set of types of accounts, toward further elucidating actors and their motives and means for promoting QAnon propaganda. While Study 1 had involved a fairly small sample, Study 2 included over 1,000 accounts of a greater variety of types than in Study 1. We further aimed to replicate and extend results from Study 1 suggesting inauthentic, unnatural, and/or coordinated Twitter account behaviors indicative of astroturfing.

Further, Study 2 tested the hypothesis that QAnon themes would show systematic blending with New Age/esoteric themes. This was predicted on at least two grounds. First, white extinction and white replacement narratives are commonly espoused by Alt-Right adherents and white supremacists, who disproportionately supported QAnon and Trump (Sainudiin et al., 2019; Cosentino, 2020). Fear of white extinction or replacement is further associated with metaphysical, occultist, and New Age themes (Bhatt, 2021), predicting their juxtaposition with QAnon. Second, New Age, esoteric, or occultist ideas and content have been proposed to promote "magical" thinking, to influence decision-making processes, and to generate "illusions of control" (Pailhès et al., 2020; Scott, 2020); Scott further argued that imbuing social media with such content could constitute a novel cyber-information warfare strategy. Finally, we aimed to establish the relative network influence of subsets of different account types (e.g., New Age/esoteric accounts) which were observed to intermix with QAnon Twitter traffic, through "benchmarking" their influence against a common set of QAnon influencers.

## Materials and Methods
### Account Selection

Twitter accounts were first selected in several key categories as follows: 1) *QAnon Influencers*. These were accounts ($N = 56$) recognizable for promotion of QAnon through content and hashtags showing evidence of influence, for instance from Study 1 (e.g., @VincentCrypt46, @CJTruth, @PrayingMedic). 2) *New Age/Esoteric Prototypical accounts*. These were accounts ($N = 23$) foregrounding New Age and/or esoteric "otherworldly" themes, many of which displayed colorful, professional-looking graphics (e.g., @MrDeeds1111, @Kabamur_Taygeta, @PetahJane). 3) *New Age Media Influencers*. These accounts ($N = 25$) corresponded to a person or organization producing and/or promoting New Age, esoteric, and/or disinformation content, e.g., UFO-themed content (e.g., @RobCounts, @Ben_Chasteen, @DavidWilcock). 4) *QAnon Insiders*. These were accounts ($N = 31$) for individuals and organizations known or suspected of previous or ongoing promotion/furtherance of QAnon propaganda, based on convergent evidence (e.g., @Bill_Binney, @StevePieczenik, @JohnMappin, @NameMySock), some of whom are affiliated with far-right organizations (e.g., @JohnBWellsCTM, who has organized for the Oath Keepers). 5) *Prominent Pro-Trump*. These were accounts ($N = 19$) for media personalities and Trump associates that appeared frequently retweeted by QAnon accounts (e.g., @DanScavino, @RudyGiuliani, @ChanelRion). (See **Supplementary Tables S7, S8** for listings of accounts in key categories and **Supplementary Material**, Section 1.2.1 for more details.)

Next, we selected: 1) $N = 24$ accounts promoting Austin Steinbart, a person with a strong social media presence who claimed to be insider "Q" from the future (e.g., @AustinSteinbart, @Michael_Rae, @Tampatha2, @Quixotry9). 2) $N = 7$ accounts associated with Jim and Ron Watkins (@RQueenInc, @CodeMonkeyZ) or individuals and groups they were associated with (e.g., @JoyInLiberty). 3) $N = 11$ accounts judged to be creating and/or promoting pro-Trump meme content (e.g., @GrrrGraphics, @Mad_Liberals, @Mil_Ops); and 4) $N = 5$ accounts associated with or promoting amnesty for Julian Assange and/or Edward Snowden (e.g., @WikiLeaks, @Unity4J).

Finally, approximately 875 additional accounts were selected for QAnon or New/Age esoteric themes and related content (e.g., gematria, #Nesara) through an iterative process of identifying accounts that had at least one connection (through a retweet, reply, mention, or occasionally following relationship) with a prior selected account. This was carried out via a combination of hand-inspection of Twitter account interactions and automated methods using NodeXL, especially through downloading data for searches of accounts of interest to identify accounts connected via high traffic and/or having high eigenvector centrality in the account's network.

---

[5]@TrustThePlan_ followed only two accounts, @realDonaldTrump and @DanScavino, until early June, 2020, following a third, @GenFlynn, shortly thereafter. The account profile suggested themes of military operations; following mid-July, 2020, QAnon account suspensions by Twitter, @TrustThePlan_ began mimicking @ETheFriend.

### Twitter Data Collection and Analysis

Node XL Pro v. 1.0.1.441 was used to download Twitter data on October 24, 2020, limited to 2,000 tweets per user, as well as to analyze network metrics and top items. (See **Supplementary Material**, Section 1.2.4 for an analysis based on retweets only.) To test for potential mixing of QAnon content and themes, e.g., New Age/esoteric themes, sentiment analysis within NodeXL was used. Character strings in accounts' description fields were compared with a set of "keywords" (i.e., words or hashtags) selected to represent political "sentiment" categories: 1) QAnon themes; or 2) New Age and/or esoteric themes; or 3) conservative-leaning identity-related terms, e.g., conveying fundamentalist views (see Supplemental Materials, Section 1.2.2).

### Creation of Data Subsets for Additional Analyses

Subsets of collected Twitter data were created to examine the network influence of each of the four key groups of accounts (New Age/Esoteric Prototypical, New Age Media Influencers, QAnon Insiders, and Prominent Pro-Trump) against the benchmark set of QAnon Influencers. To do so, we combined Twitter data for QAnon Influencers with each other group to create four separate sub-network datasets. For instance, to examine the influence of New Age/Esoteric Prototypical accounts against QAnon Influencers, we used NodeXL to recalculate network measures, "skipping" all other vertices except those for accounts in these two groups. Measures were calculated based on retweets only.

Second, we tested for coordination among accounts by isolating a subset of accounts in this network based on quantitative criteria (see below), then subsequently examined the behavior of these accounts. We first selected accounts ranking in the top 5% for eigenvector centrality across the entire network ($N$ = 53). We then excluded four accounts irrelevant to testing for coordination (@FLOTUS, @DanScavino, @RudyGiuliani, @NYPost), leaving $N$ = 49 accounts. Next, we identified additional accounts with which an account in the first set interacted "frequently," defined as an edge weight greater than or equal to 30; this gave $N$ = 62 more accounts. We then recalculated network statistics for this subnetwork of $N$ = 111 accounts, along with in-degree (how many arrows "point" to that account through, e.g., replies to or retweets of that account) and out-degree (the number of arrows that point out of that account). Finally, we categorized top-10 hashtags for accounts for which they were available as QAnon sentiment, pro-Conservative sentiment, both, or neither.

## Results
### Tweet Clusters

The download yielded Twitter data for $N$ = 1,063 accounts. Applying algorithmic clustering (Clauset-Newman-Moore) revealed nine groups; see **Figure 3**. Accounts were densely interconnected, revealing considerable crossover and interaction between groups; this is revealed by the thick grey lines connecting clusters, which represent aggregated link numbers.

Regarding sentiment analysis, of the >99% of accounts assigned to a cluster, 88% had non-blank text descriptions to which sentiment analysis could be applied. Results showed that 75% of profile descriptions displayed at least one *conservative* keyword; around a third (32%) displayed at least one *New Age/esoteric keyword*; and a quarter (25%) displayed at least one *QAnon* keyword. Many accounts featured a mixing of combinations of these categories of keywords. **Figure 3** further reveals that accounts whose profile descriptions matched QAnon keywords showed considerable interaction and mixing with those featuring New Age/esoteric and/or conservative keywords.

### Network Top Items and Overall Account Influence

QAnon hashtags were among the most frequent in the entire network: #WWG1WGA (ranked first), #QAnon (second), and #TheGreatAwakening (10th) (cf. **Supplementary Table S12**). The popularity of QAnon hashtags varied by group. #WWG1WGA was most popular for Groups 2 and 3. Group 3 also featured prominent use of #AustinSteinbart. Multiple hashtags reflected pro-Trump or anti-Democrat sentiment. Tweets were mainly in English, but Groups 7 and 8 tweeted mostly in Polish and Dutch. **Supplementary Table S13** gives top account information.

Finally, the top 10 most influential accounts across the whole network (cf. for eigenvector centrality), in descending order of influence, were: @VincentCrypt46, @CJTruth, @DanScavino, @RudyGiuliani, @John_F_Kennnedy, @HSRetoucher, @FreedPeacock10, @Neo081001, @DianaAndDennis, and @DanAuito.

### Influence of Key Account Categories Relative to "Benchmark" QAnon Influencer Set

Here, we define to be "high-ranked" those accounts in a key category which were in the top 35% ranked for eigenvector centrality in the network created through combination with the benchmark set of QAnon Influencer accounts. See **Supplementary Figures S3–S6.** Six of 23 New Age/Esoteric Prototypical accounts were high-ranked relative to QAnon Influencers: @MrDeeds1111, @StarSeed7772, @VDarknessF, @AlaraOfSirius, @TwistedSpiral2, and @TimeTravelAnon. Four of 25 New Age/Media Influencers were high-ranked: @NoRestrictions, @RobbyStarbuck, @Ben_Chasteen, and @RobCounts. Four of 19 Prominent Pro-Trump accounts were high-ranked: @DanScavino, @RudyGiuliani, @ChanelRion, and @AnnaKhait. Finally, three of 31 QAnon Insiders were high-ranked: @Saint_Germain5, @Acoustamatic, and @JohnMappin.

### Features and Connectivity Among Subnetwork Accounts

To specifically test the hypothesis of astroturfing, we considered the supplemental analysis related to organization and network statistics for the subgroup of accounts ranked in the top 5% overall for eigenvector centrality plus those connected by heavy traffic (see Method). **Figure 4** plots accounts ranked in the top 50% for eigenvector centrality within this subnetwork. Two accounts identified through our method but not in the upper 50% for eigenvector centrality, @FLOTUS and @jim_jordan, were excluded from further calculations, as they were judged irrelevant to testing for account coordination. Overall, **Figure 4** shows remarkably high interconnectedness among accounts, in spite of the Clauset-Newman-Moore clustering algorithm nominally finding two groups. The dark lines in **Figure 4** further indicate heavy, frequent traffic between accounts.

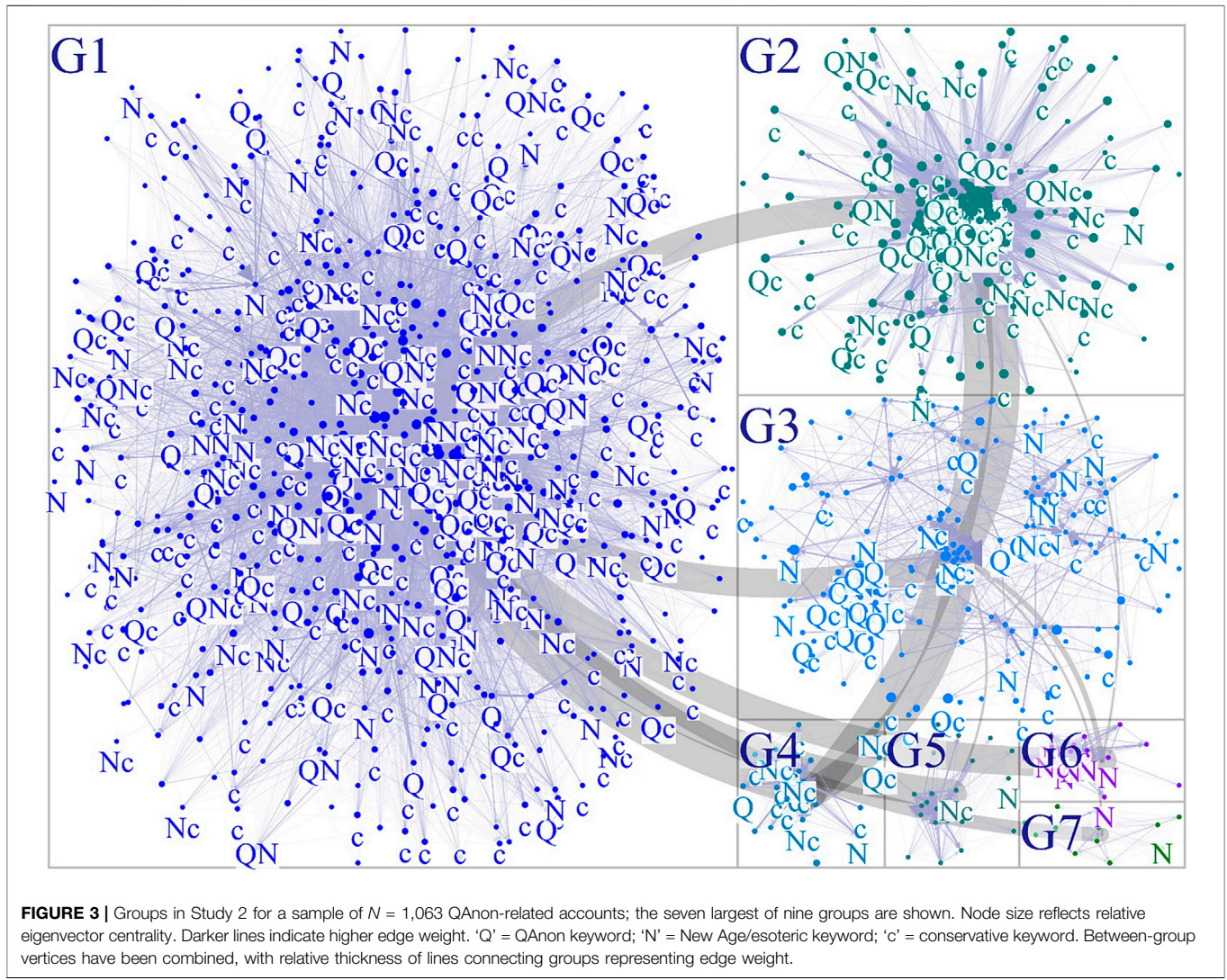

**FIGURE 3 |** Groups in Study 2 for a sample of N = 1,063 QAnon-related accounts; the seven largest of nine groups are shown. Node size reflects relative eigenvector centrality. Darker lines indicate higher edge weight. 'Q' = QAnon keyword; 'N' = New Age/esoteric keyword; 'c' = conservative keyword. Between-group vertices have been combined, with relative thickness of lines connecting groups representing edge weight.

Categorization of top-10 hashtags for accounts using hashtags (i.e., 89% of accounts) indicated that the vast majority—74% of accounts—matched QAnon themes, thus establishing the central relevance of QAnon for this subnetwork. Further, 89% matched a pro-Conservative sentiment, and 66% matched both themes. Supporting a hypothesis of coordination, the mean in-degree was $M = 46$ (range = 17–93; median = 46) out of $N = 108$; @VincentCrypt46's in-degree was 93. The mean out-degree was $M = 46$ (range = 0–98; median = 50); two accounts showed an out-degree of 98 of 108 (@vickidale12, @qbeat107).

Consistent with astroturfing, accounts' relative eigenvector centrality rank across the network was uncorrelated with metrics that typically predict influence, including number of followers, follower/followed ratio, followers/day, total tweets, or tweets/day ($p > 0.05$ for all). Further, eigenvector centrality rank within the subnetwork showed a significant *negative* correlation with number of followers ($r = -0.34$, $p < 0.001$) and followers/day ($r = -0.22$, $p < 0.05$)—precisely the *opposite* of typical behavior. Further, followers/day was significantly *negatively* correlated with account age ($r = -0.42$, $p < 0.001$); the younger the account, the faster it gained followers. While account age was significantly correlated with number of followers ($r = 0.34$, $p < 0.001$), this was driven by accounts less than 3 years old ($r = 0.35$, $p < 0.01$); for accounts older than 3 years, there was no significant relationship.

## DISCUSSION

For this sample of 1,063 accounts selected to understand influencers, presentations, and networks for QAnon propaganda on Twitter circa October 2020, the thematic centrality of QAnon was confirmed by several lines of evidence. This includes top hashtags like #WWG1WGA and #QAnon, the presence of QAnon keywords in many account profiles, and the high network influence of well-known QAnon accounts like @VincentCrypt46. Consistent with our hypothesis, keywords reflecting QAnon themes were frequently blended with those reflecting New Age and/or esoteric themes within individual accounts. Further, network traffic indicated that accounts presenting with QAnon themes readily interacted with those presenting New Age/esoteric themes. Blending of these with accounts presenting conservative themes was also observed. These data therefore confirm the generality of blending of QAnon with New Age/esoteric themes originally identified for influencers like @SnowWhite7IAm, @VeteransAlways_, and @MrDeeds1111.

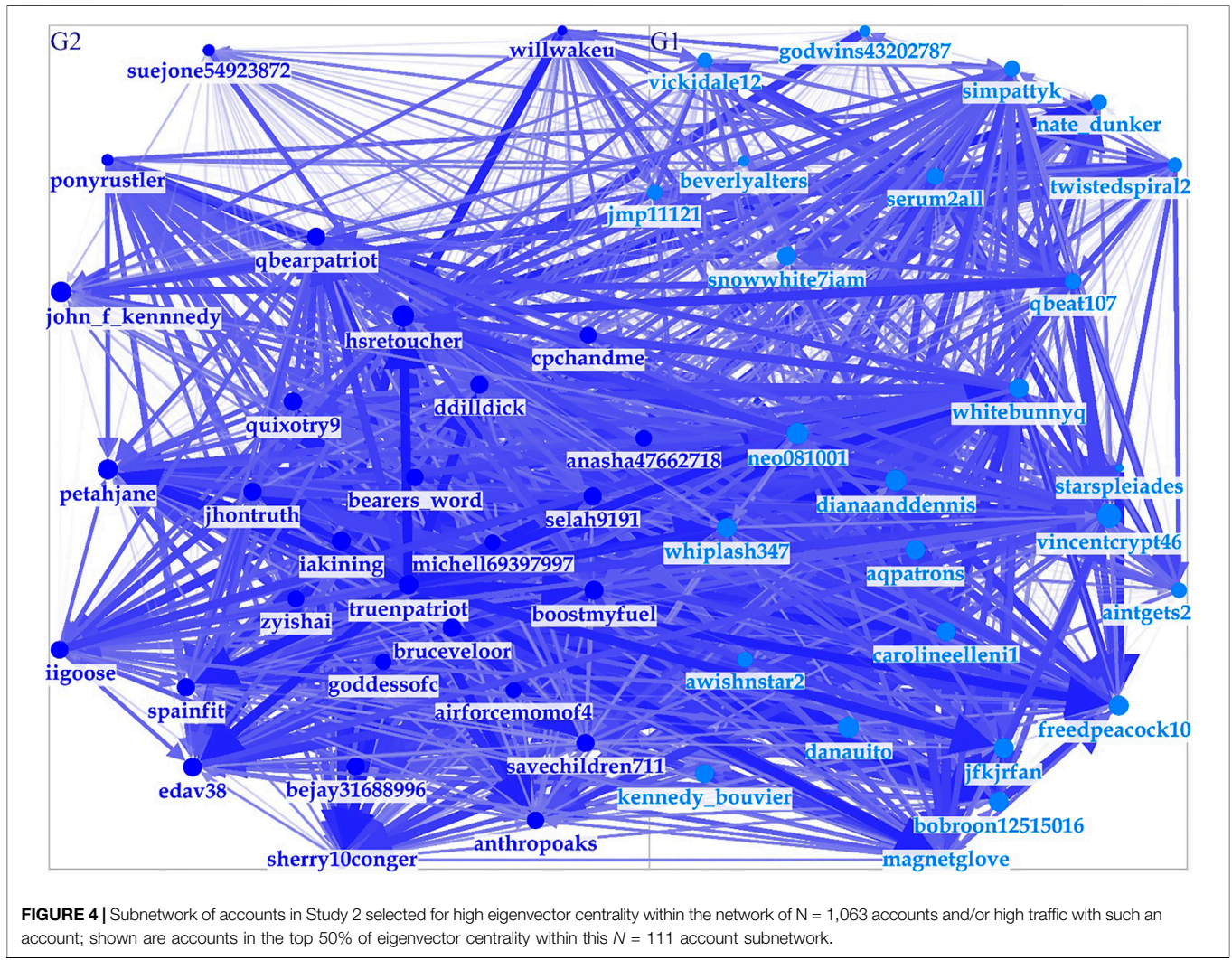

**FIGURE 4** | Subnetwork of accounts in Study 2 selected for high eigenvector centrality within the network of N = 1,063 accounts and/or high traffic with such an account; shown are accounts in the top 50% of eigenvector centrality within this *N* = 111 account subnetwork.

We also uncovered striking evidence of astroturfing in the form of account coordination among a "subnetwork" of the most influential accounts in the entire network, plus those engaging heavily with these accounts. Account coordination is considered the most important indicator of political astroturfing (Keller et al., 2020). Among odd behaviors, we found that the *higher* the network influence of accounts in the subnetwork (cf. eigenvector centrality), the *fewer* the followers, and the *slower* that accounts accrued followers, precisely the opposite pattern from authentic accounts. Further, the *younger* the account, the *faster* it gained followers; however, for accounts older than 3 years, there was no significant relationship between account age and follower numbers. For instance, @TrumperWavin was among the youngest in this dataset at just 68 days old, but it had amassed 43,233 followers, gaining 635 followers/day, thus exceeding @RudyGiuliani (555 followers/day), @NYPost (440 followers/day), and @KirstieAlley (366 followers/day). (See **Supplementary Materials** for more information.)

Further evidence of account coordination comes from at least two sources. First, we observed a "Venn diagram" relationship comparing the top 10 most influential accounts for network analyses conducted over all tweets vs only retweets. Whereas analysis of retweets showed only highly familiar accounts readily observed in QAnon-related Twitter traffic, top accounts based on all tweets showed some which were not (e.g., @DianaAndDennis,

@DanAuito); these were highly influential within the "astroturfing" subnetwork. Second, there was a high degree of interconnectedness of subnetwork accounts, with many connected to 85–90% of the others. Interestingly, many accounts in the subnetwork seemed designed to "keep a low profile" by presenting neutral content not related to QAnon. For example, @BoostMyFuel presented as a performance fuel company—featuring hashtags such as #Diesel, #Gasoline, and #EngineLife, as well as a profile picture of a canister of gasoline—yet showed an in-degree of 60 and out-degree of 67 within the subnetwork (out of 108).

Finally, we confirmed the hypothesis that New Age/esoteric accounts were indeed influential within networks of QAnon influencers. This analysis used a common set of "benchmark" QAnon influencer accounts and focused on retweets. We established that some accounts from each key category held substantial influence within QAnon-focused networks. In summary, Study 2 provided quantitative evidence from Twitter interactions that constrains further hypothesis testing and theorizing around broader origins of the QAnon propaganda "campaign".

# STUDY 3

QAnon had been argued to be an "on-ramp" to far-right extremism; however, data have so far been lacking to elucidate direct routes between QAnon and far-right extremists. Further, in some media coverage, it has been argued that far-right extremists opportunistically recruited individuals sympathetic to QAnon only after disillusionment set in following the U.S. presidential inauguration of Joe Biden in late January 2021, when the predictions of the anonymous "Q" did not materialize (Collins, 2021). An alternative possibility was that far-right ideologues might have injected extremism within QAnon narratives as part of active recruitment tactics. If so, then, at a minimum, evidence of far-right influence within QAnon networks should be apparent prior to late 2021.

Study 3 thus tested the hypothesis that QAnon accounts systematically boost far-right, extremist content. The approach to testing this hypothesis stemmed from our observations from mid-Summer 2020, through early January 2021, of a related string of accounts that claimed the persona "Wyatt". Accounts using the "Wyatt" appellation consistently promoted QAnon narratives and were recognizable for anti-Semitic themes, as well as expressions of admiration for figures like Hitler and Satan. (See **Supplementary Materials**, Section 1.3.1 for examples). "Wyatt" accounts were recognizable for using nearly-identical profile graphics depicting a man, head upturned, smoking a cigarette. We observed "Wyatt" accounts being retweeted or replied to by QAnon and New Age/esoteric accounts, including accounts identified as part of the "astroturfing" Study 2 subnetwork. "Wyatt" accounts were successively banned by Twitter, but would reappear, a cycle which happened so frequently that accounts discussed when "Wyatt" would appear next, and with what handle.

The following predictions were made. First, we predicted that accounts interacting individually with @_CEOofGenZ would show network connections and coordination with one another. Second, we predicted that, for accounts putatively interacting with one another which had individually interacted with @_CEOofGenZ, QAnon would be of central narrative importance. Third, based on a hypothesis that QAnon propaganda functioned as part of a promotion and recruitment "campaign" organized in part by far-right elements, we predicted that New Age and esoteric content would be heavily represented in networks around @_CEOofGenZ, consistent with the proposals of Bhatt (2021) that metaphysical and esoteric themes boost fear of white extinction among fascists. Finally, we predicted that a substantial proportion of accounts which had interacted with an esoteric influencer account, @SanandaEmanuel, would also show evidence of having interacted with @_CEOofGenZ.

## Materials and Methods

We first identified accounts interacting with @_CEOofGenZ. This was done by downloading this account with NodeXL on 12/31/2020. This returned $N = 1,725$ vertices. From this list, $N = 24$ vertices with over 150 k followers (e.g., @realDonaldTrump, @SeanHannity) were eliminated, due to these being deemed likely to add noise. This left $N = 1701$ Twitter accounts identified as connected to @_CEOofGenZ. To determine how these accounts were networked with one another, a Twitter users group account import was next conducted for these accounts using NodeXL on 12/31/2020 based on the last 100 tweets. Network influence and connectivity metrics were calculated as before. The relevance of QAnon and/or New Age/Esoteric themes was tested using the keyword categorization method from Study 2 to determine tweet content matching each of these themes. Finally, we determined the extent to which accounts interacting with New Age/esoteric influencer @SanandaEmanuel also interacted with @_CEOofGenZ; see **Supplementary Material**, Section 1.3.4.

## Results

### Accounts' Connections to @_CEOofGenZ and Each Other, and Relevance to QAnon

Data were downloaded from $N = 1,697$ accounts interacting with @_CEOofGenZ. Clustering algorithms revealed 16 groups, the seven largest of which are shown in **Figure 5**. 52% of accounts ($N = 876$) showed an interaction (i.e., edge) with @_CEOofGenZ in the dataset, implying that the remaining $N = 821$ accounts had established connections with @_CEOofGenZ prior to the most recent 100 tweets.

QAnon was highly relevant for the network of accounts interacting with @_CEOofGenZ. As shown in **Figure 5**, many tweets included QAnon keywords; still others displayed New Age and/or esoteric keywords or both types. Second, the most influential accounts in this network around @_CEOofGenZ, as gauged by eigenvector centrality, included numerous well-known QAnon propaganda accounts (as established in other studies); in descending order of influence, the top 15 of these were: @_CEOofGenZ, @CodeMonkeyZ, @VincentCrypt46, @PrayingMedic, @AwakenedOutlaw, @PepeMatter, @PepeNewsNow, @RichardGibb8, @HSRetoucher, @55true4u, @Mareq16, @EntheosShines, @MartinGeddes, @RedWaveWWG1, and @ValiantThor12.

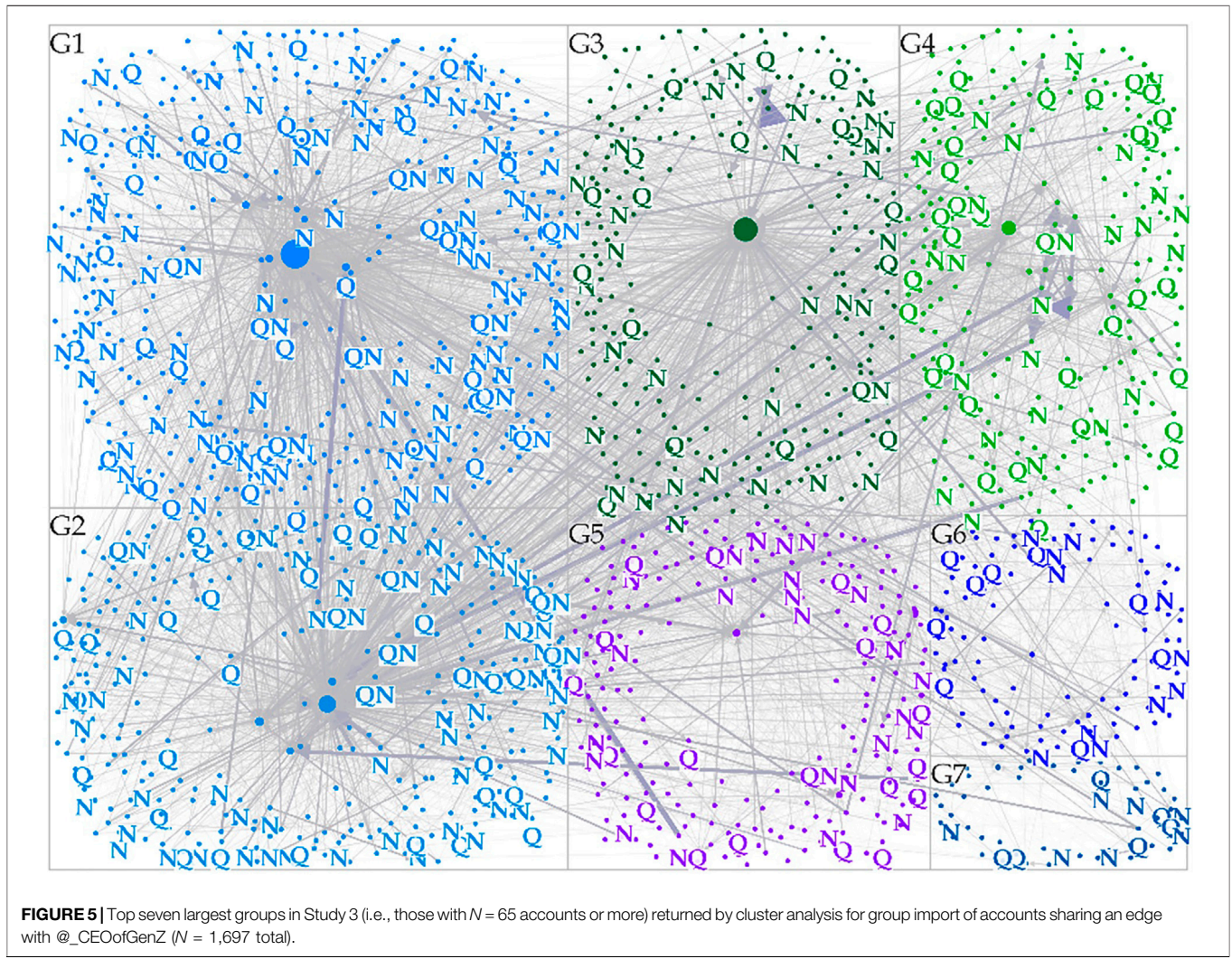

**FIGURE 5** | Top seven largest groups in Study 3 (i.e., those with $N$ = 65 accounts or more) returned by cluster analysis for group import of accounts sharing an edge with @_CEOofGenZ ($N$ = 1,697 total).

Finally, network top items revealed a predominance of QAnon themes (cf. Tables S16-S17). Overall top replied-to and mentioned accounts included well-known QAnon influencers like @VincentCrypt46 and @CodeMonkeyZ. Overall top hashtags showed QAnon themes like #Biblical, #WakeUp, and #ThePlan. Individual group top hashtags pertaining to QAnon also included #JFK (Group 2), #WWG1WGA (Group 3), and #QAnon (Group 4), among others. Additional top hashtags reflected Republican party narratives and organizing, such as #Benghazi and #StopTheSteal (Group 2), as well as #SethRich and #DigitalSoldiers (Group 4).

### Group Import From @SanandaEmanuel, Connections to @_CEOofGenZ, and Relevance to QAnon

Data for $N$ = 394 accounts derived from @SanandaEmanuel account vertices was returned. Seven groups were returned by the clustering algorithm (see **Supplemental Materials**); 86% of vertices were in the top three groups. As predicted, a substantial proportion of accounts (13%; $N$ = 53) were *also* in an "edge" with the pro-Hitler, pro-Satan @_CEOofGenZ account (i.e., through a retweet, reply-to, or mention). This indicates that many accounts which interacted with @SanandaEmanuel also had interacted with extremist @_CEOofGenZ within the prior 500 tweets, providing support for our hypotheses about network overlap and disinformation actor motivations.

Further, QAnon was relevant to this network. Of accounts with non-blank descriptions (97% of vertices), fully 57% of them contained a QAnon keyword. Further, 79% contained a New Age/esoteric keyword, and 80% contained a conservative keyword. Top hashtags also included QAnon themes, e.g., #SaveTheChildren and #QAnon (cf. **Supplementary Tables S18, S19**).

## DISCUSSION

Study 3 tested predictions of network coordination among accounts interacting with @_CEOofGenZ, an account posting anti-Semitic, extremist content. As predicted, accounts which had individually interacted with @_CEOofGenZ were well-networked with one another. Further supporting predictions, QAnon was highly relevant for the network around @_CEOofGenZ. Accounts with links to @_CEOofGenZ were among the most influential pro-QAnon accounts on Twitter, e.g., @VincentCrypt46.

We further tested the prediction that a substantial proportion of accounts linked with esoteric influence @SanandaEmanuel, would interact with @_CEOofGenZ, and that QAnon would be a central theme in this network. These predictions were based on a hypothesis that far-right extremists were injecting esoteric-themed content into QAnon propaganda narratives as a way of inducing "fear of white extinction" within a novel social media promotion and recruitment campaign. Consistent with all predictions, fully 13% of accounts connected to @SanandaEmanuel were also connected with @_CEOofGenZ in our data, and accounts were well-connected with one another. More importantly, QAnon propaganda narratives were central to themes across the network. Further evidence that the operator of @SanandaEmanuel supported far-right ideology comes from its prolific usage of the "OK" symbol in tweets, a hand gesture which sometimes signals white supremacist ideology.

This study is therefore consistent with the view that far-right adherents acted in coordination to promote far-right, extremist QAnon ideological content, based on data from late 2020 to the first half of January 2021. It, therefore, speaks against the claim that far-right recruiters opportunistically recruited QAnon believers only after Joe Biden's inauguration to the U.S. presidency.

# STUDY 4

In Study 3, we used the case study of an influential neofascist account, @_CEOofGenZ, to investigate social network links to QAnon propaganda promotion and New Age/esoteric enthusiasts. Study 4 aimed to investigate these links through collecting a larger, more extensive sample of accounts featuring QAnon and/or esoteric themes/content to better characterize the connections among accounts, as well as to extremists like @_CEOofGenZ. The events of Jan. 6, 2021 relating the Washington D.C. riots led to identifying through QAnon-related Twitter activities additional accounts connected to @_CEOofGenZ which gave impressions of promoting ideologies of accelerationism, white nationalism, and/or radical extremism. Building on these timely events and discoveries, we hypothesized that a more extensive sample of accounts not selected for connections to @_CEOofGenZ might reveal broader connections between QAnon and esoteric accounts, toward further testing the hypothesis that far-right adherents were astroturfing QAnon-centric propaganda, in part by imbuing metaphysical themes toward promoting fascist ideology (Bhatt, 2021).

Study 4 further continued our goal of investigating actors who had influenced or promoted QAnon propaganda by studying their connections on Twitter. Of particular interest in this study were Twitter accounts associated with persons identified to be central to the history of QAnon—including Thomas Schoenberger, Lisa Clapier, and Ron Watkins. Schoenberger and Clapier have been identified as a key promoter of Internet puzzle Cicada 3301 (Edel, 2021a, b; Bicks, 2021). Given Schoenberger's prominence in putative recruitment of "trolls" for, or through, the Cicada 3301 puzzle, further connections with QAnon were indicated, among other things, by the fact that QAnon influencer and recruiter @MrDeeds1111—one of the most influential Twitter accounts in our sample in Study 1—bore the Cicada 3301 puzzle insignia in its profile.

Cicada 3301 further invokes the relevance of Discordianism, a "religion" invented in the late 1950s (Robertson, 2012). While it is often dismissed as parody, interviews with Discordians in Finland (Mäkelä & Petsche, 2013) support that they see this as a worldview—one which is "liquid" (referencing the framework of Taira, 2006 for 'liquid religion', pp. 7-35) and which "intentionally 'liquifies' the boundaries between the sacred and profane" (p. 411). We hypothesized that tech-enabled QAnon propaganda, whether promoted by "trolls" or persons intent on "liquifying" boundaries between sacred institutions and profane constructs, could represent dangers to societies that were ineffectually dealt with at present. Study 4 thus aimed to expand sampling around Schoenberger and far-right accounts, while drawing in Twitter accounts linked with Cicada 3301 and/or other ARGs into the same sample along with QAnon and esoteric-themed accounts.

## Materials and Methods
### Account Selection

The first step in account selection was identifying key accounts of interest and connected accounts via account downloads of those accounts using NodeXL. From these, we selected an initial set of accounts that represented a subset of identified connections, focusing on accounts which convergent evidence suggested might be fruitful, e.g., cases where accounts were connected on Twitter over samples of key accounts taken at different times. Following the construction of the initially selected set, we iteratively identified accounts connected to these through one or more mentions, retweets, replies, or following/follower relationships through a combination of NodeXL downloaded data from these accounts and hand-inspection of Twitter account properties.

Key accounts were selected to represent the following general account types: QAnon propaganda promoters, especially accounts identified as influential in prior studies; QAnon Insiders, especially accounts attributed to Thomas Schoenberger (@FaisalLazarus)[6], Jim Watkins (@RQueenInc), and Lisa Clapier (@SnowWhite7IAm); associates of the aforementioned QAnon Insiders; accounts promoting Cicada 3301 and/or other ARG content; and far-right accounts. Note that by the time of sample construction, the prior Schoenberger account @NameMySock had been suspended and replaced with @FaisalLazarus. Two notable Schoenberger (@FaisalLazarus)

---

[6]Note that by the time of sample construction, the prior Schoenberger account @NameMySock had been suspended and replaced with @FaisalLazarus. See https://web.archive.org/web/20210126030343/https:/twitter.com/FaisalLazarus/status/1353901300056875008

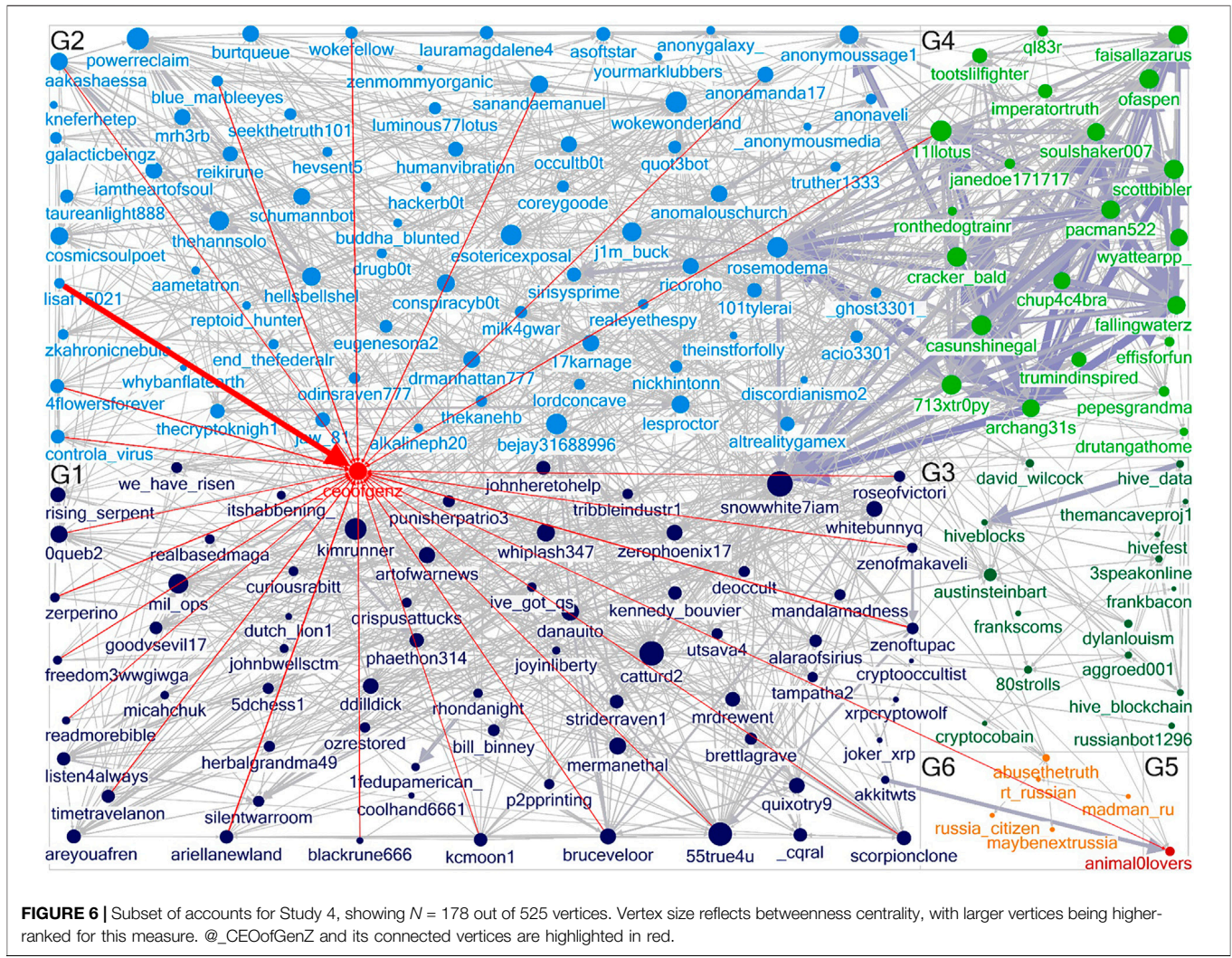

**FIGURE 6** | Subset of accounts for Study 4, showing N = 178 out of 525 vertices. Vertex size reflects betweenness centrality, with larger vertices being higher-ranked for this measure. @_CEOofGenZ and its connected vertices are highlighted in red.

associates were @TootsLilFighter and @FranksCOMS. Additional key account selection focused on evidence of far-right and/or "incel" activity, Russian-language disinformation, cryptocurrency, and references to the hacker collective Anonymous; see **Supplementary Material** Section 1.4.1 for more information.

### Data Acquisition and Analysis

Download of Twitter data using NodeXL 1.0.1.411 was initiated on January 8, 2021 for a target of N = 820 selected accounts. The general analysis approach was similar to other studies. We also determined whether any the top 10 hashtags for each account expressed themes related to QAnon, Cicada 3301, or another ARG (e.g., #TheGame23), cryptocurrency (e.g., #XRP), blockchain technology, or the Hive (e.g., #Hive). See **Supplemental Materials**.

### Results
#### Basic Network Statistics

The Twitter data download on January 8, 2021 coincided approximately with Twitter's initiating mass suspensions of QAnon accounts following the Jan. 6, 2021 Washington, D.C. riots. Reflecting high validity and specificity of our QAnon account selection criteria, Twitter mass suspensions resulted in no data being collected for 36% (N = 295) of the target sample, further illustrating QAnon's prominence in the selected sample. Data was collected from N = 525 accounts; note that many of these went on to later be suspended by Twitter. Clustering as before revealed eight groups; **Figure 6** displays select accounts for the largest six of these.

### Network Influence

**Supplementary Table S23** shows accounts ranked in the top 20 on eigenvector centrality; toward data interpretation, these accounts have been grouped according to their judged relevance to QAnon and/or New Age/Esoteric themes using Study 2 criteria, along with categories of having a Schoenberger affiliation, and/or promoting Cicada 3301 or other ARGs. We found that QAnon influencer @SnowWhite7IAm was the most influential account in the entire network. Further, at least nine of the 20 most influential accounts most influential across the network had promoted

QAnon, based on concurrent and archived data. Five accounts were judged to a have New Age/Esoteric focus, and several of these had also promoted QAnon, based on searches of archived data. For instance, @TheHannSolo had been selected for its New Age/esoteric features; however, archived data showed it previously promoted the "QMap", a diagram circulated on Twitter linking QAnon to a range of conspiracy and disinformation themes, as its banner graphic. Further, @TheHannSolo had earlier included a URL in its profile linking to documentation of Cicada 3301 and other ARGs, thereby documenting a case of at least one account which linked three key properties explored in all studies thus far (QAnon, esoteric themes, and Cicada 3301 and other ARGs). Further, nine of the 20 most influential accounts in the entire network had prominent connections to Thomas Schoenberger. Six of the top 20 accounts were identified to have a connection to Cicada331 and/or other ARGs. Although not in the top 20, @_CEOofGenZ was nevertheless influential, as gauged by its ranking of 32nd overall (94th percentile) for eigenvector centrality.

### Network Top Items

**Supplementary Table S22** shows network top items for top accounts. Reflecting a conservative measure of QAnon involvement, 19% of accounts using any hashtags used a QAnon-themed hashtag as one of its top 10 most frequent, including approximately 30 and 13% of accounts in Group 1 and 2, respectively. Cryptocurrency themes were also attested across multiple groups, particularly Groups 1 and 3. Hashtags related to the Hive were largely limited to Group 3.

Top replied-to and mentioned accounts further provide clear evidence of the relevance of QAnon for the network, as indicated by top accounts including QAnon influencers like @VincentCrypt46, @SnowWhite7IAm, @CodeMonkeyZ, and @LLinWood. Schoenberger associates made up the bulk of top engaged accounts. Several other top accounts were associated with Cicada 3301 and/or other ARGs, such as @PacMan522, @AnomalousChurch, and @Andress45303251. See **Supplementary Material**, Section 1.4.3 for more details.

## DISCUSSION

This study examined network connectivity and influence across 525 accounts selected mainly for QAnon- and esoteric-related features, as well as the far-right and links to QAnon insider Thomas Schoenberger and his associates, Internet puzzle Cicada 3301 and other ARGs. Both QAnon and esoteric accounts were influential in this network, even though data were collected just after Twitter's initiating a mass suspension of QAnon accounts which resulted in obtaining no data for around 36% of planned accounts. The crackdown likely caused overt signs of QAnon to be less prominent in our downloaded data than otherwise (e.g., reduced frequency of QAnon hashtags). For many accounts in this sample, promotion of QAnon had been established in prior studies, pilot data, and considerable observation. Also, we observed that a number of accounts had reduced the prominence of QAnon features in their profiles over time (e.g., @TheHannSolo)[7].

Accounts tied to Cicada 3301 and other ARGs were highly influential within the same network sample that included QAnon, esoteric, and far-right accounts. @SnowWhite7IAm, the account belonging to Schoenberger associate Lisa Clapier, was the most influential account of all. Further, we found that Schoenberger's account @FaisalLazarus was among the top 20 most influential accounts in the entire network; in interviews and narrative recordings, Schoenberger has denied or given misleading answers regarding his involvement with QAnon propaganda furtherance activities[8] We further found that many accounts that were observed to be in frequent two-way Twitter conversations with @FaisalLazarus were in the top 20 most influential accounts in the network and promoted QAnon, Cicada, and/or other ARGs.

Further, far-right ideology and extremism were prominent within this network sample. For instance, extremist account @_CEOofGenZ ranked at the 94th percentile for eigenvector centrality; @_CEOofGenZ also frequently promoted QAnon. Many other accounts in this sample retweeted, mentioned, or replied to @_CEOofGenZ; examples were @Freedom3WWGiWGA, @TimeTravelAnon, @55True4U, and @BlackRune666. See **Supplementary Material**, Section 1.4.3 for additional discussion.

Moreover, @AbuseTheTruth, an account tweeting in Russian with disinformation themes which interacted with esoteric influencer accounts, was among the highest for betweenness centrality. This suggestive evidence is consistent with the possibility that Russian operatives could be behind some accounts in this network. We return to this topic in the General Discussion.

Finally, we were unable to determine predominant themes within groups due to significant thematic overlap. We hypothesized that this might have been due in part to undersampling brought about in part by the timing of Twitter account suspensions. Study 5 therefore aimed to replicate and extend these findings but with a larger, more balanced sample.

## Study 5

Study 5 built on evidence across studies to further understand links among QAnon propaganda, and New Age- and esoteric-themed accounts, and notably, far-right and extremist ideologies. To this end, we drew on a larger sample of far-right accounts than in Study 4, while also repeating the account selection process to partly offset missing data from Study 4. We aimed to further elucidate which actors' convergent media evidence had continued to implicate—namely Schoenberger, Clapier, and R. Watkins. To this end, we also expanded our sample of Cicada 3301 and other ARG-themed accounts in Study 5.

---

[7]As of 2/9/21, Twitter had suspended N = 118 additional accounts, corresponding to 22% of accounts downloaded for this Study 4a sample. Further, around 3% of accounts had deactivated.

[8]Account @NameMySock, whose operator self-identified as Thomas Schoenberger in an archived tweet (cf. https://archive.is/GL5xz), retweeted a QDrop and referenced "information warfare" and "death" on Oct. 18, 2020. See https://archive.is/ndrFW

## Materials and Methods
### Account Selection Procedure

Our selection procedure was similar to Study 4. While a goal of Study 5 was to investigate network connectivity of accounts espousing far-right and/or extremist ideology, affiliations of accounts with known neo-Nazi and/or militia groups (e.g., Proud Boys, Oath Keepers) were not known or not observed. Thus, we instead used proxy cues for account selection and emphasized selection of accounts which: 1) were associated with known Alt-Right figures (e.g., @RichardBSpencer); 2) gave the appearance of promoting far-right political views, such as swastikas, pepe, or "Groyper" memes (Hawley, 2021), and/or seemingly approving references to Satan (e.g., @a2ktribal); 3) presented esoteric-themed content as well as pro-QAnon, pro-Trump, and/or pro-Republican content; 4) displayed imagery of guns, tanks, or explosions; 5) promoted content or hashtags related to Cicada 3301 and/or other ARGs; 6) tweeted in Russian; 7) appeared to be "resurrected" versions of prior influencer accounts (e.g., @VincentCrypt46's suspension was followed by evidence of migration to @CryptoKoba); or 8) demonstrated either by heavy traffic with a key account and/or connections with influential accounts of interest. Finally, a handful of Republican influencer accounts observed to be frequently retweeted were included (e.g., @JackPosobiec).

### Data Collection and Analysis

NodeXL was used to attempt download of Twitter data for $N = 887$ selected accounts on 1/24/21. Analysis procedures were similar to prior studies.

## Results
### Basic Network Data and Statistics

We obtained Twitter data for 93% of our target sample, i.e., $N = 824$ accounts. A handful of selected accounts were suspended prior to download, including @_CEOofGenZ, @SnowWhite7Iam, @bejay31688996, @55True4U, and @0queb2. Clustering of data as before revealed 11 groups. Five of these contained three accounts or fewer, and one contained accounts ($N = 23$) not connected to any others in this sample.

### Network Organization: Connectivity and Influence

**Figure 7** shows groups and connectivity for a subset of accounts chosen to maximize visibility and clarity. Thematic convergences were identified for major clusters; descriptive labels have been appended to clusters consistent with a predominance of accounts contained therein representing account selection features of interest.

The Schoenberger account @FaisalLazarus and its connections are highlighted in red in **Figure 7**. @FaisalLazarus was high-ranked for eigenvector centrality (87th percentile). **Supplementary Table S26** lists accounts top-ranked for influence based on eigenvector centrality. Top-ranked QAnon influencers were @CryptoKoba (i.e., the apparent replacement for @VincentCrypt46), @KimRunner123, and @Iakining (cf. Study 2). A Cicada 3301/ARG-linked account, @Quasarcasm47, was also top-ranked, along with two Schoenberger-linked accounts, @11llotus and @22DubTrip333. The high influence of esoteric and conspiracy themes is apparent by the top-ranked status of accounts @EsotericExposal and @Conspiracyb0t.

Importantly, the high influence of far-right ideology is apparent in top-ranked billing of @Caleidoscope11 (>99th percentile); while the name of the account was suggestive of esoteric themes, it featured a pepe meme in its header and presented as "Pepe's Girlfriend". This account was connected to multiple "groyper", "fren," and white supremacist accounts: e.g., @chaosfren, @micro__benis, and @DiamondGunProd. Notably, the account also displayed connections to QAnon and esoteric accounts, including @CryptoKoba (cf. the old VincentCrypt46), @Kabamur_Taygeta, @TheHannSolo, and @occultb0t.

Further, several top-ranked accounts fall within the quadrant labeled "Insurrectionist/Other"—including @Shorty56167141, @TweeTweetBeyoch, @hb04920973, and others. These accounts had connections in the present data or pilot data to accounts observed to promote insurrection and show imagery of weapons of war (e.g., @MrDrewEnt, @TrumpsOtherSon1, @ZeroSum24, all of which ranked at the 87–88th percentile for eigenvector centrality). Additionally, relatively heavy traffic is apparent between these accounts and @Patrioness (92nd percentile), an account engaged in account coordination with QAnon Insiders like @FaisalLazarus circa January 1, 2021, around the creation of account @QL83R (see **Supplementary Material**). Finally, several right-wing influencers were among the most influential in this network (e.g., @JackPosobiec, @MrAndyNgo). Network top items are presented in **Supplementary Material**, Section 1.5.3.

## DISCUSSION

This study investigated over 800 Twitter accounts selected through iterative methods to understand relationships among accounts promoting QAnon propaganda, New Age and esoteric themes, far-right and insurrectionist themes, and Cicada 3301/ARG themes. A remarkable finding was that multiple accounts linked to Schoenberger were among the most influential across the entire 800-account network although these represented a small fraction of accounts in this sample. Moreover, multiple accounts selected for promoting far-right ideology and insurrection—through displaying pepe and Groyper memes, imagery of weapons of war, and retweeting participants in the January 6, 2021, U.S. Capitol insurrection—were highly influential within this network sample (85th percentile and above). Importantly, we demonstrated, within the same network sample, that accounts for Schoenberger himself (@FaisalLazarus), along with multiple Schoenberger- and/or Cicada 3301/ARG-linked accounts—including @11Llotus, @PacMan522, @TheHannSolo, @RoseModema, and @SirisysPrime—were among the very most influential across the entire network, and that these were among the most frequently engaged and referenced through network traffic.

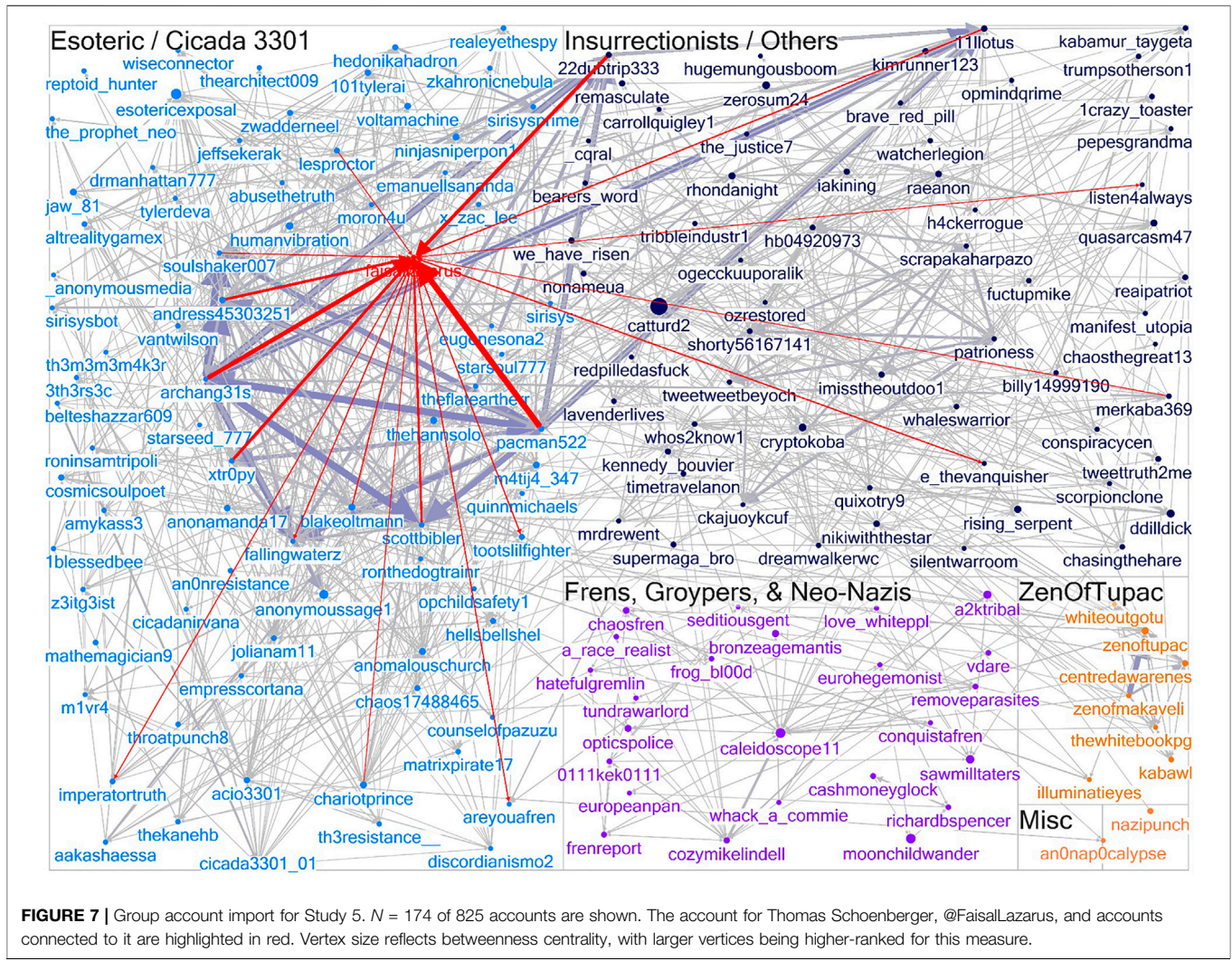

**FIGURE 7 |** Group account import for Study 5. $N$ = 174 of 825 accounts are shown. The account for Thomas Schoenberger, @FaisalLazarus, and accounts connected to it are highlighted in red. Vertex size reflects betweenness centrality, with larger vertices being higher-ranked for this measure.

## GENERAL DISCUSSION

In five studies, rich Twitter datasets were constructed and analyzed with the aim of shedding light on actors, motivations, and strategies in promoting QAnon propaganda on Twitter in the months before and after the 2020 U.S. Presidential election. Clear evidence was obtained of significant digital astroturfing activities in the promotion of QAnon propaganda on Twitter. This included abnormal follower growth rates, coordinated network behavior, and inconsistencies between established metrics of account influence and network influence measures like eigenvector centrality. Coordinated account activity has been argued to be among the most robust indicators of astroturfing (Kellar et al., 2020). These findings provide evidence against claims of some journalists and disinformation researchers that the popularity of QAnon resulted solely from "grassroots" political activities of authentic individuals acting independently.

Remarkable quantitative evidence of highly unusual account behaviors was demonstrated. Numerous pro-QAnon influencer accounts showed unnaturally high follower acquisition rates and follower:followed ratios, which was striking due to the anonymity of most profiles. While celebrities' and politicians' accounts may gain followers rapidly due to personal brand renown, by contrast, the largest, most influential QAnon accounts presented largely generic profiles. For instance, @VincentCrypt46 and @MajorPatriot presented anonymous personas with graphics obscuring faces, together with cues suggestive of military backgrounds. @CJTruth presented a generic profile featuring gladiator imagery and themes of religious piety. @M2Madness presented a mysterious hacker persona. @TrumperWavin featured a generic profile of a man until around the U.S. presidential election when it claimed to be a 17-year-old boy.

Evidence demonstrated here of political astroturfing to promote QAnon propaganda raises critical questions about individuals or groups that may have participated in campaign coordination and use of technology to target susceptible individuals (Youyou et al., 2015; Baimel, 2019; Bronstein et al., 2019; Bakir, 2020; Kriel, 2020; Moskalenko, 2021). Systems justification theory (Jost & Hunyady, 2005) states that actors operating within human institutions that perpetuate social injustices tend to justify continued participation in those institutions. Hypothesis generation around actors was constrained and informed by the novel conceptual framework of Bakir et al. (2019), which highlighted roles of deception, coercion, and incentivization in organized persuasion campaigns. Russia is a nation-state actor of primary interest for a potential role in observed QAnon astroturfing activities and the broader QAnon propaganda campaign. Here, we found numerous lines of evidence consistent with Russia being a major actor behind QAnon propaganda astroturfing on Twitter. Numerous Russian-speaking accounts were observed to interact with influencer accounts in these studies, including: @MoonChildWander, which promoted white supremacist Richard Spencer and other extreme far-right accounts; @AbuseTheTruth, a Russian-language disinformation account interacting with numerous influencer accounts; and @NoNameUA, an account which interacted with numerous Russian-speaking accounts as well as @ZeroSum24, which promoted QAnon and violent, insurrectionist themes.

Observational evidence collected during these studies also supports probable Russian involvement in promoting QAnon propaganda on Twitter. Spontaneous Russian language was used on multiple occasions by pro-QAnon influencer accounts that primarily tweeted in English. For instance, influential QAnon account @VincentCrypt46 tweeted the Russian word "#пожар" (fire) on August 13, 2020, prompting both @PetahJane and @gapatriot3 to reply, ostensibly notifying @VincentCrypt46 of the "slip-up"[9] (see **Supplementary Materials** for more examples). Multiple reports documented Russian promotion of QAnon (Davis, 2020; Menn, 2020a, b; Smith, 2020). That U.S. far-right ideological adherents tend to be sympathetic to Russian narratives is well-documented (Michael, 2019). Understanding Russia's role in promoting QAnon will need updated as ongoing relevant evidence emerges.

In this study, we investigated the relationship of Twitter far-right accounts to pro-QAnon accounts over five studies with increasingly specific predictions. Notably, Study 3 tested predictions that accounts which interacted with @_CEOofGenZ, who regularly tweeted praise for QAnon, Hitler, and Satan, not only would be networked with each other but also that these accounts would show the centrality of QAnon narratives and influencers. Both predictions were strongly upheld. Studies 4 and 5 likewise demonstrated the influence of QAnon propaganda for samples constructed of far-right accounts on Twitter.

The relevance of esoteric and New Age themes to QAnon propaganda networks was replicated multiple times with increasing specificity of predictions. Notably, Studies 4 and 5 provided insight into New Age and esoteric content provided identifiable "on-ramps" to the far-right. Our data indicate that accounts promoting far-right ideologies were already well-organized and prominent within QAnon-related Twitter networks well in advance of Biden's inauguration in late January 2021. The combination of narratives around esoteric themes may be understood as a persuasion and recruitment tactic aimed at individuals prone to mindsets sympathetic to "white extinction" narratives common on the far-right (Cosentino, 2020; Bhatt, 2021).

Overall, we argue that QAnon propaganda on Twitter has been one facet of a large-scale, broad-based media campaign toward "engineering" an entirely new conspiracy-based social category: "QAnon" (cf. Sternisko, Cichocka, & Van Bavel, 2020). Individual psychographic profiling from social media is not only technically possible but also achieves high accuracy (Youyou et al., 2015). We suggest that starting points for evidence-based consideration of possible architectures of this industrial-scale cyber social-engineering campaign may be influencers like Russian propagandist Alexander Dugin and Trump campaign manager Steve Bannon. Dugin and Bannon's shared admiration for, and promotion of, Italian fascist political theorist Julius Evola is noteworthy in light of Evola's advocacy of esotericism (Lachman, 2018; Bhatt, 2021). In light of systems justification theory, prior activities around creation of "cultural weapons" to change politics and culture by Bannon in collaboration with the Mercers are noteworthy (Briant, 2018; Wylie, 2019; Bakir, 2020).

Theoretical and empirical work in language, cognition, and social psychology helps explain how social categories might be "engineered" through the misuse of social and mass media. Linguistic stimuli and visual imagery influence thought patterns (Vygotsky, 1997; Thibodeau et al., 2017). Neural processing may be affected through mere statistical exposure to complex stimuli, even when perceptibility thresholds are not reached (Henry et al., 2016), elaborating mechanisms entailed in conceptual proposals regarding meme theory (Dawkins, 1989). Further, Foucault's (1975/2012) work on the "panopticon" illustrates how cognitive control through social norms can foster political power.

Consideration of individual differences in cognition is important for understanding and theorizing how large-scale misuse of technological innovations for communication might be engineered on grand scales toward influencing socio-political human behaviors in the aggregate. Effective persuasion tactics would presumably aim to identify, on the one hand, personalities susceptible to QAnon narratives (Yi & Tsang, 2020) through psychographic profiling.

On the other hand, effective tactics would presumably also involve seeking strategic alliances with groups of individuals disposed toward furthering QAnon. This raises the interesting issue of connections to alternative reality games, such as Cicada 3301. Magical thinking and ARGs are conceptually related to Discordianism, which some adhere to as a world-view intent on "liquifying" boundaries between the sacred and profane. Robertson (2012:421) argues that "Wilson's (1969/1979) exegesis of Discordianism appealed to groups who would never have come into contact with it those utilizing emergent technologies, such as computer hackers and dance musicians."

Journalistic sources have documented that Cicada 3301 attracted "notorious" Internet trolls; trolling "4daLulz" reflects everyday sadism (Buckels et al., 2014; see also Tuters, 2019.) Manipulation of others' gifts and vulnerabilities may be consistent with Machiavellian trait-dominance (Rauthmann and Kolar, 2013). Such manipulation of others toward propagandistic ends may be exemplified in far-right media personality Jack Posobiec's advancement of "weaponized autism" as a strategic means of "troll-[ing] the Left" (Deplorable News Daily, 2017; see also; Klee, 2018), facilitated by recruitment from the /pol/message board on 4 chan. Related activities undertaken to promote QAnon propaganda have included LARPing (for "live action role play"; Tuters, 2019) and/or memetic warfare (Giesea, 2015).

---

[9]See https://archive.is/6MD9h

Further, the socially divisive, damaging effects of some trolling and extreme "othering" behaviors invoking, e.g., moral disgust (Giner-Sorolla et al., 2018) can be understood in the context of the framework of Bakir et al. (2019), as when QAnon "supporters" have labeled celebrities like Oprah Winfrey and Chrissy Teigen as pedophiles. Bakir et al.'s framework elucidates roles of both coercion and incentiviation in organized persuasion campaigns; within this framework, hashtags like #SaveTheChildren can be understood as forms of "soft" coercion constituting implicit threats, e.g., through status loss for not supporting the "norms" co-opted by QAnon narratives. Threats of corporal injury (as implied by the #PushingUpDaisies hashtag in Study 1) or claims that businesses are "grooming children" through toys advanced by QAnon propaganda as age-inappropriate can likewise be considered forms of coercion, due to implications of material loss, social ostracism from moral disgust, and/or bodily harm.

The framework of Bakir et al. (2019) allows for understanding the role of incentivization in deceptive, organized persuasion campaigns. Some elements of networks studied hinted at financial incentives, including common themes of cryptocurrencies, as well as gematria, which is linked to gambling. Further, some accounts promoting ARGs and Discordianism referred to blockchain art, which far-right elements developed innovations for.

Finally, these studies documented Twitter activities of individuals identified in media as "persons-of-interest" in having furthered QAnon propaganda, including Thomas Schoenberger, who has been associated with accounts @NameMySock and @FaisalLazarus. Frequent online associates of these Schoenberger accounts showed signs of far-right views and promoted QAnon narratives. For instance, @11Llotus ranked at the 99th percentile for influence in Study 5 and frequent contact on Twitter with @FaisalLazarus, retweeted @_CEOofGenZ's tweet about baby sacrifice and adrenochrome. @Archang31s, who was at the 80th percentile for influence in Study 5, promoted itself as "Ancient Aryan" with a picture of an Iron Cross—a symbol used by the Nazis—and replied to "groyper" accounts like @ChaosFren. @Archang31s was likewise in many conversations with @FaisalLazarus. @FallingWaterz and @ImperatorTruth—two frequent online associates of Schoenberger—interacted with @DrutangAtHome, who posted about the runic symbols for "SS" and Nazis' interest in esoterics. Many Cicada 3301-themed accounts interacted with @NickHintonn, who started the Institute for Folly (@theInstforFolly) to promote Discordianism. @Acio3301s profile referenced "psyops", "cyberwarriors soldiers" [sic], "neural network", "lulz", and "!?fnord?!", a reference to Discordianism. Collectively, evidence presented here is a strong confirmation that Schoenberger-linked accounts were in often frequent contact with accounts that promoted extreme far-right content, QAnon propaganda, and Discordianism.

In summary, these studies showed strong evidence of astroturfing within QAnon-related propaganda activity on Twitter. Promotion of extreme far-right ideology and Discordianism were active dynamics within this QAnon propaganda campaign, where this was well-organized and occurred well prior to the inauguration of President Joe Biden. These studies shed light on novel ways in which large-scale, deceptive organized influence campaigns can be conducted on social media by strategically aligned and/or malign actors. Finally, the current study lends insight into how social media cyber-spaces increasingly are domains for cutting-edge information warfare campaigns.

## DATA AVAILABILITY STATEMENT

The datasets generated for this study are available at: https://osf.io/rztes.

## ETHICS STATEMENT

Written informed consent was not obtained from the individual(s) for the publication of any potentially identifiable images or data included in this article.

## AUTHOR CONTRIBUTIONS

LD designed the studies, conducted analyses, and wrote the paper. WW wrote computer programs to assist analyses by the team, and both WW and FF conducted analyses and contributed to paper writing and editing.

## FUNDING

We gratefully acknowledge support from the Dept. of Communicative Sciences and Disorders and College of Communication Arts and Sciences at Michigan State University.

## ACKNOWLEDGMENTS

We gratefully acknowledge the assistance of Zachary Ireland with pilot analyses and editing, Arturo Tafoya for research background and feedback on an earlier version of the paper, and MM and LM for general research support.

## SUPPLEMENTARY MATERIAL

The Supplementary Material for this article can be found online at: https://www.frontiersin.org/articles/10.3389/fcomm.2021.707595/full#supplementary-material

# Supplementary Material

## 1    Study 1

The network analysis of downloaded Twitter data showed 128 vertices with 120,476 total edges with 790 unique edges (104,392 self-loops). The reciprocated vertex pair ratio was 0.20, and the reciprocated edge ratio was 0.34. Betweenness centrality rank and eigenvector centrality rank were strongly correlated, $r(126) = 0.77$, $p < .001$, reflecting dense interconnectedness of these accounts.

Network top items for Study 1 are shown in Table S1. The top URLs, ranked highest to lowest, were: Five-by-Five network YouTube channel; www.gematriaclub.com; 5x5.carrd.co; 8ch.net/index.html; www.therevolutionnetwork.com; grrrgraphics.com; dlive.tv/WTPI_1776; a tweet about the Schumann resonance; YouTube episode for @intheMatrixxx; and the "Fall Cabal" YouTube channel of Dutch conspiracist Janet Ossebaard. Quantitative characteristics of accounts ranked in the top 10 for various influence metrics are shown in Table S2. Finally, see Table S3 for a repeat of analyses based on retweets only.

Descriptions of the 15 accounts designated as primary are shown in Table S4, while relative influence metrics for these 15 are reported in Table S5. Twelve of 15 primary accounts showed follower:followed ratios of greater than 4:1, including @OSSRobertSteele, who had a follower:followed ratio of 90:1. @MrDeeds1111, @RQueenInc, and @JasonSullivan_ exhibited influence for other metrics. For instance, @MrDeeds1111 showed 136 tweets/day (87th percentile across all accounts), a gain of 371 followers/day (97th percentile), and high eigenvector centrality relative to the 128-node network (83rd percentile).

Account connectivity and network traffic for the 15 primary accounts is shown in Figure S1. Accounts are ordered in the figure top-to-bottom and left-to-right as a function of their relative eigenvector centrality within the entire 128-account sample. Of these 15, QAnon accounts @Elenochle, @TrustThePlan_, @MrDeeds1111, and @SnowWhite7IAm exerted the most influence within the 128-node network. Substantial network traffic is observed in the lower left quadrant of the figure, as well as among @SnowWhite7IAm, @ETheFriend, and @AmborellaWWG1, not to mention between @AmborellaWWG1 and @MrDeeds1111.

Network metrics were recalculated for the 15-node primary account network alone. The network of primary accounts had 15 vertices with 8 unique edges (of 854 total edges). The reciprocated vertex pair ratio was 0.37 and the reciprocated edge ratio was 0.54, indicating a fairly high proportion of of reciprocal connections for this small network.

Accounts' betweenness centrality was significantly correlated across the 15-node primary account network analysis (Table S5, rightmost column) and the 128-node network, $r(13) = .60$, $p < .05$. However, eigenvector centrality for these accounts was not correlated across the 15-node primary network analysis vs. the 128-node analysis, $r(13) = 0.21$, $p > .05$. (See Figure S2.) The most influential nodes within the 15-node primary account network were @ETheFriend, @JasonSullivan_, CodeMonkeyZ, and @OzRevealed (vs. @Elenochle, @TrustThePlan_, @MrDeeds1111, and @SnowWhite7IAm for the 128-node network). See Table S5 for further details.

**Table S1**. Study 1 network top items. Accounts with one or two asterisks were accounts selected for the Study 1 sample, where two asterisks designated one of the 15 "primary" accounts.

| Rank | Top replied-to | Top mentioned accounts | Top hashtags |
|---|---|---|---|
| 1 | @realDonaldTrump | @realDonaldTrump | #QAnon |
| 2 | @SSG_Pain* | @POTUS | #WWG1WGA |
| 3 | @VincentCrypt46* | @YouTube | #MAGA |
| 4 | @ETheFriend** | @GenFlynn | #Gematria |
| 5 | @KarluskaP | @JoeBiden | #Trump2020 |
| 6 | @AdjunctProfesor* | @SSG_Pain* | #SaveTheChildren |
| 7 | @Listening4His* | @SantaSurfing17* | #SaveOurChildren |
| 8 | @realJamesWoods | @DonaldJTrumpJr | #Unrig |
| 9 | @JuliansRum* | @VincentCrypt46* | #ObamaGate |
| 10 | @Elenochle** | @m_o_t_h_3_r* | #PushingUpDaisies |

**Table S2**. Top 10 accounts for influence measures in Study 1. Primary accounts are shown with an asterisk. See text.

| Rank | Account | # Followers | Account | # Followers/Day | Account | Follower/Followed Ratio | Account | # Total Tweets | Account | # Tweets/Day |
|---|---|---|---|---|---|---|---|---|---|---|
| 1 | QAnon76 | 552,787 | JudithRose91* | 2,521 | X22Report | 17,842 | KLSouth | 616,358 | Tampatha2 | 578 |
| 2 | X22Report | 410,366 | ETheFriend* | 637 | InfiniteChan | 4,565 | SuperYayadize | 604.818 | SuperYayadize | 401 |
| 3 | PrayingMedic | 387,665 | QAnon76 | 559 | IsItWetYet | 2,547 | Darhar981 | 600,667 | Trudi21140637 | 335 |
| 4 | M2Madness | 226,561 | QohnG | 411 | ETheFriend* | 2,518 | ZeusFanHouse | 449,481 | Dovetail1221 | 317 |
| 5 | GrrrGraphics* | 222,585 | MrDeeds1111* | 371 | TrustThePlan17 | 2,024 | AdjunctProfesor | 321,380 | FreedPeacock10 | 277 |
| 6 | JuliansRum | 209,560 | BVoice_P | 353 | WarNuse | 1,617 | QTAnon1 | 228,136 | MagaMedium2 | 277 |
| 7 | InTheMatrixxx | 199,393 | BardsFM | 333 | MajorPatriot | 1,486 | Listening4His | 182,716 | QTAnon1 | 274 |
| 8 | MartinGeddes | 195,082 | Kate_Awakening | 322 | CoreyGoode | 1,023 | EsotericExposal | 182,267 | Majestic691 | 238 |
| 9 | CJTruth | 190,121 | VincentCrypt46 | 321 | QtheWakeUp | 709 | EntheosShines | 164,332 | 4HeartAndSoul | 179 |
| 10 | QTAnon1 | 183,466 | 3Days3Nights* | 313 | LauraMagdalene4 | 665 | CJTruth | 127,114 | AwakenedOutlaw | 173 |

Table S3. Network top items for Study 1 based on retweets only. Accounts with two asterisks (**) indicate accounts in the Study 1 sample designated primary, and accounts with a single asterisk (*) indicate ones that were also part of the sample.

| Rank | Top replied-to | Top mentioned accounts | Top hashtags | Top domains | Top URLs |
|---|---|---|---|---|---|
| 1 | @realDonaldTrump | @realDonaldTrump | #QAnon | Twitter.com | Five-by-Five network YouTube channel |
| 2 | @ETheFriend** | @GenFlynn | #Trump2020 | YouTube.com | 5x5.carrd.co |
| 3 | @RQueenInc** | @YouTube | #StopGroomingKids | CoreysDigs.com | www.grrrgraphics.com |
| 4 | @MrDeeds1111** | @FivexFive2019* | #WWG1WGA | Pscp.tv | Tweet from @KarluskaP about Chicago protests |
| 5 | @DanScavino | @TheSpeaker2018* | #SaveTheChildren | GrrrGraphics.com | salemnow.com/trump-2024 |
| 6 | @Jordan_Sather_ | @CoreysDigs* | #SaveOurChildren | Breitbart.com | Google Map link to Guantanamo Bay Detention Center |
| 7 | @KarluskaP | @POTUS | #Digit | co.uk | Tweet from @realDonaldTrump |
| 8 | @Listening4His* | @RedPill78* | #ObamaGate | FoxNews.com | YouTube channel |
| 9 | @Kevin_Shipp | @GoJackFlynn | #GodWins | TheGatewayPundit.com | grrrgraphics.com |
| 10 | @MajorPatriot* | @SpirituallyRaw* | #DisneyGate | Zerohedge.com | Tweet from @ElijahSchaffer |

**Table S4**. Descriptions of 15 accounts of primary interest in Study 1.

| Account | Description |
| --- | --- |
| @JasonSullivan_ | A Roger Stone associate posting frequent pro-QAnon slogans who had operated the Power 10 botnet on behalf of the Republican Party for boosting pro-Trump propaganda on Twitter through March, 2020. |
| @RQueenInc | The account for Jim Watkins, owner/operator of the 8chan message board and the qmap.pub site, which were used for "Q drops", i.e., the cryptic messages supposedly from the anonymous government insider known as "Q". |
| @CodeMonkeyZ | Account for Jim Watkins' son Ronald Watkins, operator of the 8kun message board used for Q drops after the New Zealand massacre. |
| @ETheFriend | An influential QAnon organizer account speculated in media and on Twitter to be operated by a government insider. |
| @OzRevealed | QAnon account mentioned by, and in frequent communication with, @JasonSullivan_ displaying magical and mysterious imagery. |
| @3Days3Nights | Another QAnon account mentioned by, and in frequent communication with, @JasonSullivan_. |
| @TrustThePlan_ | An influential QAnon account with a high number of followers whose profile appeared to mimic that featured presentation of QAnon being military operations and later mimicked @ETheFriend. |
| @SnowWhite7IAm | Account featuring otherworldly, professional-style graphics attributed to Lisa Clapier and noted for promoting the QAnon hashtags #FollowTheWhiteRabbit and #FollowSnowWhite from early 2018 and for. |
| @GrrrGraphics | Account for Ben Garrison, an influential pro-Trump cartoonist with a strong on-line presence. |
| @OSSRobertSteele | Account for Robert David Steele, a former intelligence official known for openly advocating QAnon, claiming that colonies of enslaved children exist on Mars, and the dismantling of the Deep State. |
| @JudithRose91 | Account for a viral TikTok user advancing a persona of a former liberal voter now promoting QAnon and Trump. This account replaced a prior suspended account. |
| @Elenochle | Influential pro-Trump organizer account which assisted organizing videos during the QAnon #TakeTheOath campaign posted by "patriots" in Summer, 2020. |
| @VeteransAlways_ | QAnon account displaying religious and esoteric themes (alien races, numerology/gematria, the coming apocalypse), as well as mysterious, striking, esoteric, and colorful imagery. |
| @MrDeeds1111 | An influential pro-QAnon account organizing "follow trains" which featured colorful, eye-catching graphics and whose profile featured the logo for internet puzzle Cicada 3301 for many months |
| @AmborellaWWG1 | QAnon organizer account displaying striking otherworldly imagery and ornate fonts advertising QAnon; this account promoted itself as a "backchannel". |

**Table S5**. Account influence statistics for Study 1 for 15 accounts of primary interest. Accounts are presented in order of descending value of follower:followed ratio. Column 1 lists the primary account. Column 2 and 4-7 list several influence metrics and days since account creation (column 3); for these columns, the percentile rank with respect to the 128 accounts is listed. Columns 8 and 9 list eigenvector and betweenness centrality ranks relative to the 128-node network; the value in parentheses lists the rank for these metrics of the account relative to the 15 other primary accounts within this dataset only. Columns 10 and 11 list eigenvector centrality rank and betweenness centrality rank, respectively, for the 15-node network of primary accounts; the value in parentheses lists the change in rank for the metric of the same type relative to the 128-node network.

| Account | # Followers (%ile) | Account Age (Days) | # Followers /Day (%ile) | Follower: Followed Ratio (%ile) | # Total Tweets (%ile) | Tweets /Day (%ile) | Eigenvector centrality rank,128-node network | Betweenness centrality rank, 128-node network | Eigenvector centrality rank,15-node network | Betweenness centrality rank, 15-node network |
|---|---|---|---|---|---|---|---|---|---|---|
| ETheFriend | 55,390 (65) | 87 (6) | 637 (99) | 2,518 (98) | 2,643 (24) | 30 (65) | 57 (9) | 85 (4) | 1 (-8) | 1 (-3) |
| OzRevealed | 7,491 (30) | 275 (20) | 27 (47) | 576 (91) | 709 (7) | 3 (19) | 13 (14) | 13 (15) | 4 (-10) | 12 (-3) |
| TrustThePlan | 34,876 (53) | 3,221 (76) | 11 (24) | 513 (89) | 8,539 (35) | 3 (20) | 87 (2) | 92 (2) | 10 (+8) | 3 (+1) |
| CodeMonkeyZ | 126,342 (86) | 2,543 (70) | 50 (65) | 430 (87) | 1,975 (14) | 1 (7) | 46 (11) | 78 (8) | 3 (-8) | 10 (+2) |
| JudithRose91 | 26,115 (47) | 10 (1) | 2,521 (>99) | 237 (84) | 206 (2) | 20 (57) | 60 (8) | 80 (6) | 13 (+5) | 8 (+2) |
| OSSRobertSteele | 2,065 (14) | 1,017 (46) | 2 (12) | 90 (78) | 1,487 (10) | 1 (14) | 6 (15) | 21 (14) | 15 (0) | 15 (+1) |
| Elenochle | 134,504 (87) | 981 (43) | 137 (83) | 76 (76) | 32,409 (69) | 33 (67) | 91 (1) | 83 (5) | 7 (+6) | 5 (0) |
| SnowWhite7IAm | 28,215 (50) | 1,020 (48) | 28 (49) | 56 (71) | 62,836 (80) | 62 (76) | 77 (4) | 76 (9) | 8 (+4) | 4 (-5) |
| 3Days3Nights | 58,241 (66) | 186 (17) | 313 93) | 52 (70) | 7,174 (30) | 39 (68) | 76 (5) | 68 (11) | 9 (+4) | 7 (-4) |
| GrrrGraphics | 222,585 (97) | 3,800 (86) | 59 (70) | 8.7 (55) | 43,936 (76) | 12 (43) | 63 (7) | 94 (1) | 6 (-1) | 2 (+1) |
| AmborellaWWG1 | 5,234 (24) | 1,820 (59) | 3 (17) | 6.0 (51) | 17,923 (56) | 10 (41) | 69 (6) | 72 (10) | 11 (+5) | 6 (-4) |
| VeteransAlways | 49,702 (64) | 902 (41) | 55 (68) | 4.0 (42) | 83,726 (84) | 93 (83) | 40 (12) | 57 (12) | 12 (0) | 14 (+2) |
| MrDeeds1111 | 37,540 (56) | 101 (8) | 371 (97) | 1.7 (28) | 13,752 (48) | 136 (87) | 83 (3) | 87 (3) | 14 (+11) | 13 (+10) |
| RQueenInc | 6,859 (28) | 244 (19) | 28 (50) | 1.5 (24) | 8,041 (33) | 33 (66) | 50 (10) | 80 (7) | 5 (-5) | 11 (+4) |
| JasonSullivan | 15,409 (40) | 2,366 (65) | 7 (23) | 1.1 (16) | 28,827 (65) | 12 (45) | 30 (13) | 50 (13) | 2 (-11) | 9 (-4) |

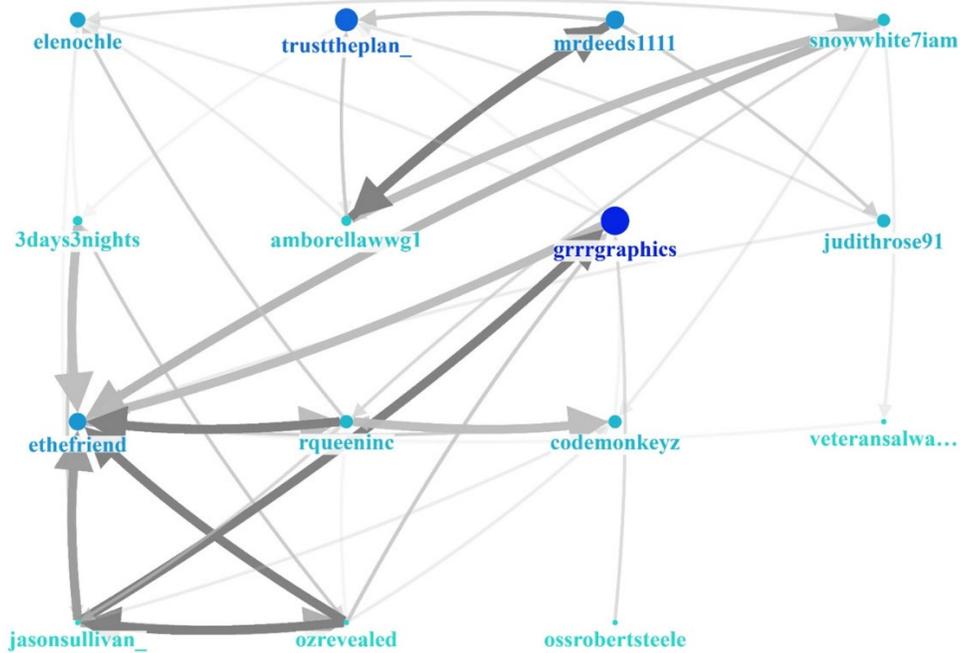

**Figure S1.** Network for 15 primary accounts in Study 1. Vertex layout left-to-right and top-to-bottom indicates rank ordering of eigenvector centrality based on 128-node analysis. Vertex color and size indicate relative rank ordering of betweenness centrality from 128-node analysis; darker blue and/or larger vertices represent higher values. Edges connecting nodes are directional. Line darkness and width indicate edge weight.

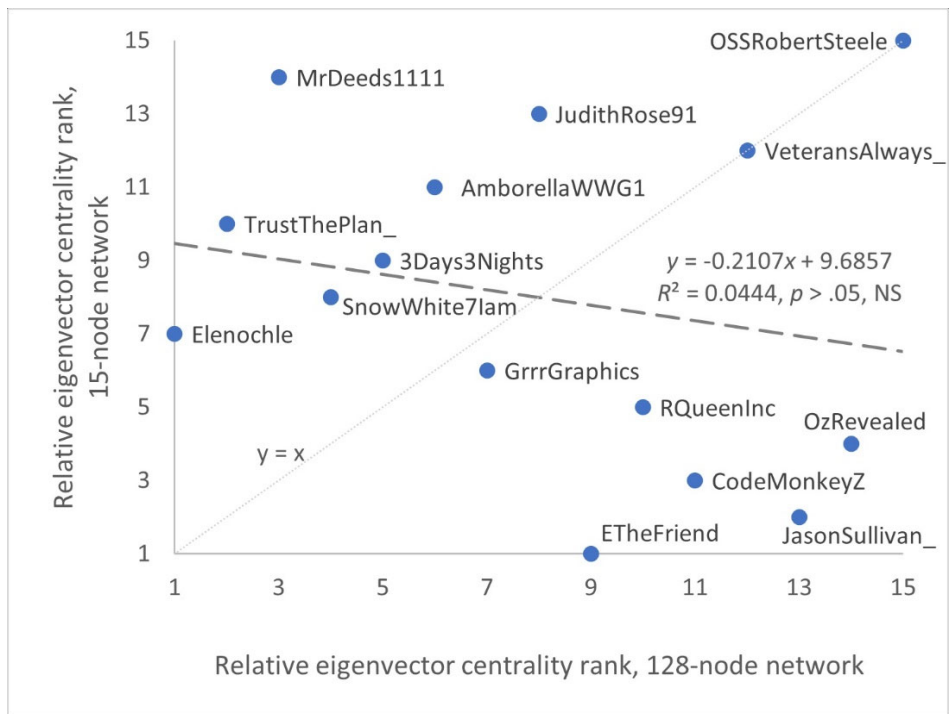

**Figure S2.** Eigenvector centrality for Study 1 for 15 primary accounts calculated relative to the 128-node network and for the 15-node primary account subnetwork alone.

## 2 Study 2

The network sample showed 1,063 vertices forming 16,245 unique edges (699,670 total edges), with 393,347 self-loops. The reciprocated vertex pair ratio was 0.099 and the reciprocated edge ratio was 0.179. Data were graphed using the Fruchterman Reingold algorithm with groups laid out in different panels. Details about keywords and hashtags used in sentiment analysis and found in the various clusters can be found in Tables S6-S8.

Accounts ranked 11-20 for eigenvector centrality were @MagnetGlove, @BobRoon12515016, @TrumperWavin, @PrayingMedic, @NYPost, @BruceVeloor, @Qtah17, @Whiplash347, @JFKJrFan, and @Iakining. The top 15 accounts for betweenness centrality, in descending order, were: @NYPost, @RudyGiuliani, @CJTruth, @VincentCrypt46, @FoxNews, @JohnMappin, @DanScavino, @FLOTUS, @PrayingMedic, @AnnaKhait, @Qtah17, @Grzegosh21, @M2Madness, @MajorPatriot, and @InTheMatrixxx.

(1) *Follower acquisition rate*. Subnetwork accounts acquired $M = 40$ followers/day (SD: 91; range: 0.2 – 635; 1st quartile = 2.4, median = 9; 3rd quartile = 35). On average, accounts were $M = 1,281$ days old (*SD*: 1,363; range: 15 – 4,488; 1st quartile = 157; median = 666; 3rd quartile = 2013).
(2) *Follower:followed ratio*. Subnetwork accounts had a mean follower:followed ratio of $M = 91:1$ (*SD*: 590:1; range: 0 – 5,869:1; 1st quartile = 0.95:1; median = 1.3:1; 3rd quartile = 4.7:1).
(3) *Tweet activity*. Subnetwork accounts showed $M = 35,059$ total tweets (*SD*: 68,234; range: 87 – 569,736; 1st quartile = 6,685; median = 12,748; 3rd quartile = 32,969).
(4) *Tweet rate*. Accounts sent a mean $M = 63$ tweets/day (*SD*: 91; range: 1 – 515; 1st quartile = 8.2; median = 26; 3rd quartile = 85).
(5) *Eigenvector centrality*. The top 20 accounts for eigenvector centrality based on calculations for the subnetwork were: @dianaanddennis, @magnetglove, @neo081001, @qbeat107, @nate_dunker, @vickidale12, @twistedspiral2, @simpattyk, @zyishai, @truenpatriot, @aqpatrons, @iakining, @freedpeacock10, @carolineelleni1, @airforcemomof4, @bobroon12515016, @anasha47662718, @quixotry9, @serum2all, and @suejone54923872.

Significant bivariate correlations not already reported are as follows: eigenvector centrality rank across the entire network vs. eigenvector centrality within subnetwork [$r = .67, p < .001$]; number of followers vs. follower:followed ratio [$r = .66, p < .001$]; number of followers vs. number of total tweets [$r = .57, p < .001$]; number of followers vs. followers/day [$r = .67, p < .001$]; follower:followed ratio vs. followers/day [$r = .51, p < .001$]; total tweets vs. tweets/day [$r = .44, p < .001$]; followers/day vs. tweets/day [$r = .56, p < .001$]. All other bivariate correlations were nonsignificant ($p$'s > .05).

Illustrating some notable account properties (ref. Tables S9-S11), @Phaethon314 was less than three months old but accrued 255 followers/day – the 5th highest for this subnetwork. This account had a highly generic and anonymous profile consisting of pictures of clouds and an aerial landscape, yet it had accrued over 21,500 followers and followed 365. @JFKJrFan and @Okabaeri9111 were next for rate of follower accrual at 129 and 123 followers/day, respectively. @JFKJrFan was only about 6 months old; yet it was followed by 24,996 and followed 23. @Okabaeri9111 was followed by 79,388 and followed 54, 624; the Twitter profile for this account advertised "QMapJapan" and linked to the "QarmyJapanFlynn Official Website". These data are consistent with account coordination and astroturfing in the QAnon propaganda campaign.

**Table S6**. Keywords and hashtags for Study 2 sentiment analysis. Alternative spellings were accepted.

| QAnon themes | New Age/Esoteric themes | | Conservative themes | |
|---|---|---|---|---|
| 17 | #Aquarius | heart | 3:16 | forgiven |
| #AustinSteinbart | #Aries | hippie | 45 | freedom |
| #AustinSteinbartIsQ | #blueavian | interstellar | #2A | freedoms |
| #BabyQ | #Cancer | Leo | #AmericaFirst | God |
| #FollowSnowWhite | #Capricorn | Libra | #ConservativePatriot | God's |
| #FollowTheWhiteRabbit | #clairvoyant | lifestyle | #constitution | heaven |
| #JFKJR | #Galactic | lightbearer | #DeepState | holy |
| #Q | #Gemini | lightworker | #DigitalSoldier | Israel |
| #QAnon | #goddess | love | #DigitalSoldiers | Jesus |
| #Qarmy | #Leo | medium | #DrainTheSwamp | John |
| #SaveourChildren | #Libra | mindfulness | #FamilyIsEverything | KAG |
| #SaveTheChildren | #lyracommander | organic | #Fightback | kek |
| #SpiritualWarfare | #NewAge | paranormal | #HonorVets | kill |
| #TheGreatAwakening | #phoenixrising | Pisces | #JesusIsLord | King |
| #TrustThePlan | #Pisces | Pleiadian | #Kag | liberty |
| #WWG1WGA | #Sagittarius | portal | #KAG2020 | libtard |
| afraid | #Scorpio | priestess | #LatinasForTrump | libtards |
| Anon | #starseed | prophet | #MAGA | MAGA |
| Anons | #Taurus | psychic | #Navy | Marine |
| Austin | #UFO | sacred | #PresidentT | meme |
| AustinSteinbart | #UFOs | Sagittarius | #Trump2020 | memes |
| awakening | #VictoryOfTheLight | Scorpio | #womenfortrump | military |
| biblical | #Virgo | seer | #YeshuaIsLORD | Navy |
| cabal | alien | shaman | @realDonaldTrump | open |
| Cue | aliens | spiritual | 45th | patriot |
| drops | Angel | spiritualist | aequitas | POTUS |
| gematria | Aquarius | starseed | American | praise |
| GESARA | Arcturian | Taurus | believer | prayer |
| Kennedy | Aries | UFO | blessed | President |
| NESARA | artist | ufologist | born | Pro-Israel |
| pedophile | ascendant | UFOs | bornagain | redeemed |
| pedophiles | ascension | unidentified | Christ | religious |
| pilled | Cancer | vibration | Christian | Republic |
| Q | Capricorn | Virgo | communist | Rev1:8 |
| QAnon | channeled | wellness | conservative | righteous |
| Q-Anon | clairvoyant | wisdom | constitution | scripture |
| Q-Team | consciousness | | country | SEAL |
| queue | Earth | | deplorable | semper |
| rabbit | earthling | | devil | soldier |
| redpill | ECETI | | Ephesians | soul |
| redpilled | energy | | evil | souls |
| Satan | enlightenment | | eyes | spirituality |
| Satanic | frequency | | faith | swamp |
| Snow | Gemini | | fascism | Trump |
| Steinbart | geometry | | father | United |
| storm | harmonic | | fidelis | US |
| W|W|G|!1|W|G|A | healer | | flag | veritas |
| wake | health | | Flynn | veteran |
| WWG1WGA | healthy | | | veterans |
| | | | | warrior |

Table S7. Top-10 hashtag categorization for accounts in subnetwork in Study 2.

| Category | Match description |
| --- | --- |
| QAnon | #Adrenochrome; #Anons; any references to JFK Jr.; #Biblical; #Bqqm; #CabalTakeDown; casting Democrats as pedophiles or cannibals (e.g., #BillClintonIsAPedo, #Pedogate, #Pedowood, #Pizzagate, #Frazzledrip); #DarkToLight; #DeepState; #DoItQ; #DoItQArmy; #DoYourOwnResearch; #FutureProvesPast; #GalacticReunion; #Gematria; #Gitmo; #Godwins and variants (#Godwon, etc.); any hashtags with '17' (e.g., #DigitalSoldiers17); #HuntersBecomeTheHunted; #Neioh; #Nesara; #NothingCanStopWhatIsComing and variants; #Q, 'Q' followed by anything (e.g., # QArmyJapanFlynn, #QAnon), #SaveTheChildren and variants (e.g., #SaveOurChildren); #TheGreatAwakening and variants; #TheMoreYouKnow; #ThesePeopleAreSick and variants (e.g., #ThesePeopleAreEvil); #TrustThePlan & variants (#TrustTheDivinePlan); #WWG1WGA and variants |
| Pro-Conservative sentiment | #AllLivesMatter; neutral, positive, or gloating allusions to Trump and/or Trump victory, presidency, or praise (e.g., #PresidentDonaldTrump2020, #BestPresidentEver45, #PresidentT, #Trump2020, #Trump2020mf); #AmericaFirst; references to any other Kennedy family member than JFK Jr.; #BendedKnee; #BibleCode; #DigitalArmy; #FightBack; #Fooked; #GenFlynn; #Hamashia; #LawAndOrder; #MAGA and variants (e.g., #MAGA2020); #OctoberSurprise; #Patriot and variants; references to Biden denigration narratives (e.g., #BumblingJoe); references to Biden family or Obama scandal or crime narratives (e.g., #HunterBidenEmails, #LaptopFromHell, #ObamaGate, #TheMediaLiesObamaSpies and variants, #BidenCrimeFamily and variants); references to Clinton crime narratives other than pedophilia; references to family values (#FamilyIsEverything, #ForTheLoveOfOurChildren); references to protecting human life (e.g., #HumanLife); references to Republican victory (e.g., #RedWave2020, #VoteRedToSaveAmerica); references to the Seth Rich conspiracy/Republican exoneration narratives (e.g., #SayHisName); references to Trump narratives (#hydroxychloroquine); #TrustGod; #Truth; #USMC; 'X' forTrump (e.g., #BoatersforTrump); #Yashua and variants |

**Table S8**. For Study 2 top hashtags and languages by group, from most to least frequent. QAnon hashtags are shown in italics. Group 9 had no hashtags.

| Group | Top Hashtags |
|---|---|
| All | *#WWG1WGA*, #MAGA, *#QAnon*, #Trump2020, #HunterBiden, #WarRoomPandemic, #BidenCrimeFamily, #SteveBannon, #Debates2020, *#TheGreatAwakening* |
| 1 | #MAGA, #HunterBiden, #WarRoomPandemic, #SteveBannon, #BidenCrimeFamily, #Debates2020, *#QAnon*, #Trump2020, #Trump, #JoeBiden |
| 2 | *#WWG1WGA*, #Trump2020, #PresidentT, #MAGA, #HoHoHo, #SCOTUS, *#Q*, #GodWins, #Fooked, *#QAnon* |
| 3 | *#WWG1WGA*, *#AustinSteinbartIsQ*, *#TheGreatAwakening*, *#QAnon*, *#Assange*, *#FreeAustinSteinbart*, *#FreeAustin*, #FreeAssange, #AssangeCase, #F2B |
| 4 | #MAGA, #POTUS, #MAGARollerCoaster, #Trump2020, #EnationTrains, #VoteRedToSaveAmerica, #WeLoveYouTrump, *#SaveOurChildren*, #Trump2020LandslideVictory, #Trump2020Landslide |
| 5 | #FootSoldierOrigins, #British, #RiseOfTheFootSoldier, #Movies, #NeroVodka, #RiseOfTheFootSoldier3, #OnceUponATimeInLondon, #Essex, #Film, #RiseOfTheFootSoldierOrigins |
| 6 | #JoyTrain, #Love, #GoldenHearts, #LightUpTheLove, #ChooseLove, #IamChoosingLove, #FamilyTrain, #StarfishClub, #JesseLewisChooseLoveMovement, #Lutl |
| 7 | #FF, #KoronawirusPolska, #Eutanazjaplus, #Zajob |
| 8 | #StopDeVirusHooligans, #StopWillemEngel, #RUQR, #Fabeltjesfuik, #RUQRExposed, #Nietinmijnnaam, #Zondagmetlubach, #Lamlultinus, #Viruswappies, #CovidIdioten |

Table S9. Select influence metrics for Study 2: number of followers, followers/day, and followers/Followed ratio. Percentages in parentheses reflect the percentile rank in the subnetwork of N = 109 accounts and the overall network of all N = 1,063 accounts, respectively.

| Rank | Followers | | Followers/Day | | Followers:Followed Ratio | |
|---|---|---|---|---|---|---|
| | Account | # Followers | Account | # Followers /Day | Account | Ratio |
| 1 | PrayingMedic (8%, 99%) | 416,448 | Trumperwavin (26%, 99%) | 635 | Striderraven1 (6%, 79%) | 5868.8 |
| 2 | John_F_Kennnedy (50%, >99%) | 388,704 | John_F_Kennnedy (50%, >99%) | 420 | MajorPatriot (5%, 97%) | 1631.2 |
| 3 | M2Madness (7%, 97%) | 256,152 | VincentCrypt46 (78%, >99%) | 376 | JFKJrFan (63%, 98%) | 1086.8 |
| 4 | VincentCrypt46 (78%, >99%) | 234,507 | 3Days3Nights (36%, 94%) | 365 | VincentCrypt46 (78%, >99%) | 366.4 |
| 5 | CJTruth (34%, >99%) | 228,543 | Phaethon314 (3%, 79%) | 255 | Qtah17 (13%, 98%) | 174.2 |
| 6 | MajorPatriot (5%, 97%) | 156,599 | JFKJrFan (63%, 98%) | 129 | AQTime (32%, 98%) | 92.9 |
| 7 | InItToWinIt007 (4%, 80%) | 122,165 | okabaeri9111 (1%, 59%) | 123 | 3Days3Nights (36%, 94%) | 81.0 |
| 8 | 55True4U (35%, 96%) | 108,008 | Dovetail1221 (40%, 93%) | 120 | DoqHolliday (22%, 96%) | 79.6 |
| 9 | AQTime (32%, 98%) | 103,098 | Stormis_Us (38%, 92%) | 111 | SnowWhite7IAm (55%, 97%) | 66.8 |
| 10 | 3Days3Nights (36%, 94%) | 89,873 | M2Madness (7%, 97%) | 106 | Phaethon314 (3%, 79%) | 59.0 |

**Table S10**. Select influence metrics for Study 2: total tweets and tweets/day. Percentages in parentheses reflect the percentile rank in the subnetwork of N = 109 accounts and the overall network of all N = 1,063 accounts, respectively.

| Rank | Tweets | | Tweets/Day | |
|---|---|---|---|---|
| | Account | # Tweets | Account | # Tweets/Day |
| 1 | InItToWinIt007 (4%, 80%) | 569,736 | AQPatrons (90%, 97%) | 515 |
| 2 | Pat_Shanks (15%, 83%) | 270,892 | Goosesotherhalf (50%, 93%) | 365 |
| 3 | era_pero_ya_no (43%, 91%) | 254,898 | neo081001 (97%, 99%) | 360 |
| 4 | Serum2all (83%, 95%) | 152,382 | InItToWinIt007 (4%, 80%) | 333 |
| 5 | CJTruth (34%, >99%) | 134,110 | Dovetail1221 (40%, 93%) | 302 |
| 6 | suejone54923872 (82%, 93%) | 115,200 | Sommertime812 (30%, 85%) | 298 |
| 7 | Bestnaunieever (19%, 87%) | 113,786 | FreedPeacock10 (88%, 99%) | 257 |
| 8 | p0a_triot23 (23%, 88%) | 99,922 | TwistedSpiral2 (94%, 95%) | 245 |
| 9 | VeteransAlways (5%, 87%) | 91,497 | kat14kountry (17%, 79%) | 215 |
| 10 | 55True4U (35%, 96%) | 90,020 | Carolinaouest (24%, 90%) | 205 |

**Table S11**. Study 2 top replied-to, top mentioned, and top tweeters. Accounts with an asterisk (*) indicate accounts selected for the Study 2 sample.

| Rank | Top replied-to | Top mentioned accounts | Top tweeters |
|---|---|---|---|
| 1 | @hsretoucher* | @hsretoucher* | @mitchiepoo46* |
| 2 | @vincentcrypt46* | @realdonaldtrump | @ocraziocornpop* |
| 3 | @ethefriendly* | @neo081001* | @drawandstrike* |
| 4 | @cjtruth* | @vincentcrypt46* | @inittowinit007* |
| 5 | @realdonaldtrump | @goosesotherhalf* | @elgatoweebee* |
| 6 | @trumperwavin* | @dianaanddennis* | @duxgirl27* |
| 7 | @doqholliday* | @freedpeacock10* | @foxnews* |
| 8 | @phaethon314* | @whitebunnyq* | @wavetossed* |
| 9 | @kendavi80404473* | @quixotry9* | @uncjay1* |
| 10 | @m2madness* | @danauito* | @kirchbabe* |

## 2.1 Comparison of QAnon Influencer accounts with other key account types in Study 2

Subsets of collected Twitter data were created to evaluate relative network influence of each of four key groups of accounts (New Age/Esoteric Prototypical, New Age Media Influencers, QAnon Insiders, and Prominent Pro-Trump) against a benchmark set of QAnon Influencers (Tables S12, S13). To do so, we combined Twitter data for QAnon Influencers with each other group to create four separate sub-network datasets. For instance, to examine the influence of New Age/Esoteric Prototypical accounts against QAnon Influencers, we used NodeXL to re-calculate network measures, "skipping" all other vertices except those for accounts in these two groups. Measures were calculated based on retweets only. Technical details of the subnetworks created for these analyses can be found in Table S14.

Here, we define to be "high-ranked" those accounts in a key category which were in the top 35% ranked for eigenvector centrality in the network created through combination with the benchmark set of QAnon Influencer accounts. See Figures S3-S6. Six of 23 New Age/Esoteric Prototypical accounts were high-ranked relative to QAnon Influencers: @MrDeeds1111, @StarSeed7772, @VDarknessF, @AlaraOfSirius, @TwistedSpiral2, and @TimeTravelAnon. Four of 25 New Age/Media Influencers were high-ranked: @NoRestrictions, @RobbyStarbuck, @Ben_Chasteen, and @RobCounts. Four of 19 Prominent Pro-Trump accounts were high-ranked: @DanScavino, @RudyGiuliani, @ChanelRion, and @AnnaKhait. Finally, three of 31 QAnon Insiders were high-ranked: @Saint_Germain5, @Acoustamatic, and @JohnMappin.

We first compared the relative influence of QAnon Influencer accounts against that of New Age/Esoteric Prototypical accounts by conducting a network analysis combining these two subsets of data. We focused on retweets only, which typically reflect endorsement. Results revealed that several New Age/Esoteric Prototypical accounts showed considerable influence within the combined network. Figure S3 depicts eigenvector centrality for the accounts scoring highest for eigenvector centrality (N = 36); the layout top-to-bottom and left-to-right reflects' accounts relative rank for this measure. Several New Age/Esoteric Prototypical accounts – including @StarSeed7772, @VDarknessF, @AlaraOfSirius, @TwistedSpiral2, @TimeTravelAnon, and @MrDeeds1111 – were among these top-ranked accounts. This analysis approach was repeated for each of the other groups; see Table S14 and Figures S3-S6 for more information.

**Table S12**. Accounts selected for Study 2 representing "benchmark" QAnon Influencers.

| QAnon Influencers (N = 56) | |
|---|---|
| 1naasty | qanon76 |
| 3days3nights | qohng |
| 55true4u | qtah17 |
| 99freemind | qthewakeup |
| an0n661 | ru_awake_yet |
| andweknow | santasurfs17 |
| anonpatriotq | savagedystrophy |
| aqtime | seancordanon |
| awakenedoutlaw | snowwhite7iam |
| awakening5d | ssg_pain |
| awishnstar2 | stoppedago |
| beer_parade | stormis_us |
| benosey | thesharpedge1 |
| cjtruth | thespeaker2018 |
| doqholliday | trusttheplan_ |
| elenochle | vfuska |
| ethefriendly | vincent__fusca |
| freedpeacock10 | vincentcrypt46 |
| hsretoucher | wokesocieties |
| inthematrixxx | x22report |
| ipot1776 | |
| john_f_kennnedy | |
| judithrose91 | |
| juliansrum | |
| kate_awakening | |
| ladyqanuck | |
| littllemel | |
| llinwood | |
| m_o_t_h_3_r | |
| m2madness | |
| majorpatriot | |
| martingeddes | |
| ozreturns_ | |
| p0a_triot23 | |
| prayingmedic | |
| qanon_report | |

Supplementary Material

**Table S13**. Accounts selected for representing "comparison" key category types (New Age/Esoteric Prototypical, New Age/Media Influencer, Insiders, Prominent Pro-Trump).

| New Age/Esoteric Prototypical (N = 23) | New Age/Media Influencers (N = 25) | Insiders (N = 31) | Prominent Pro-Trump (N = 19) |
|---|---|---|---|
| alaraofsirius | ben_chasteen | acoustamatic | annakhait |
| amborellawwg1 | benchasteen | actondavid | atensnut |
| ariellaonewland | blueavians | annvandersteel | chanelrion |
| i_am_liqht | david_wilcock | bill_binney | charliekirk11 |
| kabamur_taygeta | davidicke | docdhj | danscavino |
| lightworkercain | drstevengreer | drcharlieward1 | debbieaaldrich |
| marshal45872251 | eceti | fieldmcc | devinnunes |
| mrdeeds1111 | emerysmith33 | georgwebb | flotus |
| petahjane | janetossebaard | immappin | foxnews |
| phaethon314 | jchurchradio | itnj_committee | georgepapa19 |
| pharaoh_aten_ | johnxdesouza | jerome_corsi | hawleymo |
| siriusbshaman | lauramagdalene4 | johnbwellsctm | jim_jordan |
| starseed521 | michaelsalla | johnmappin | lindseygrahamsc |
| starseed7772 | michaeltelling2 | kevin_shipp | markdice |
| starsoul777 | nickhintonn | kirkwiebe | mchooyah |
| thejennifermac | norestrictions | laralogan | realtina40 |
| timetravelanon | richdolan | listen4always | rudygiuliani |
| twistedspiral2 | robbystarbuck | namemysock | sentedcruz |
| utsava4 | robcounts | ossrobertsteele | warroompandemic |
| valiantthor12 | sachastone | patton6966 | |
| vdarknessf | stevebassett | raymcgovern | |
| veteransalways_ | ubuntumichael | realmattcouch | |
| xoana_ra | watchingsean | saint_germain5 | |
| | yourmothergaia | steveouttrim | |
| | | stevepieczenik | |
| | | stranahan | |
| | | suzi3d | |
| | | t_s_p_o_o_k_y | |
| | | tracybeanz | |
| | | trevorfitzgibb1 | |
| | | wyattearpp_ | |

**Table S14**. Subnetworks in Study 2 created by comparing our benchmark set of QAnon Influencer Accounts and other account types.

|  | Combined with N = 23 Prototypical Esoteric Accounts | Combined with N = 25 New Age Media Influencer Accounts | Combined with N = 31 Insider Accounts | Combined with N = 19 Prominent Pro-Trump Accounts |
|---|---|---|---|---|
| Vertices | 79 | 81 | 87 | 75 |
| Unique Edges | 322 | 252 | 294 | 316 |
| Edges with Duplicates | 2096 | 1596 | 2137 | 1834 |
| Total Edges | 2418 | 1848 | 2431 | 2150 |
| Self-Loops | 904 | 589 | 815 | 744 |
| Reciprocated Vertex Pair Ratio | .099 | .100 | .092 | .086 |
| Reciprocated Edge Ratio | .181 | .182 | .169 | .158 |

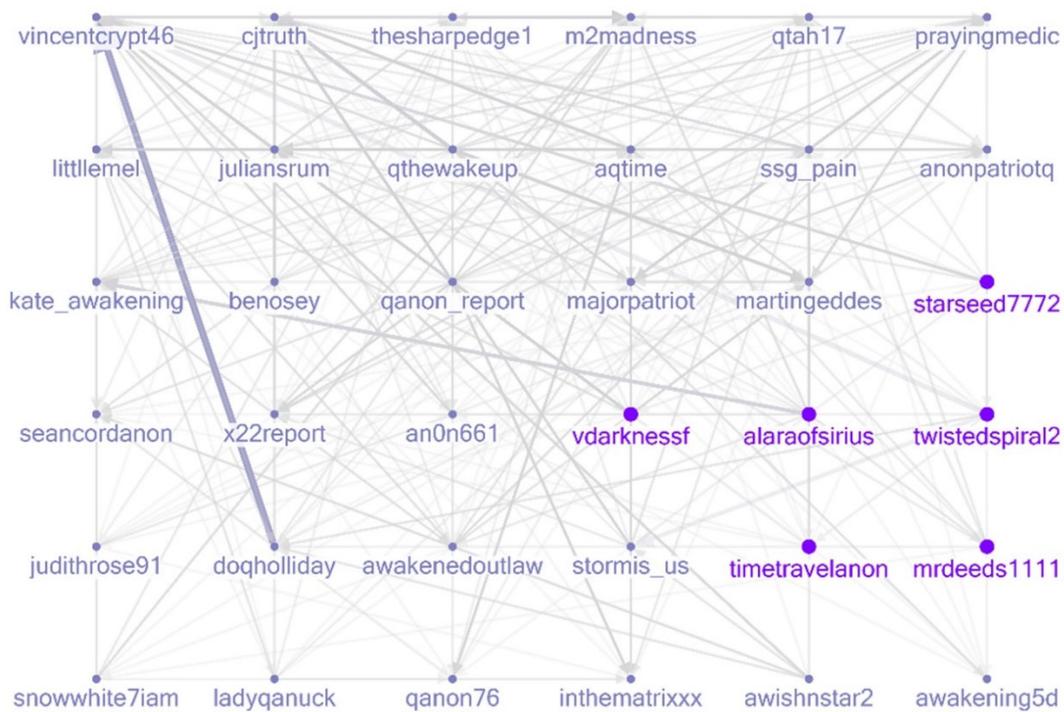

**Figure S3.** For Study 2, the top 36 most influential accounts (based on eigenvector centrality) in a subnetwork formed by accounts designated QAnon Influencers (grey-purple) and New Age/Esoteric Prototypical accounts (bright purple).

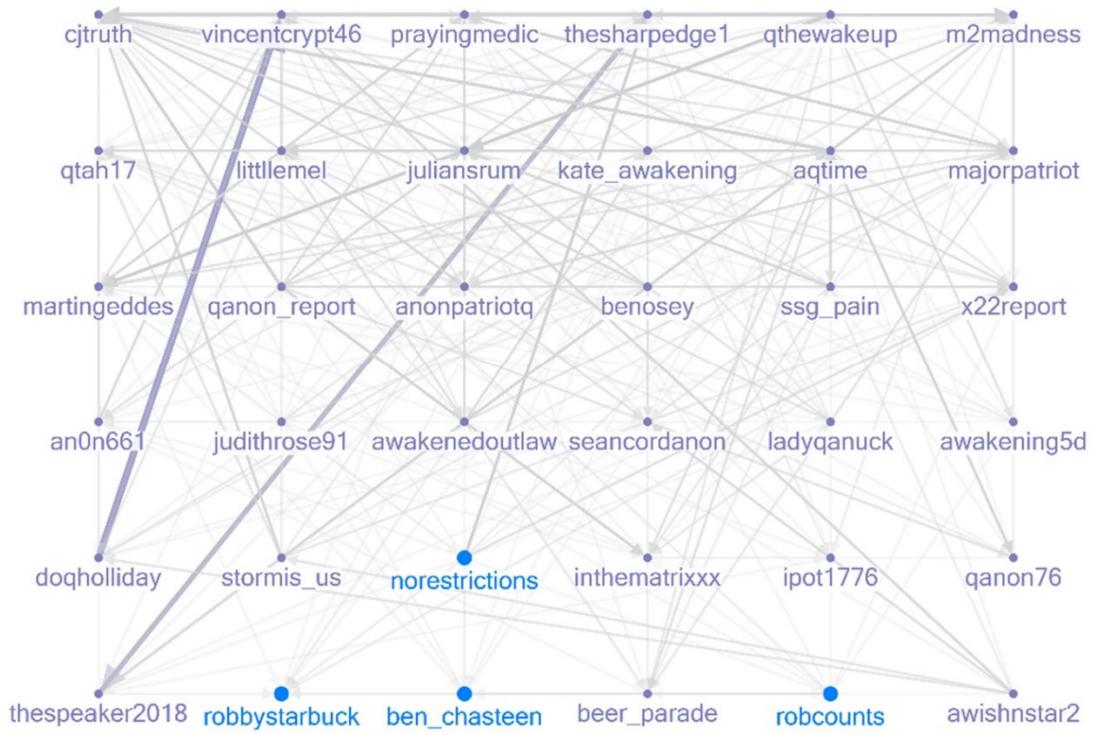

**Figure S4.** For Study 2, the top 36 most influential accounts (based on eigenvector centrality) in a subnetwork formed by QAnon Influencers (grey-purple) and New Age Media Influencers (blue).

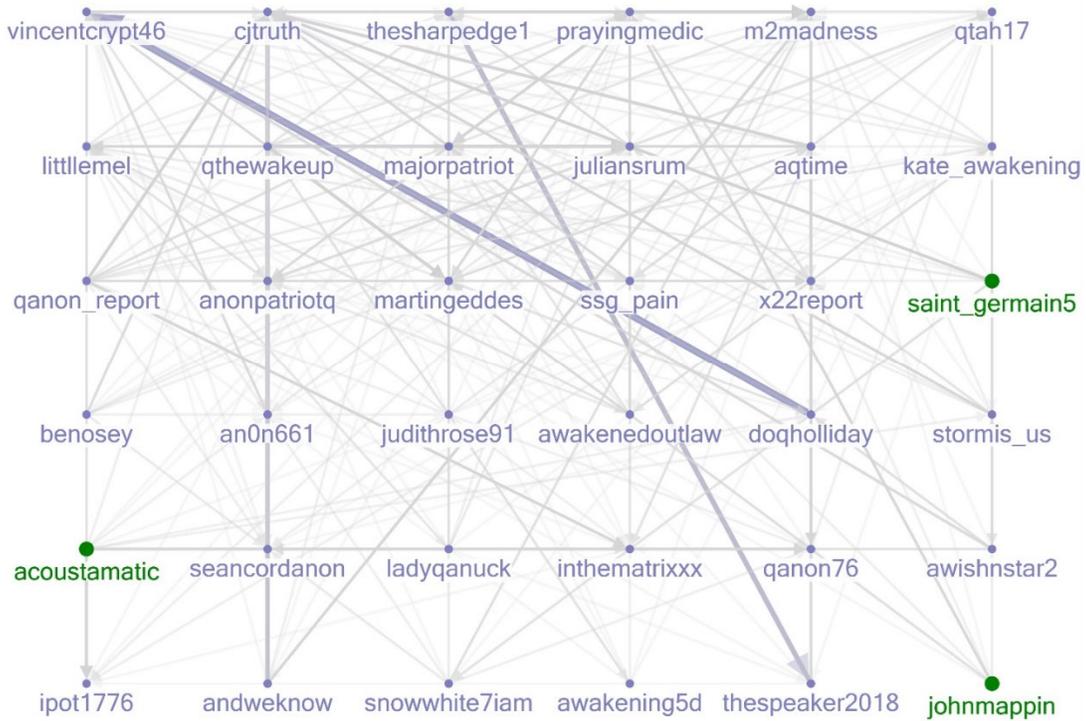

**Figure S5.** For Study 2, the top 36 most influential accounts (based on eigenvector centrality) in a subnetwork formed by accounts designated QAnon influencers (grey-purple) and QAnon Insider accounts (bright purple).

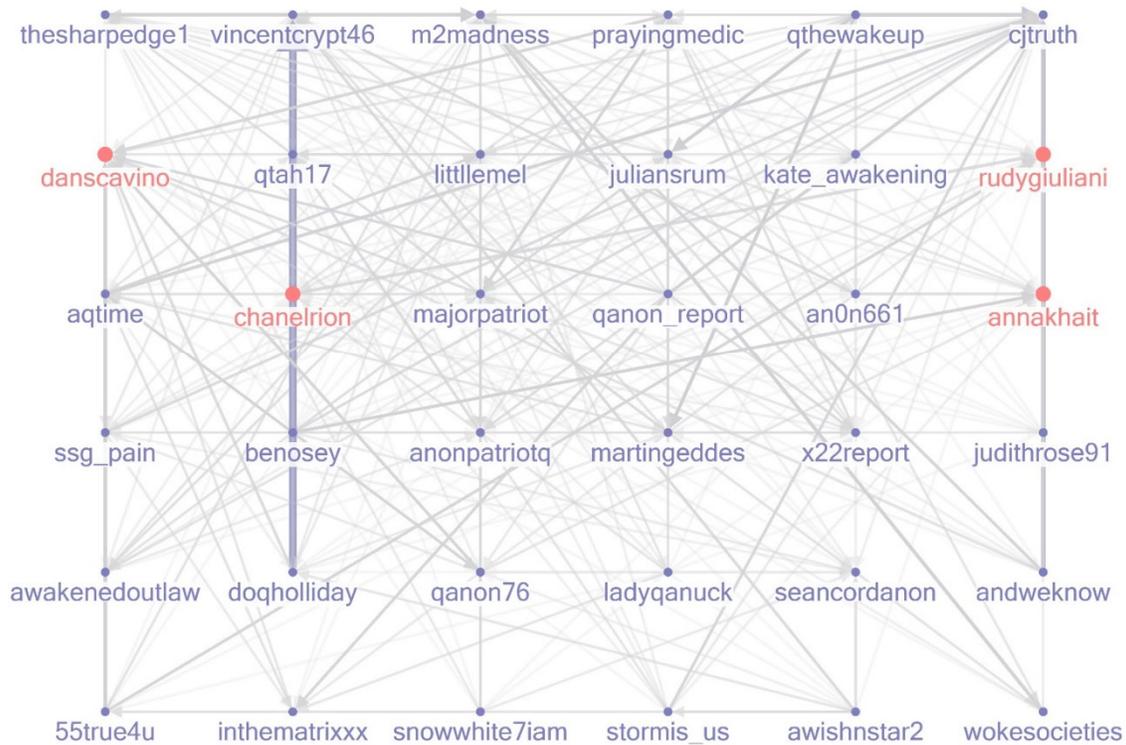

**Figure S6.** For Study 2, the top 36 most influential accounts (based on eigenvector centrality) in a subnetwork formed by accounts designated QAnon influencers (grey-purple) and Prominent Pro-Trump accounts (bright purple).

### 2.2 Replication based on retweets: Network clusters and properties for Study 2 data

The analysis was repeated based on retweets only, excluding self-loops. The analysis showed 997 vertices forming 9,359 unique edges (46,720 total edges). The reciprocated vertex pair ratio was 0.035 and the reciprocated edge ratio was 0.068. Clusters were identified from retweet traffic using the Clauset-Newman-Moore algorithm, placing all neighborless vertices into one group. A total of 14 clusters were identified.

Based on retweets only, the 20 accounts top-ranked for eigenvector centrality were: @VincentCrypt46, @CJTruth, @DanScavino, @M2Madness, @PrayingMedic, @RudyGiuliani, @Qtah17, @MajorPatriot, @TheSharpEdge1, @JuliansRum, @AnnaKhait, @LittlleMel, @AnonPatriotQ, @DoqHolliday, @All50News, @InTheMatrixxx, @TrumperWavin, @John_F_Kennnedy, @Tafkag, and @MartinGeddes.

Based on retweets only, the 20 accounts top-ranked for betweenness centrality were: @CJTruth, @VincentCrypt46, @RudyGiuliani, @DanScavino, @CharlieKirk11, @M2Madness, @MartinGeddes, @PrayingMedic, @AnnaKhait, @Harrow1984, @HSRetoucher, @InTheMatrixxx, @MajorPatriot, @ConorNigel, @55True4U, @SSG_Pain, @Qtah17, @KirstieAlley, @DrCharlieWard1, and @AngelHealingArt.

### 2.3 Additional observations

@TrumperWavin was an influential account in Study 2. It is noteworthy in light of the fact that around the time of the U.S. presidential election, @TrumperWavin changed its profile presentation to claim to be 17 years old (recognizing the significance of the 17[th] letter of the alphabet for QAnon narratives). In the days immediately following the U.S. Presidential election, @TrumperWavin was heavily retweeted and tagged into media in promotion of the narrative that Trump had secretly seeded ballots with watermarks that would reveal Democrats' ballot forgery. @TrumperWavin's top 10 hashtags included #Qfamily, #SaveTheChildren, #SatanicElite, #PizzaGate, and #LaptopFromHell.

## 3 Study 3

### 3.1 Examples of tweets from "Wyatt" accounts, including @_CEOofGenZ

Below, we present some examples of tweets from four accounts presenting the Twitter "Wyatt" persona described in the main text to illustrate their themes, as well as observed behaviors of other accounts.

(1) *@CEOofGenZ*. In late June, 2020 the account @CEOofGenZ with the banner "Wyatt, Austere 17" tweeted pictures of a Nazi insignia and Nazi eagle, battleships, planes, and UFO's with the description "#OperationHighJump Antarctica 1940s". This tweet was retweeted by @WhiteBunnyQ, while @TimeTravelAnon replied to the tweet with pictures featuring Hitler, Nazi insignias, and UFOs and the caption "Roswell UFO 'wasn't aliens – it was a top-secret spacecraft built by Nazi scientists.'" @TimeTravelAnon's tweet, in turn, was engaged with (retweeted or replied to) by five accounts featuring QAnon-related messages, esoteric imagery, religious themes (including Crusaders-related imagery), memes made popular during Trump's campaign applied to Russian soldiers, and/or retweeting cryptocurrency.

(2) *@1perSonOfEarth7*. On 10/13/2020 this account posted simply "Hello :)" to which esoteric influencer account @TheHannSolo posted "What up dawg" in reply, attaching a colorful gif of an angelic figure in a Buddha pose. @1perSonOfEarth7 later posted "The Truth of Satan will not be suppressed any longer…It's time We (Father Satan's Own) put a stop to it!" Another tweet was, "True Satanism is working to advance one's mind and soul and personal powers." Further, @1perSonOfEarth7 tweeted: "There was a coup in this country under the Obama admin, and with Hillary sitting there as Secretaryof [sic] State, the world was about to be their oyster, a world in which at last pedovores could frolic, diddle, And [sic] terrorize all they wanted to in a global communist slave state."

(3) *@AperSunOfEarth*. On 10/16/2020, Wyatt account @AperSunOfEarth tweeted: "The Six Million lie has been a powerful propaganda tool that enabled them to steal Palestine from and set up the bandit state of Israel. Thanks to the Six Million lie they had the support and blessing of the befuddled world, especially the J-infested United States." Further, on 10/17/2020 QAnon and Esoteric influencer @PetahJane commented "Boom" with a fire emoji then quote-tweeted @AperSunOfEarth as follows: "Fools claim Satan is a 'Christian invention.' All of the Old Gods were made into Devils and Demons in order to destroy relationships and many other things, it was all supposed to be about elevating and empowering humanity, which was our True Creator (Satan's) intention."

(4) *@_CEOofGenZ*. This account was a focus of Studies 3 and 4 in the main text. On 1/6/21 Wyatt account @_CEOofGenZ sent multiple tweets that included the following (with typeface in original): (1) "RT if you know HITLER was a GOOD MAN"; (2) "RT if you know the JEWISH PLAN TO ENSLAVE HUMANITY AND CREATE A ONE WORLD COMMUNISt [sic] GOVERNMENT."; (3) "RT if you know that the SATANIC GLOBAL ELITE work for ISRAEL"; (4) "RT if you want JUSTICE for SETH RICH, ISAAC KAPPY, AVICII, ANDREW BREITBART etc.. [sic]"; and (5) "RT if you know about the TRUTH of SATAN". @PetahJane and @DanAuito were two accounts identified as network influencers in Studies 1 and 2; it is notable that each of these accounts responded to _CEOofGenZ's very first tweet from November, 2020.

The above brief anecdotes illustrate the continuity of thematic content and perspectives across all these Wyatt accounts. These observations further illustrate how pro-QAnon propaganda influencer accounts showed convergent Twitter behaviors consistent with the importance for networks of Wyatt accounts like @_CEOofGenZ. Other Twitter handles which "Wyatt" appeared to go by from 2020-2021, based on convergent evidence, included @CE0ofGenZ, @_SayWhen2A_, @NTeslaIsKing, @17maga666, @CEOofGenZ17, @Wyatt168882, @Wyatt439917, @WyattQ17, @Qr8Awakening, @QreatAwakening1, @Qatriot89, @SatanIsGod17, @Austere_Wyatt, @Wyatt00231, @Wyatt11702, @Wyatt2103, @JohnWyatt76, @PolishPatriot76, @AustereWyatt16, @AustereWyatt1, @AustereWyatt16, @Poiishamerican1, @Polishpatriot1, @polishpatriot35, @polishpatriot45, and @magajames1. The Twitter User ID for @_CEOofGenZ was 1323635721081659392, with others for "Wyatt" including, but not limited to, 1369037937552330752, 1386159920874237954, and 1316916238472368128.

### 3.2 Network properties for group import of accounts connected to @_CEOofGenZ

The network of accounts described in the main text showed N = 1,697 vertices with 5,097 unique edges (169,784 total edges). There were 149,975 self-loops. The reciprocated vertex pair ratio was 0.029 and the reciprocated edge ratio was 0.057. Clustering was carried out using a Clauset-Newman-Moore algorithm, which revealed 16 groups. The number of vertices per group was as follows: Group 1, 401; Group 2, 290; Group 3, 272; Group 4, 270; Group 5, 144; Group 6, 115; Group 7, 65; Group 8, 51; Group 9, 28; Group 10, 15; Group 11, 12; Group 12, 10; Group 13, 10; Group 14, 9; Group 15, 3; Group 16, 2. Group 8 consisted of vertices not connected to any other component. Data in Figure 4 of the main text were graphed using the Fruchterman Reingold algorithm with groups laid out in different panels. Tables S15 and S16 show network top items.

In addition to the top 15 accounts ranked for eigenvector centrality in the main text, the following were ranked 16-25: @JohnHereToHelp, @VDarknessF, @FabioPatriot17, @Carmen79809708, @GCanty3, @TweetWordz, @PetahJane, @MikePompeo, @JeffreyParamo1, and @Amissyia. It can be observed that many of these most influential accounts were well-known pro-QAnon accounts.

Accounts ranked highest for betweenness centrality, in descending order, were: @_CeoofGenZ, @CodeMonkeyZ, @VincentCrypt46, @PrayingMedic, @PepeMatter, @AwakenedOutlaw, @PepeNewsNow, @RichardGibb8, @HSRetoucher, @Mareq16, @55true4u, @MartinGeddes, @EntheosShines, @p8r1ot, @ValiantThor12, @MSAvaArmstrong, @jwq2020, @johnheretohelp, @redwavewwg1, @ZenOfTupac, @FabioPatriot17, @OcrazioCornPop, @mikepompeo, @ezralevant, and @VDarknessF.

## 3.3 Analytic methods and network properties for group import of accounts connected to @SanandaEmanuel plus @_CEOofGenZ

Esoteric account @SanandaEmanuel had previously been observed to interact with @_CEOofGenZ. @SanandaEmanuel had 8,791 followers, a follower:followed ratio of 6.4, and had tweeted 14,320 times. We determined the extent to which accounts interacting with @SanandaEmanuel also interacted with @_CEOofGenZ. To do so, we first downloaded Twitter data for @SanandaEmanuel on 1/11/21 using NodeXL; this returned $N = 440$ accounts. $N = 32$ accounts with over 30k followers deemed likely to add noise to the analysis were removed, leaving $N = 408$ accounts. Next, @_CEOofGenZ was added to this list, yielding $N = 409$ accounts targeted for import. A Twitter users group import was then conducted from this list on 1/12/21 in NodeXL, with a specified maximum of 500 tweets per account. Sentiment analysis was applied to downloaded tweets to determine content matches to QAnon, New Age/esoteric, or conservative themes using keywords from Study 2.

The search returned N = 394 vertices. There were 1708 unique edges (178,259 total edges), along with 155,465 self-loops. The reciprocated vertex pair ratio was 0.149, and the reciprocated edge ratio was 0.259. A Clauset-Newman-Moore algorithm was used to group vertices into clusters. Seven clusters were returned; see Figure S7. The number of vertices per group was as follows: Group 1, 189; Group 2, 121; Group 3, 30; Group 4, 27; Group 5, 12; Group 6, 11; Group 7, 4. All edges with @_CEOofGenZ represented retweets of, replies to, or mentions of @_CEOofGenZ, rather than the other way around. Edges shared with @_CEOofGenZ are shown in Figure S7. Of the 52 vertices that shared an edge with @_CEOofGenZ, 30 were from Group 1, 20 were from Group 2, and one each were from Groups 3 and 5.

Accounts showing the highest influence as gauged by eigenvector centrality were, in descending order: @SanandaEmanuel, @PowerReclaim, @SiriusBShaman, @TheHannSolo, @HellsBellShel, @MLealcala, @LesProctor, @IowaStarseed, @Feathers711, @MrH3rb, @WokeWonderland, @GiuliaMGuerrero, @VantWilson, @JolianaM11, @EmilyLinnMille1, @IAmTheArtofSoul, @AnyaManifested, @Electric_Mage, @Glickman_Daniel, and @DrManhattan777. Accounts showing the highest betweenness centrality were, in descending order: @SanandaEmanuel, @SiriusBShaman, @PowerReclaim, @TheHannSolo, @WokeWonderland, @HellsBellShel, @Phaethon314, @Feathers711, @GalacticBeingz, @IowaStarseed, @MrH3rb, @LesProctor, @GiuliaMGuerrero, @MLealcala, @VantWilson, @JolianaM11, @SamuelBradbur11, @IAmTheArtofSoul, @_CeoofGenZ, @jzblove, @startra44830631, @EmilyLinnMille1, @ChiefLord999, @Electric_Mage, and @MagentaPixie.

Finally, Tables S17 and S18 present network top items and top hashtags for this network. It should be noted that accounts in Table S18 included some of the most influential and visible esoteric-themed influencer accounts.

Table S15. Network top replied-to, top-mentioned, and top tweeters overall for Study 3 for the group import of accounts connected to @_CEOofGenZ.

| Rank | Top replied-to | Top mentioned accounts | Top Tweeters |
|---|---|---|---|
| 1 | _ceoofgenz | realdonaldtrump | msavaarmstrong |
| 2 | vincentcrypt46 | vincentcrypt46 | ocraziocornpop |
| 3 | codemonkeyz | codemonkeyz | hlaurora63 |
| 4 | hsretoucher | llinwood | winegirl73 |
| 5 | prayingmedic | pepenewsnow | beank511 |
| 6 | 55true4u | genflynn | dawalf |
| 7 | richardgibb8 | _ceoofgenz | lyndaatchison |
| 8 | pepematter | johnheretohelp | bryan700 |
| 9 | awakenedoutlaw | awakenedoutlaw | celesteherget |
| 10 | pepenewsnow | prayingmedic | trumpcat04 |

Table S16. Network top hashtags for Study 3 for seven largest clusters identified by algorithms in the group import of accounts connected to @_CEOofGenZ.

| Group | Top Hashtags |
|---|---|
| All | #Hackers, #notpregnant, #fakebaby, #gaslighter, #boycottWalmart, #May1st2020, #Biblical, #blackmailed, #WakeUp, #ThePlan |
| 1 | #Hackers, #notpregnant, #fakebaby, #gaslighter, #May1st2020, #blackmailed, #WakeUp, #LawOfOne, #Monster, #Tweet |
| 2 | #hashtag, #BoycottWalmart, #Walmart, #SpecialDelivery, #OSSPosse, #JFK, #OperationWetBlanket, #Benghazi, #StopTheSteal, #1776Hammer |
| 3 | #Winning, WeThePeople, #USA, #Soros, #InformationWar, #JulianAssange, #ElectionFraud, #WWG1WGA, #DOJ, #UK |
| 4 | #hernamewasmollymacauley, #recallgavinnewsom, #sethrich, #digitalsoldiers, #boycottwalmart, #qanon, #greatawakeningalchemists, #ossposse, #q, #hbd |
| 5 | #Biblical, #recallgavinnewsom, #unmaskamerica, #flynnflashback, #treason, #santasurfing, #fightfortrump, #bidenwillneverbepresident, #americafirst, #science |
| 6 | #cdnpoli, #boycottwalmart, #greatreset, #goteam, #magaparty, #saskatoon, #saskatchewan, #yxe, #endthelockdowns, #lockdownsforever |
| 7 | #fightlikeaflynn, #fightfortrump, #holdtheline, #stopthesteal, #benghazi, #hailcaesar, #savetherepublic, #godblessamerica, #crosstherubicon, #wwg1wgaworldwide |

**Table S17**. Select network top items for group import of accounts connected to @SanandaEmanuel and @_CEOofGenZ in Study 3.

| Rank | Top replied-to | Top mentioned accounts | Top Tweeters |
|---|---|---|---|
| 1 | SanandaEmanuel | PowerReclaim | Searcher9090 |
| 2 | PowerReclaim | TheHannSolo | CannaFrom |
| 3 | SiriusBShaman | SanandaEmanuel | NewsGirly |
| 4 | HellsBellShel | SiriusBShaman | CarrollQuigley1 |
| 5 | TheHannSolo | HellsBellShel | AuCommander |
| 6 | WokeWonderland | WokeWonderland | End_TheFederalR |
| 7 | LesProctor | Xoana_Ra | NicPaints |
| 8 | MrH3rb | realDonaldTrump | K4rlGruen |
| 9 | Joliana11 | JolianaM11 | GianniProvz |
| 10 | VantWilson | JimWhispers | Right558 |

**Table S18**. Top hashtags by group in Study 3 for the group import of accounts derived from @SanandaEmanuel plus @_CEOofGenZ, ordered from most to least frequent. Hashtags reflecting QAnon themes are shown in italics. No network hashtags were returned for Group 7.

| Dataset | Top Hashtags |
|---|---|
| Entire Network | #Jesus, #LoveWins, #Christians, #Christ, #TransitionToGreatness, #Go, #Christianity, #LTTL, #WJDR, *#SaveTheChildren* |
| Group 1 | #Go, #InJoyTheShow, #Redemption, #PeaceIsHere, #WeDeserveIt, #Starseeds, #TheBestIsYetToCome, *#SaveOurChildren*, #TransitionToGreatness, #Endgame |
| Group 2 | #Jesus, #Christians, #Christ, #Christianity, #LTTL, #LoveWins, #WJDR, #TransitionToGreatness, #JesusChrist, #Christmas2020 |
| Group 3 | #LoveWins, #Aura, #Namaste, #Devil, #Fire, #Biblical, #Family, #10Bulls, #TenBulls, #BTO |
| Group 4 | #LoveWins, *#QuantumMattrixz*, #Ashtar, #DivinePlan, #TheBestIsYetToCome, *#FightLikeAKennedy*, #ThePatriotParty, #GuardiansOfTheGalaxy, #PatriotsAreInControl, #9Lives |
| Group 5 | #GroundCrew, #Tyler, #HybridTheory, #TheGiftThatKeepsOnGiving, #HurtPeopleHurtPeople, #FamilyOfMichael, #VictoryOfTheLight |
| Group 6 | #TwentyTwentyOne, *#QAnon*, *#QAnon172021*, #Trump2020, #Sark |
| Group 7 | N/A |

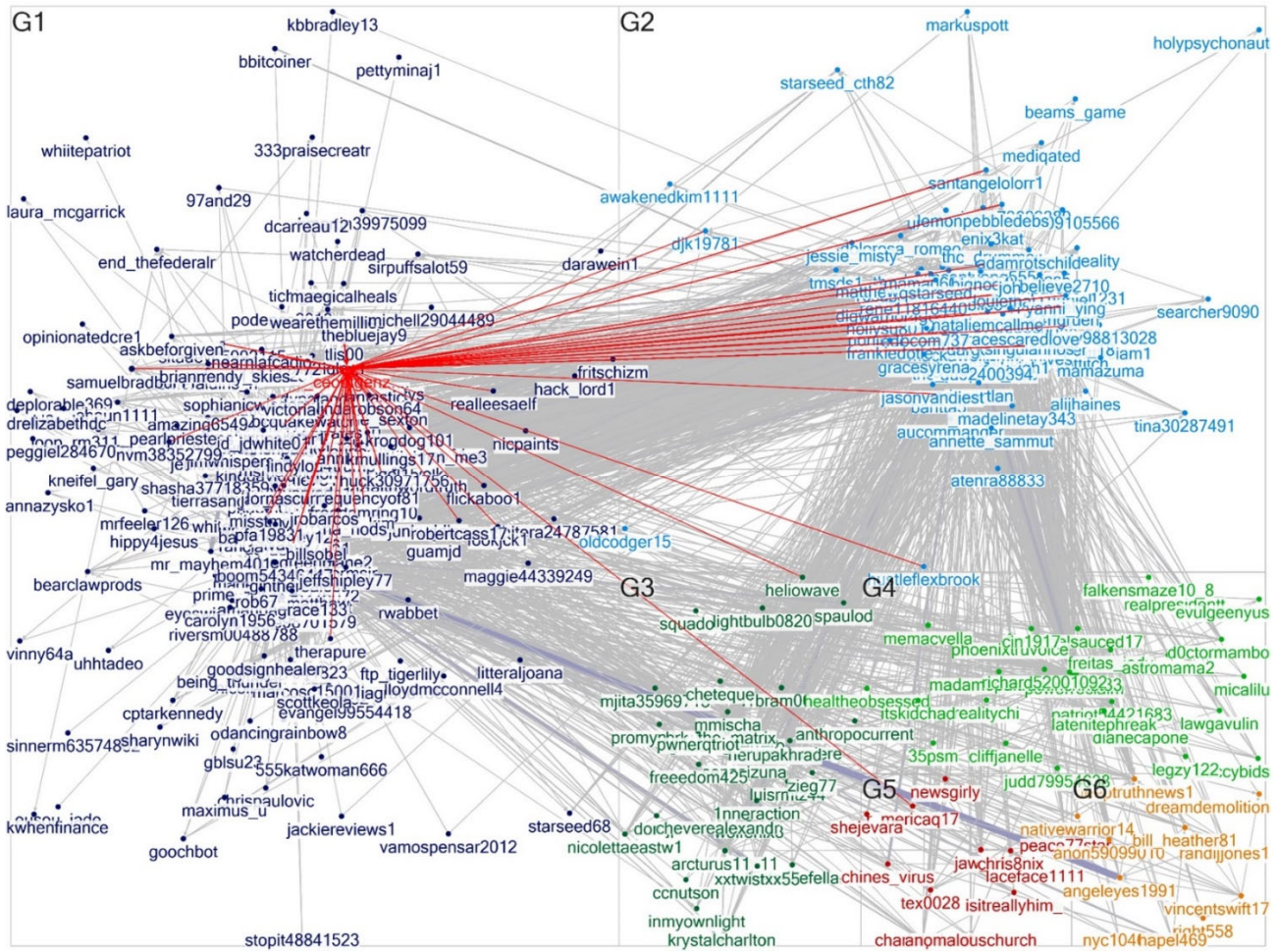

**Figure S7.** Network of accounts for Study 3 based on group import of vertices from download of @SanandaEmanuel, plus @_CEOofGenZ. @_CEOofGenZ and its edges are highlighted in red.

# 4 Study 4

## 4.1 Further account selection details

For far-right accounts, we selected key accounts like @MrDrewENT, @_CEOofGenZ, @Ag3nt_chaos, and @DanAuito; we also sought through convergent methods the identification and inclusion of accounts promoting insurrection, imagery of guns and tanks, and/or extremist views. In addition, we discovered a Russian-language disinformation account that interacted with esoteric influencer accounts, @AbuseTheTruth, which was further identified as a key account. Additional selected accounts emphasized the following features toward systematic and exploratory investigation of actors, means and motives in QAnon propaganda promotion: QAnon or esoteric themes; cryptocurrency or "the Hive" (e.g., @hive_blockchain); "incel" and/or extremist content; and/or presentation as a member of Anonymous, the hacker collective. Finally, we selected a handful of connected accounts showing content with positive sentiment (cute animals, vintage actress Ethel Merman), due to observations that these were frequently retweeted by QAnon and esoteric accounts. Below we offer some additional information supporting selection of key accounts.

### 4.1.1 Schoenberger-linked accounts and associates

Here, we present additional details supporting and motivating our selection of particular key accounts and account features for the study's sample. Both @NameMySock and @FaisalLazarus engaged in numerous conversations with @TootsLilFighter (a.k.a. "JAnon" from banner profile information). @FaisalLazarus and @TootsLilFighter discussed that @TootsLilFighter would be the next "Q" in late 2020. Circa January 1, 2021, @TootsLilFighter was involved in that involved the sharing of a six-digit code with other accounts. Shortly thereafter we observed coordination among an overlapping set of accounts that involved posting to @VincentCrypt46's timeline, the creation of an account @QL83R that was rapidly followed within hours by @FaisalLazarus, and @QL83R's sending a message to @SnowWhite7IAm. These Twitter events are described further in Supplemental materials. A subset of accounts associated with Twitter activities before and after the creation of @QL83R were included in the sample. @FranksCOMS was an account attributed to internet troll Frank Bacon, recognized for an early role in QAnon.

### 4.1.2 Selection of Cicada 3301 and other ARG accounts

We identified key Cicada 3301 and ARG accounts via inspection of properties of accounts directly connected to @NameMySock and/or @FaisalLazarus; (ii) Twitter searches of hashtags such as #Cicada3301 and #C3301; and/or (iii) searches on Twitter and the internet for reported ARG enthusiast and artificial intelligence expert Quinn Michaels, who was an early collaborator of QAnon promoter and Cicada 3301 enthusiast Manuel "Defango" Chavez, who was a former Schoenberger associate. Three key accounts selected were @_ghost3301_, @Acio3301, and @AnomalousChurch.

### 4.1.3 Additional details supporting account selections

Circa the Jan. 6, 2021, riots in Washington, D.C., we observed @MrDrewENT, an which presented itself as "Digital Bodyguard 4 @VinceMcMahon…Digital Secret Service 4 @Linda_McMahon" interacting with other accounts in a manner suggesting promotion and/or organizing of insurrectionist activity. On January 5, @MrDrewENT retweeted a map from @KCMoon1 of how to get to the Capitol. Early on Jan. 6, @BrettLaGrave tagged @MrDrewENT and other accounts (e.g., @Scorpionclone, @_CQRAL) – into a tweet that stated "Today we WIN." Later that day, @BrettLaGrave tweeted pictures of himself – or a person matching the account's profile picture – standing in a crowd at the Trump rally. Later, @BrettLaFave sent tweets with messages that included "Pence is a TRAITOR!" and "We are storming the Capitol." @MrDrewENT retweeted multiple tweets that day from @BrettLaGrave. Inspection of the @MrDrewENT account revealed it promoted QAnon influencers' content in months prior , including retweeting influencers like @VincentCrypt46, @KimRunner, and @TrumperWavin. Further, on Jan. 2, 2021, @MrDrewENT tweeted simply "17"; 'Q' is the 17th letter of the alphabet. @MrDrewENT had even promoted @_CEOofGenZ; on January 6, 2021, @MrDrewENT retweeted @0queb2's 12/25/2020 message to @_CEOofGenZ: "Merry Christmas, Wyatt! I love you brother!"

In addition, @FranksCOMS was an account featuring a description "Frank Bacon's COMs channel." Frank Bacon has been identified as an internet troll who played a role in the history of QAnon.

## 4.2   Case study: QAnon insider coordination around Jan. 1, 2021 creation of account QL83R

The events surrounding the creation of account @QL83R demonstrate clear coordination of accounts linked to Thomas Schoenberger, Ron Watkins, Lisa Clapier, and close on-line associates of Schoenberger. @QL83R was thus selected as a key account for Studies 4 and 5. In interviews Thomas Schoenberger has repeatedly denied involvement in QAnon and evaded questions and/or presented misleading answers on the matter of his role in this propaganda campaign. Twitter events occurring circa the creation of an account @QL83R on 1/1/21 present clear evidence of Schoenberger's involvement in ongoing aspects of coordination of QAnon. We briefly sketch these events below.

The broader context of interactions related to @QL83R appeared to relate to a transfer of power of who would be the new "Q", as conveyed through Twitter engagements with @FaisalLazarus. Identification of the @QL83R account began with observations of @FaisalLazarus engaging in Twitter "conversations" (e.g., alternating tweet-reply sequences) with @TootsLilFighter suggesting that @TootsLilFighter would take over as Q. @FaisalLazarus also engaged in conversations with @Patrionesss (who listed as its backup account @Patrioness; interactions between @Patrionesss and @Patrioness were documented, e.g. through retweets). (See below for more information on @Patrionesss.)

We then witnessed coordination among @Patrionesss, @TootsLilFighter, and a small group of other users which involved posting of a code of some kind. @TootsLilFighter then posted a "wink" emoji. @Patrionesss then engaged in Twitter interactions with @ShadowBan_01, @JANEDOE171717, and @GoodVsEvil17. @JANEDOE171717 and @GoodVsEvil17 then posted "." on VincentCrypt46's timeline.

At 6:44 p.m. on 1/1/21, @QL83R tweeted, "@JANEDOE171717 hi :)". By noon on 1/3/21, @QL83R had already been followed by @TootsLilFighter (the first follower), @FaisalLazarus, @SnowWhite7IAm, @DragonFlyGlitta (an account interacting with @RQueenInc), and other influencer accounts.

At 9:12 a.m. on 1/3/21, @QL83R tweeted, "SnowWhite7IAM SW I got pretty much blocked from accessing two accounts I had… [EffisforFUn] that's one of them. Well a while back I created thus [sic] custom Google search engine and well if you or anyone wants to use it by all means its [sic] on that account page." @QL83R attached to the tweet a picture of account @EffisforFun. Convergent evidence suggested the other account that was referred to was @Chup4C4bra.

Further evidence supports that @QL83R, @Chup4C4bra, and @EffisforFun possibly all had the same account operator. @Chup4C4bra had as its pinned tweet a link to @EffisforFun and the account banner said "Eff's pet Chup4c4br4". Both accounts were visible after clicking a Twitter acknowledging that the account had shown suspicious activity, indicating that the accounts had had user privileges suspended.

Leading up to the events described above, there had been notable interactions between Schoenberger's account @FaisalLazarus and @Chup4C4bra which further assist contextualizing the creation of @QL83R. On December 12, 2020, user @b_wolgast, whose banner was "Fuck the Q Cult" was observed in a thread with disinformation researchers. @b_wolgast tweeted: "I love to obtain publicly available documents. I make them even more publicly available too." @FaisalLazarus then replied on 12/21/20 at 9:44 PM: "Me too   Meet my friend @Chup4C4bra  Cheers…." A Twitter web site search for 'From:FaisalLazarus "chup4c4bra"' on Jan. 2, 2021 showed that @FaisalLazarus had replied to chup4c4bra directly with all of the following: "Nice !" (12/31/20), "Nice !!" [with hat-wearing emoji] (12/30/20), "Brooo….)" [with link to archive.org] (12/30/20), a link to the Talking Heads' song "Road to Nowhere" (12/30/20), "Denn du wirst meine Seele nicht in der Hölle lassen" (12/30/20), and other posts. @FaisalLazarus also replied to tweets that included @Chup4C4bra and at least one of the other accounts: @OfAspen, @TracyBeanz, @Patrionesss, @RonTheDogTrainR, @babyfist, and others which appear in the main text. @Chup4C4bra also interacted with Schoenberger's then-suspended account, @NameMySock.[1] The operator of @NameMySock had identified himself as Thomas Schoenberger on Oct. 12, 2020.[2]

Finally, on 1/24/21 we documented @babyfist2, who displayed a pepe meme for its account profile picture, to send a tweet which mentioned @FaisalLazarus, QL83R, @TootsLilFighter which included the hashtag #cicada3301 and the word "kek". Schoenberger's account @FaisalLazarus then engaged in conversational turns with @babyfist; the topic appeared to be the movie. However, the interaction also featured cryptic comments such as "'Mr. Black' any comments?" from @babyfist2, to which @FaisalLazarus replied "Can't say much..yet…" Taken together, the events surrounding the creation of @QL83R circa January 1, 2021 illustrate clear evidence of Schoenberger engaging in coordination activities around Cicada 3301 in interactions with at least one far-right account.

### 4.3  Betweenness centrality and connectivity features

Table S23 shows the top 20 accounts ranked for betweenness centrality. Extremist @_CEOofGenZ ranked 2nd. The Russian-language disinformation account @AbuseTheTruth ranked 4th, while QAnon influencer @SnowWhite7IAm ranked 5th. Several additional QAnon influencer accounts were top-ranked, including @Catturd2, @55True4U, @KimRunner, @Mil_Ops. Multiple accounts selected for their esoteric features were also top-ranked. Further, multiple accounts top-ranked for this analysis also showed evidence of Alt-Right, extremist, or suspected insurrectionist affiliations. For example, @DrutangAtHome had been selected for featuring colorful graphics, cryptocurrency content, and connections to other accounts; yet, the account tweeted about QAnon themes (e.g., "red pilling"), as well as occult- and Nazi-themed topics (e.g., yellow-highlighted pictures of Wikipedia pages describing Heinrich Himmler's interest in esotericism and the runic symbols resembling 'SS').

As another example illustrative of our findings, we consider @3speakonline, whose profile advertised it as "The Home of Free Speech!...Powered by #Hive #Bitcoin". In one interaction, this account posted a reply to @jimmy_dore, which involved an intermediate "mention" of @DrutangAtHome, implying that @3speakoline received @jimmy_dore's tweet via @DrutangAtHome first retweeting it. @DrutangAtHome then posted a reply to @3speakonline which said "check dm

---

[1] Archived tweet evidence shows @Chup4C4bra interacted with @NameMySock: https://archive.is/mqLmO.

[2] Archived tweet evidence shows the operator of @NameMySock identified as Thomas Schoenberger: https://archive.is/GL5xz. 7

dan" in the same thread; @3speakonline subsequently liked this tweet. QAnon influencer @AustinSteinbart mentioned @3speakonline approvingly several times.

## 4.4 Additional analysis details

Based on a top hashtag analysis (cf. Table S21), tweets across the network were predominantly in English; however, over 90% of tweets in Group 6 were in Russian. Overall, top hashtags related to themes of the Hive, cryptocurrency, Julian Assange, and Cicada 3301 and other ARGs. Group 1 hashtags also included right-wing narratives, e.g., #StopTheSteal, as well as themes of cryptocurrency and Assange. Group 2 hashtags were related to ARGs, Assange, and the hacker group Anonymous. Group 3 hashtags focused on the Hive, cryptocurrency, and blockchain technology. Group 4 themes related mostly to Assange and right-wing politics. Group 5 hashtags related primarily to Canadian politics. Group 6 hashtags related primarily to Putin and were mostly in Russian. Group 7 presents a variant QAnon hashtag, #WWGiWGA, and otherwise presents a variety of themes, as was the case for Group 8. Further, a supplemental analysis showed 27 accounts had mentioned "cicada" specifically; 16 of these were in Group 2, 10 were in Group 4, and one was in Group 1.

While accounts had been selected for one or more key canonical properties or connections – such as QAnon content, we examined active and/or archived Twitter data for top-ranked account results to identify additional relevant properties to provide fuller interpretation of results, particularly in an attempt to identify alt-right and/or suspected insurrectionist accounts in absence of evidence of overt affiliation with far-right groups in account descriptions. To do so, we used searches of Twitter's site and/or our archives to identify additional applicable categories as follows, from profile presentations and content matching the following criteria: (i) *QAnon*. The account (a) displayed QAnon graphics, hashtags, or keywords in the profile or pinned tweet; (b) had retweeted tweets containing QAnon keywords, and/or (c) was an influencer who tweeted about QAnon themes (e.g., pedophilia) while also being retweeted by other QAnon influencers. (ii) *New Age/Esoteric*. The account displayed New Age, esoteric, or occult hashtags or descriptors or graphical content matching the list for Study 2. (iii) Schoenberger direct associate. The account had a direct connection to an account linked to Thomas Schoenberger. (iv) *Cicada 3301 or other ARG*. The account (a) presented content associated with Cicada 3301 or another ARG in its profile, tweets or retweets (e.g., web sites or hashtags or associated with Cicada 3301 or another ARG in tweets or retweets, e.g., #TheGame23); (b) was identified to have a direct connection to a key account for this category – a mention, reply or retweet; or (c) was identified as linked to Cicada 3301 or ARGs based on journalistic reports. (v) *Alt-Right/Extremist/Suspected Insurrectionist Affiliation*. We did not detect direct signs of accounts' affiliation with far-right groups (e.g., declaring membership in the Proud Boys). Therefore, we used alternative criteria: The account (a) had a direct connection to an account associated with insurrectionist or bellicose themes (e.g., @MrDrewENT, @_CEOofGenZ); (b) interacted with accounts featuring graphics, themes, or handles consistent with alt-right political affiliation (e.g., pepe or toad memes, the word "fren") or extremist views (e.g., imagery of Satan); or (c) tweeted content, hashtags or emojis associated with the alt right ("OK" emoji symbol, "kek" or "fren"). (vi) *Other*. The account did not meet criteria for any of the other categories.

Finally, note that QAnon influencer @VincentCrypt46 was also observed to be in direct communication with @_CEOofGenZ. On 12/27/20 @_CEOofGenZ tweeted a claim that the Christmas Day Nashville explosion was an FBI false flag operation. In response, @VincentCrypt46 produced a cryptic, single-character reply to @_CEOofGenZ's tweet: "."; esoteric account @Aakashaessa then replied to @VincentCrypt46 with the same response: "." On another notable occasion @VincentCrypt46 produced similar coordinated tweeting patterns with insider accounts, including some directly linked with the (Schoenberger-linked) account @Patrioness in an extraordinary example of QAnon insider account coordination around the creation of account @QL83R circa January 1, 2021; see Supplementary Materials.

## 4.5 Accounts in @FaisalLazarus network

@FaisalLazarus presented as Thomas Schoenberger on Jan. 25, 2021.[3] The @FaisalLazarus account was downloaded using NodeXL on 1/3/21; it was connected to N = 261 accounts. To identify accounts of interest and examine their relevance to QAnon and/or Cicada3301, the name and description fields were matched against keywords and phrases representing pro-QAnon ideology with similar criteria as Study 2 for "QAnon" and as referencing 3301 for Cicada3301. Accounts associated with government agencies, politicians or officials, journalists or media, well-known personalities, and/or disinformation researchers were grouped separately. See Supplemental text.

---

[3] See https://web.archive.org/web/20210126030343/https://twitter.com/FaisalLazarus/status/1353901300056875008.

N = 49 accounts consisted of government agencies, media personalities, etc.; importantly, the connections featured multiple pro-QAnon accounts, such as @LLinWood, @SidneyPowell1, @TracyBeanz, and @GenFlynn. Because of the high follower counts of these accounts, they were not considered further. Of the remaining accounts which had non-blank descriptions (N = 193), N = 27 (or 14%) displayed QAnon hashtags or slogans, and an additional N = 5 were frequently retweeted by QAnon accounts (e.g., @schumannbot). Further, by approximately 10 days after data download, following two purges by Twitter of QAnon accounts, a total of N = 38 accounts (or 20%) had been suspended. An additional N = 5 had deactivated themselves, and N = 2 accounts had changed handles. Finally, two accounts were identified with Cicada3301 references. This analysis confirms that Thomas Schoenberger's @FaisalLazarus account was in touch with many QAnon and disinformation accounts, supporting his involvement in QAnon.

**Table S19**. Hashtag classification criteria for Study 4.

| Category | Match/criteria description |
|---|---|
| "QAnon" theme | #WWG1WGA and variant spellings, anything with 8kun, any hashtag referencing Austin Steinbart, #byebyecabal, #childsacrifice, #darktolight, #enjoytheshow, #deepstate, #freeanons, #greatawakening, references to Isaac Kappy, #linwood, #opchildsafety, #oppedohunt, #pedogate, #pedophile, #pizzagate, #q, #qanon, #qpharisees, #redpill78, #robertdavidsteele, #santasurfing, references to saving the (or 'our' or 'god's') children, references to sex trafficking, references to the great awakening, references to these people being sick/evil, references to trusting the plan |
| "Cryptocurrency" theme | any hashtag beginning with "xrp", #ripple, any hashtag beginning "bitcoin", #btc, any hashtag beginning with "crypto", #eth, #ethereum |
| "Blockchain" theme | Any hashtag beginning with "blockchain" |
| "Julian Assange" theme | Any hashtag containing "Assange", e.g. #FreeAssange |
| "Cicada 3301" theme | #Cicada3301, #C3301, mention of "Cicada" |
| "Other ARG" theme | #HiveMind23, #TheGame23, #TG23, #kstxi, #00ag9603, #chaos, #Virus23 |
| "The Hive" theme | Hashtags beginning with "hive" except #HiveMind23 |

**Table S20**. Study 4 top hashtags.

| Dataset | N | Top Hashtags |
|---|---|---|
| Entire Network | 525 | #Hive, #crypto, #XRP, #TheGame23, #bitcoin, #WeAreAllAssange, #blockchain, #hivechat, #OpPSION, #00ag9603 |
| Group 1 | 220 | #xrp, #xrpcommunity, #FreeAssangeNow, #MAGA, #WeAreAllAssange, #StopTheSteal, #ElectionFraud, #Ripple, #Assange, #demmink |
| Group 2 | 176 | #TheGame23, #OpPSION, #00ag9603, #dataplex, #kstxi, #OpMimic, #Anonymous, #Postneonism, #galdrux, #WeAreAllAssange |
| Group 3 | 49 | #hive, #crypto, #bitcoin, #blockchain, #hivechat, #hivefixesthis, #hiveUA, #posh, #splinterlands, #btc |
| Group 4 | 42 | #Syria, #FreeAssangeNow, #StopTheSteal, #AssangeCase, #Assange, #WeThePeople, #COVID, #FreeAssange, #MAGA, #DC |
| Group 5 | 12 | #CdnPoli, #COVID19, #OnPoli, #Ableg, #TrudeauMustGo, #WeScandal, #AbPoli, #RudysCommonSense, #BCPoli, #Canada |
| Group 6 | 12 | #COVID19, #мальцев [Maltsev], #желаюпутину [I wish Putin], #япрезираюпутина [I despise Putin], #COVID__19, #госдума [State Duma], #StopCallingHimPresident, #яуважаюпутина [I respect Putin], breaking, #шулипа [šulipa] |
| Group 7 | 8 | #fail, #newprofilepic, #freegrimes, #butachild, #elonMusk, #Grimes, #FreeCosby, #*WWGiWGA*, #RIP, #FriedTurkey |
| Group 8 | 6 | #TheVenusProject, #TVP, #RBE, #ResourceBasedEconomy, #JacqueFresco, #RoxanneMeadows, #SocialDesign, #ScientificMethod, #DoYouSpeakFuture, #COVID19 |

**Table S21**. Top replied-to and mentioned accounts in Study 4. Accounts marked with an asterisk (*) were included in the sample.

| Rank | Top replied-to | Top mentioned accounts |
|---|---|---|
| 1 | realDonaldTrump | realDonaldTrump |
| 2 | VincentCrypt46 | SnowWhite7IAm* |
| 3 | LLinWood | SteveOuttrim |
| 4 | DualDuels* | IloMagyar |
| 5 | FallingWaterz* | Wfinalle57 |
| 6 | CodeMonkeyZ | Blogjam_net |
| 7 | PacMan522* | Andress45303251 |
| 8 | FaisalLazarus* | OdilonRoss |
| 9 | PowerReclaim* | RandThompson16 |
| 10 | AnomalousChurch* | RoseModema* |

Table S22. Top 20 accounts for eigenvector centrality for the network in Study 4. Each account is grouped into categories according to properties leading to its selection ("Criteria Identified at Selection"), and/or verified after selection. Accounts selected for more than one category of properties are listed with "*".

| Category | Criterial Categories Identified at Selection | Additional Account Characteristics |
|---|---|---|
| QAnon | SnowWhite7IAm (1)* <br> Catturd2 (2) <br> 55True4U (3) <br> KimRunner (5) | WokeWonderland (6) <br> 11Llotus (7) <br> Bejay31688996 (8) <br> Mil_Ops (11) <br> TheHannSolo (15) |
| New Age / Esoteric | SnowWhite7IAm (1)* <br> PowerReclaim (4) <br> WokeWonderland (6) <br> EsotericExposal (9) <br> TheHannSolo (15) | |
| Direct association to Schoenberger account | 11Llotus (7) <br> SnowWhite7IAm (1)* <br> RoseModema (10)* CASunshineGal (12) <br> 713xtr0py (13) <br> OfAspen (14) <br> Cracker_Bald (16) <br> ScottBibler (17) <br> FaisalLazarus (18) | |
| Cicada 3301 or other ARG | Bejay31688996 (8)* <br> J1m_buck (19) <br> RoseModema (10)* | TheHannSolo (15) <br> 55True4U (3) <br> SnowWhite7IAm (1) |
| Alt-Right / Suspected Insurrectionist Affiliation | Bejay31688996 (8)* | |
| Other | Bejay31688996 (8)* <br> Mil_Ops (11) <br> AnonymousSage1 (20) | |

**Table S23**. Top 20 accounts for betweenness centrality for the network in Study 4; number in parenthesis indicates rank. Each account is grouped into categories according to properties leading to its selection ("Criteria Identified at Selection"), and/or verified after selection ("Additional Account Characteristics"). Accounts selected for more than one category of properties are listed with "*".

| Category | Criterion Identified at Selection | Additional Account Characteristics |
|---|---|---|
| QAnon | Catturd2 (1)<br>SnowWhite7IAm (5)<br>55True4U<br>Whiplash347 | WokeWonderland (6)<br>Bejay31688996 (19)*<br>Drutangathome (20) |
| New Age / Esoteric | EsotericExposal (3)<br>WokeWonderland (6)<br>Conspiracyb0t (13) | |
| Cicada 3301 or other ARG | Bejay31688996 (19)* | PowerReclaim (11) |
| Alt-Right / Suspected Insurrectionist Affiliation | _CEOofGenZ (2)<br>Zerperino (8)<br>Rising_serpent (14)<br>17karnage (16)<br>Bejay31688996 (19)* | Whiplash347<br>PowerReclaim (11)<br>3speakonline (17)<br>Drutangathome (20) |
| Other | AbuseTheTruth (4)<br>Bejay31688996 (19)*<br>80strolls (15)<br>3speakonline (17)<br>Aggroed001 (18)<br>Drutangathome (20) | |

# 5 Study 5

## 5.1 Descriptive properties of key accounts

Our list of selected accounts was updated based on apparent replacements of accounts known by that time to have been suspended (e.g., @KimRunner123 for @KimRunner; @CryptoKoba and @Billy14999190 for @VincentCrypt46; @AmyKass3 for AnnieLaulainen). Judgments that an account was a "successor" to a suspended QAnon or esoteric influencer account were made based on: presentation of similar or identical profile pictures to the original account, interactions with similar accounts as the influencer had done, similar handles to the suspended account, similar patterns of followed accounts, and/or explicit claims that the account had the same operator as a suspended influencer account. In addition to key accounts discussed in the main text, the following were also key accounts (and additional details on those mentioned):

(1) **@Billy14999190**. This was an account discovered Jan. 9, 2021 following Twitter's suspension of QAnon influencer @VincentCrypt46. @Billy14999190 bore an apparently identical picture to that @VincentCrypt46 had used and interacted with similar influencers. @Billy14999190 demonstrated multiple retweets of, and responses to, @MrDrewENT, and @MrDrewENT was the second account followed by @Billy14999190.

(2) **@CryptoKoba**. This account appeared to be one that the former operators of @VincentCrypt46 transitioned to using a few days after @VincentCrypt46's suspension, leaving @Billy14999190 again as a backup account.

(3) **@Patrioness.** @Patrioness and @Patrionesss were two accounts each involved in QAnon insider coordination around @QL83R, which is discussed above for Study 4. @Patrioness was the backup account for @Patrionesss; @Patrionesss had been suspended by the time of Study 5 data sampling. @Patrionesss engaged in numerous interactions with Schoenberger's account @FaisalLazarus. Both @Patrioness and @Patrionesss posted frequent pepe memes. The

@Patrionesss profile from Sept. 16, 2020 contained the following in its header: mention of @RQueenInc (Jim Watkins), hashtags #TH3D3N and #WWG1WGA, a website thadragonsden.com, and other information. @RQueenInc also retweeted @Patrionesss.

(4) **@TRUMPSOTHERSON1**. This account retweeted posts about Q drops and showed extensive mutual interactions with @Billy14999190 in months prior to January, 2021. Further, it interacted with an account @OpArmageddon. @OpArmageddon was observed at the beginning of a reply by @TRUMPSOTHERSON1 to an account, @Theresa95661410 (formerly @Theresa1111) with a gif saying "I'm over here" (https://archive.is/GCUkc). These Twitter "breadcrumbs" suggested that @OpArmageddon had received @Theresa95661410's tweet and retweeted it, resulting in its receipt by @TRUMPSOTHERSON1.

(5) **@ZeroSum24** was retweeted by @MrDrewENT and followed by @Billy14999190. Archived files showed that @SnowWhite7IAm, @JudithRose91, and @SomeBitchIKnow had all retweeted this account, also.

(6) **@1Crazy_Toaster** had replied to a tweet from @MrDrewENT on 3/20/20; Twitter details showed that @Utsava4 was an intermediary account. We had previously identified @Utsava4 as an esoteric influencer account. Also, 1Crazy_Toaster had replied in tweet threads with Schoenberger's account @FaisalLazarus. The profile for @1Crazy_Toaster featured a picture of the lead female character Trinity in the 1999 movie *The Matrix*. Our archived web pages also showed @1Crazy_Toaster had interacted with @_CEOofGenZ through a tweet he wrote 11/14/2020 that stated (in all capital letters): "'Satan' means 'truth' in Sanskrit, one of the oldest languages in the world. Satan stands for freedom and rebellion against tyranny! Stop being pussies and learn about your creator!" @1Crazy_toaster responded: "This doesn't resonate unless what they believe is truth 'to them'. If anything that team is about inversion, confusion, & dominion (pun intended)".

(7) @**NonameUA** was an account with banner in Cyrillic characters tweeting in Russian and English which replied to @ZeroSum24's Jan. 8 tweet of "Add my Parlar now in case I got Snowdened" (sic) with "45 followers/min" and later "Right! Great Exodus from Twitter. It must be better late night." (sic).

Other notable accounts are briefly described in Table S24.

### 5.1.1 Betweenness centrality

The top 30 accounts for betweenness centrality in Study 5, in descending order, were: @JackPosobiec, @Catturd2, @Conspiracyb0t, @Esotericexposal, @Archillect, @MrAndyNgo, @Caleidoscope11, @MoonChildWander, @AnonymousSage1, @SawmillTaters, @PrisonPlanet, @A2kTribal, @DDilldick, @occultb0t, @ZeroSum24, @HueyPilled, @CryptoKoba, @ghost5011, @HumanVibration, @Atensnut, @EvanAKilgore, @Rising_Serpent, @BronzeAgeMantis, @ZenOfTupac, @QckDaddyCoze, @OpticsPolice, @RhondaNight, @Quasarcasm47, @Blompf2020, and @TheHannSolo.

## 5.2   Network top items

Network top items are shown in Table S25. Far-right account @micro__benis was among top replied-to accounts; this account ranked at the 74[th] percentile for eigenvector centrality. Schoenberger-linked accounts constituted most top replied-to and top mentioned accounts overall. These including @FallingWaterz and @PacMan522, while @RoseModema had interacted with Schoenberger through his prior account, @NameMySock. Top accounts @PacMan522 and @RoseModema were further both linked to Cicada 3301/ARGs, along with @thehannsolo and @sirisysprime, @SnowWhite7IAm, @EsotericExposal, and @EmpressCortana.

Top hashtags reflected themes of ARGs including #TheGame23, #TG23, and #HiveMind23 (cf. Table S26) The salience of New Age/esoteric themes was likewise reflected in the appearance of accounts that had been selected for these properties being among network top items, including @TheHannSolo, and hashtags such as #LookingGlass and #SynchroMysticism.

## 5.3   Other details

Tweets by @Acio3301 circa 2/7/21 confirmed suspected connections among @Acio3301, @FaisalLazarus, and his associates; in these tweets, @Acio3301 tagged in @FaisalLazarus and acccounts like @Archang31, @11Llotus, @FallingWaterz, @PacMan522, and @SoulShaker007.

**Table S24**. Additional details and descriptions of notable accounts selected for Study 5.

| Account(s) | Brief description / Rationale for selection |
|---|---|
| 0queb2 | Observed to connect to _CEOofGenZ, SanandaEmanuel, bejay31688996, CryptoKoba, and others |
| jaw_81 | Interacted with a2ktribal, "groyper" accounts, QAnon influencers and esoteric accounts, and posting disturbing content (e.g., Jeffrey Dahmer) |
| a2ktribal | Featured a swastika with hashtags #PowerdbySatan (sic) and #666, interacted with far-right account @SawMillTaters |
| 0111kek0111 | Observed connected to multiple accounts interacting with @FaisalLazarus (including @TruMindInspired, @PepesGrandma, @CASunshineGal) as well as @ZeroSum24 and "groyper" accounts |
| Acio3301 | Account bearing a reference to Cicada ("3301") which posted about QAnon, Cicada, and other ARGs |
| Bejay31688996 | Selected for its connections to both @DanAuito and @MrDrewENT and its frequent presence in long tweet threads with QAnon influencers. This account presented a picture of a middle-aged man wearing a tuxedo and lacked hashtags or descriptors. |
| Caleidoscope11 | Account interacting frequently with both esoteric accounts and "groyper" accounts which described itself as "Pepe's Girlfriend" and featured a pepe meme |
| Chaos17488465 | Account observed to interact with many accounts noted for their posting content about Cicada3301 and other ARGs |
| ChaosFren | "Groyper" account shown to be influential in a pilot group account analysis |
| Imisstheoutdoo1, onelildoe, hb04920973, z3itg3ist | Accounts connected to one or more accounts suspected to be accelerationists or insurrectionists (e.g., 0queb2, MrDrewENT, bejay31688996, WhalesWarrior, OpArmageddon, ZeroSum24, Whos2Know1) |
| Moonchildwander | Account featuring Cyrillic characters in its banner interacting with white supremacist Richard Spencer, @RichardBSpencer. |
| pdf00587301 | Account connected to _CEOofGenZ which had high betweenness centrality in an analysis of _CEOofGenZ's connected accounts |
| Rhondanight | Observed to have conversed with _gHOST3301_, an account that interacted with FaisalLazarus, and to suspected insurrectionist account hb04920973 |
| Quasarcasm47 | Account observed connected to accounts promoting Cicada 3301 and/or other ARGs, as well as to rhondanight, who connected to _ghost3301_ |
| SawMillTaters | Far-right influencer account |
| wokefeiiow | Account observed to connect to AbuseTheTruth interacting with hb04920973 and esoteric influencer accounts |
| ZenOfTupac | Observed to be connected to _CEOofGenZ and showing the highest betweenness centrality after _CEOofGenZ |
| zkahronicnebula | Connected to many QAnon and esoteric influencers (@kabamur_taygeta), as well as @ImperatorTruth and @1crazy_toaster (associates of FaisalLazarus) |

Supplementary Material

**Table S25**. Top 30 accounts for eigenvector centrality for the network in Study 5; number gives rank of N = 824 accounts. Those marked with * had been suspended as of 2/9/21.

| Category | Criterial Categories Identified at Selection |
|---|---|
| QAnon | CryptoKoba* (3), KimRunner123* (14), Iakining (29) |
| Schoenberger associates | 11llotus (6), 22DubTrip333 (25) |
| Cicada 3301/ARGs | Quasarcasm47 (13) |
| New Age/Esoteric | EsotericExposal (7), Conspiracyb0t (21) |
| Far-Right | Caleidoscope11 (5) |
| Anonymous | 1reborn_* (15) |
| Conservative/MAGA | JackPosobiec (1), Catturd2 (2), Atensnut (8), Rising_serpent (9), BrandonStraka (26), SolMemes1 (12), MrAndyNgo (24), EvanAKilgore* (19) |
| Possible Insurrectionists | Shorty56167141 (4), Tweetweetbeyoch (10), 3d574685425 (11), Bean9986 (16), RondaGLarson (17), hb04920973* (18), Scrapakaharpazo* (22), RioDeJello (20), NikiWithTheStar (23), JC4Truthnumber2 (27), MaryOrTwerth (28), Candicemack123 (30) |

**Table S26**. Network top items for Study 5. Those included in the sample are marked with one asterisk (*).

| Top replied-to | Top mentioned | Top tweeters | Top hashtags |
|---|---|---|---|
| fallingwaterz* | wfinalle57 | reptoid_hunter* | #thegame23 |
| thehannsolo* | snowwhite7iam* | esotericexposal* | #zen |
| dualduels | blogjam_net | tweettruth2me* | #tg23 |
| jolianam11* | randthompson16 | listen4always* | #thepeoplesbridge |
| twinklecrooksh1 | yamwasher | lavenderlives* | #primenumber |
| sirisysprime* | ilomagyar | raeanon* | #hivemind23 |
| pacman522* | actondavid | hb04920973* | #symbolism |
| soulshaker007* | odilonross | empresscortana* | #lookingglass |
| micro__benis* | rosemodema | whiteoutgotu* | #synchromysticism |
| beedub53572007 | aughraobserves | pepesgrandma* | #backtothefuture |

# 6 Additional details across studies: Spontaneous usage of Russian or indication of Russian Federation affiliation

In addition to examples of Russian language disinformation account investigated and spontaneous Russian language usage discussed in the main text, we documented numerous other examples of presentation of Russian heritage and affiliation of accounts in our sample and those they interacted with. For instance, @RicoRoho, an account promoting the benefits of AI (including through a YouTube channel) and engaging with ARG accounts who occasionally used the #WWG1WGA hashtag and retweeted pepe memes, spontaneously tweeted a Russian-language media clip and Russian-language commentary, on January 14, 2021. Also, the account @Schumannbot, which displayed pictures of frequency information about the earth, several times displayed as having a URL ending in .ru. Further, Thomas Schoenberger's account @FaisalLazarus interacted in mutual Russian language conversation with @OfAspen – an account frequently included in conversational threads with other Schoenberger-linked accounts, including @11Llotus, @Archang31s, @PacMan522, and others. In February, 2021, Schoenberger-linked account @FallingWaters changed its main profile handle description (i.e., information displayed before the handle in each tweet) to display the Belarusian word for "love" in Cyrillic characters. Moreover, @Schumannbot, an account giving regular updates on the Earth's frequencies which was widely retweeted by QAnon and esoteric accounts, displayed URLs for its update ending in .ru on two occasions we detected. The first author was also "trolled" by an account called @Corvaxin8tor which used colloquial Russian and Putin memes when interacting with Twitter "friends"; the first author drew the troll's ire for her tweeting observations about two accounts, @TrumperWavin and @DrawAndStrike, promoting disinformation; however, the trolling. Accounts that @Corvaxin8tor was connected with included @Tuscon4Warren, which had recently been suspended and which interacted with Azerbaijani accounts, and @SatanicLulz, whose profile picture featured an upside-down pentagram who appeared to be involved in coordinating a "team" of other accounts around unknown activities, many of whom engaged positively (e.g., through likes) of non-ironic memes and content related to Satan posted by @SatanicLulz.

# 7 Supplemental study: Validation of account selections related to Cicada 3301 and #TheGame23 across Studies 4 and 5

Finally, we describe here supplemental analyses aimed at determining the relative importance of accounts selected in Studies 4 and 5 to one another, and the representativeness of these to the constructs of Cicada 3301 and #TheGame23, a notable ARG frequently posted about on Twitter with references to Discordianism. This was done by examining follower and followed account relationships for accounts matching Twitter searches related to these themes, then calculating eigenvector centrality metrics, as described below.

## 7.1 Methods and Materials

### 7.1.1 Software and tools

To identify followers and followed accounts for a target account, the following software was used: (i) Neo4j 4.2.2 w/APOC 4.2.0.1 & Graph Data Science Library 1.4.1; (ii) twarc 1.12.1; and (iii) python 3.8.5. All accounts were collected using the twarc command line tool by first calling 'friends' command to collect all the user ids followed by the 'users' command to gather individual account information. Custom commands were created for inserting these into Neo4j, notably "create constraint on (z:screen_name) ASSERT z.screen_name IS UNIQUE" and "call db.index.fulltext.createNodeIndex("BiosNThings", ["screen_name"], ["description", "location", "screen_name"])". Refer to [*Link*] for the source code of the python scripts used in this study which used Neo4j for data insertion and analysis.

This data insertion operation resulted in a total of $N = 39{,}506$ accounts with a corresponding $N = 49{,}547$ edges.

### 7.1.2 Account selection

The list of accounts selected for this analysis was derived from inspection of accounts @SIRISYSPrime, @Acio3301, @AnomalousChurch, @SheCait85_23, and @m1vr4, together with accounts identified through searches of the hashtag #TheGame23. This resulted in an initial list of $N = 56$ accounts on Jan. 17, 2021. An additional $N = 28$ accounts were added after searching the initial list of followers and following from the initial N = 56 accounts and looking for keywords "Tyler", "LetsCodeTyler", "Indra", "feecting", "00AG9603", and "Cicada23" in the bios, account names, and locations using Neo4J's full text search function on Jan. 19, 2021. This resulted in a total network of $N = 84$ accounts. These accounts were selected based upon researcher judgment of moderate-high aggregated participation levels in #TheGame23 from account activity around this theme. This included the account's discussion of media involving artwork (where "lead" accounts like @Acio3301 advanced the goal that others should make artwork as part of game activities) and related material mentioning or involved in #TheGame23. Judgment was based on recent posts showing a high level of participation in #TheGame23 and/or interacting with other accounts involved in #TheGame23.

## 7.2 Results

Using the functionality of the Graph Data Science Library in Neo4j, we performed eigenvector analysis with max normalization, implying that each account was given a score between 0 and 1 for importance in the network of N=84 accounts, limiting the results to the first 1000 results sorted by highest (1) to lowest (~0.12) score.

The account @RealEyeTheSpy, which has been argued by some disinformation researchers on Twitter to possibly have the same operator as @ETheFriend, showed a score of ~0.45 at rank 68. This account is a major influencer within QAnon, and it is of significant note that this account shows up within the top 100 eigenvector scores of TheGame23's following in the $N = 84$ network of accounts.

There are several accounts of interest that were directly mentioned in Study 4. These accounts are limited to ranking within the first 1,000 and sorted by highest to lowest: @Acio3301, with a score of ~0.59 and a rank of 32; @Conspiracyb0t with a score of ~0.49 and a rank of 53; @Schumannbot, with a score of ~0.39 and a rank of 87; @PacMan522, with a score of ~0.36 and a rank of 102; @Occultb0t, with a score of ~0.32 and a rank of 141; and @Archang3ls, with a score of ~0.14 and a rank of 743. There are a total of $N = 60$ accounts from Study 4's network of $N = 525$ accounts that appear in the following of thegame23's network of $N = 84$ accounts (i.e., 71% of TheGame23's network).

Moreover, there are several accounts of note that were directly mentioned in Study 5. These accounts are, limited to ranking within the first 1,000 sorted by highest to lowest: @Acio3301, with a score of ~0.59 and a rank of 32; @QuinnMichaels, with a score of ~0.50 and a rank of 48; @Conspiracyb0t, with a score of ~0.49 and a rank of 53; @PacMan522, with a score of ~0.36 and a rank of 102; @Occultb0t with a score of ~0.32 and a rank of 141; and @Archang3ls, with a score of ~0.14 and a rank of 743. There are a total of $N = 73$ accounts from Study 5's network of $N = 824$ accounts that appear in the following of TheGame23's network of $N = 84$ accounts (i.e., 87% of TheGame23's network).

## 7.3 Discussion

"TheGame23" appears to be based upon several key aspects of the occult involving Chaos Magick, including evidence that participation goals include the construction of "magickal" sigils and the creation of artwork. Chaos Magick can be described as a system of Magick in which personal power is derived from from beliefs in abilities to manipulate reality at the will of the "wizard". During investigations, the overall majority of materials and communications put out concerning TheGame23 appeared nonsensical in nature, which is consistent with a Discordian "world-view" (Mäkelä & Petsche, 2013); creating nonsensical content appeared to be an overarching goal of such materials and communications, due to the lack of coherency involved. We also investigated the contents of TheGame23 discord server; directives to the group were to make artwork, gain famous recognition, get a famous person to make TheGame23 artwork, and to make magickal sigils.

It is worth noting that TheGame23 network involves a large number of "Tyler" themed accounts, based upon the character Tyler Durden from the movie Fight Club. This character is related to the far-right (Bhatt, 2021). These accounts appear to follow the "human programming language" known as "feecting" created and described by Quinn Michaels (@QuinnMichaels); this is consistent with the fact that the hashtag #feecting is included in the bio along with behavior similar to other "feecting" directives located on the "indraai" Github account.

We speculate that creator of TheGame23 indeed may be @QuinnMichaels, who achieved a score of ~0.50 and rank of 48 in the $N = 84$ network of accounts. He is the creator of "feecting" as mentioned and referenced in relation to TheGame23 from the $N = 84$ network of accounts. From our current understanding of "feecting", it appears to be a list of directives that are intended to be followed by a human being in a structured format that resembles a programming language for computers. The Github account named "indraai"[4] has specifications and directives both for "feecting", as well as for the operation of several Twitter accounts, some of which were not included in the $N = 84$ network of accounts and analyses. Overall, usages of advanced social media and internet-based technologies to promote a Discordian worldview which "intentionally 'liquifies' the boundaries between the sacred and profane" (Mäkelä & Petsche, 2013: 411) through ARGs like TheGame23 as enabled via a "human programming language" appear to represent largely unexamined, unmitigated risks to the stability of societies.

---

[4] See https://github.com/indraai.

Formatting Notes and Other Corrections/Edits (Submitted to *Frontiers in Communication* on Oct. 14, 2021):

Edits to the publisher's typset version that were communicated to *Frontiers in Communication* have been made to enhance readability and/or to foster compliance with *arXiv* formatting; for instance, line numbers, publisher logo, and publisher's query annotations (Q1, Q2, etc.) have been removed in the present document. Line numbers appeared on pages as follows: p. 1 - Lines 1-114; p. 2 – Lines 115-228; p. 3 – Lines 229-342; p. 4 – Lines 343-456; p. 5 – Lines 457-570; p. 6 – Lines 517-684; p. 7 – Lines 685-798; p. 8 – Lines 799-912; p. 9 – Lines 913-1026; p. 10 – 1027-1140; p. 11 – Lines 1141-1254; p. 12 – Lines 1255-1368; p. 13 – Lines 1369-1482; p. 14 – Lines 1483-1596; p. 15 – Lines 1597-1710; p. 16 – Lines 1711-1824; p. 17 – Line 1825-1938; p. 18 – Lines 1939-2052; p. 19 – Lines 2053-2166; p. 20 – Lines 2167-2280; p. 21 – Lines 2218-2394; p. 22 – Lines 2395-2508; p. 23 – Lines 2509-2622. Queries appeared in margins on the following pages and approximate line locations: p. 1 – Q1: Line 13, Q2: Line 12, Q6: Line 40, Q7: Line 76, Q8: Line 106, Q9: Line 108; p. 5 – Q27: Line 497; p. 20 – Q10: Line 2197, Q11: Line 2240, Q12: Line 2247, Q13: Line 2254, Q14: Line 2262, Q16: Line 2276; p. 21 – Q17: Line 2287, Q18: Line 2321, Q19: Line 2326, Q20: Line 2359, Q21: 2382; p. 22 – Q22: Line 2412, Q23: Line 2473, Q24: Line 2483, Q25: Line 2502; p. 23 – Q26: Line 2522, Q15: Line 2573, Q28: Line 2582; remaining publisher's queries did not appear in margins.

Additional corrections/edits communicated to *Frontiers in Communication* in October, 2021 are as follows: p. 1, "4chan" has no space (throughout); p. 2, Line 120: Change date for Friedberg and Donovan to "forthcoming"; Line 303: Hyperlink Scott (2020); Line 324: replace "and groomed" with "and may have groomed"; Line 372: Add hyperlink to Sohn and Choi (2019); p. 4, Line 389: No italics on "inclusion criteria"; Line 414: replace "was" with "were"; Line 452: Replace "5k" with "5,000"; p. 5, Line 506: Replace "2020" with "in press"; p. 6, Line 606: Italicize "Number of followers"; Line 610: Italicize "Follower acquisition rate"; Line 663: Italicize "Follower:followed ratio"; Line 667: "8chan" has no space (throughout); Line 677: Italicize "Tweet activity"; Line 680: Italicize "Tweet rate"; p. 7, Line 685: Italicize "Eigenvector centrality"; Line 747: Hyperlink Supplementary Figure S1; Line 2085: Hyperlink Michael (2019); p. 20, Line 2171: Remove space before "["; Line 2172: remove extra semicolon, change "2018" to "2017", add space before /pol/; Line 2199: replace "innovation in" with "innovations for"; Lines 2242-2244: add "Code is available at https://github.com/DisinfoResearch."; References: see date format changes throughout; italicize only titles of books and names of journals and media outlets throughout; reverse order of Edel (2021b) and Edel (2021a); Friedberg and Donovan – add "Routledge: Abingdon-on-Thames, Oxfordshire, UK"; HBO (2021) – amend to "Hoback, C. (director) and McKay, A. (producer)" and add URL https://www.hbo.com/q-into-the-storm; Kristiansen – list of editors should read "J. Simonsen, C. Svabo, S. M. Strandvad, K. Samson, M. Hertzum, and O. E. Hansen"; switch order of Menn (2020b) and Menn (2020a); Scott (2020) – should read "Scott, K. (2020). "'Nothing up my sleeve': Information warfare and the magical mindset," in *Cyber Influence and Cognitive Threats*. Editors V. Benson and J. Mcalaney (Cambridge, MA: Academic Press), 53-76."